\documentclass{aa}

\usepackage{graphicx}
\usepackage[varg]{txfonts}
\usepackage{natbib,twoopt}
\bibpunct{(}{)}{;}{a}{}{,}             
 
\usepackage{color}
\usepackage{paralist}

\usepackage{epstopdf}
\makeatletter
 \newcommandtwoopt{\citeads}[3][][]{\href{http://adsabs.harvard.edu/abs/#3}%
    {\def\hyper@linkstart##1##2{}%
     \let\hyper@linkend\@empty\citealp[#1][#2]{#3}}}
  \newcommandtwoopt{\citepads}[3][][]{\href{http://adsabs.harvard.edu/abs/#3}%
    {\def\hyper@linkstart##1##2{}%
     \let\hyper@linkend\@empty\citep[#1][#2]{#3}}}
  \newcommandtwoopt{\citetads}[3][][]{\href{http://adsabs.harvard.edu/abs/#3}%
    {\def\hyper@linkstart##1##2{}%
     \let\hyper@linkend\@empty\citet[#1][#2]{#3}}}
  \newcommandtwoopt{\citeyearads}[3][][]%
    {\href{http://adsabs.harvard.edu/abs/#3}
    {\def\hyper@linkstart##1##2{}%
     \let\hyper@linkend\@empty\citeyear[#1][#2]{#3}}}
\makeatother

\newcommand{\zMUSE}{\ensuremath{z_{\mathrm{MUSE}}}}
\newcommand{\zphot}{\ensuremath{pz}}
\newcommand{\zBPZ}{\ensuremath{\zphot_{\mathrm{BPZ}}}}
\newcommand{\zEAZY}{\ensuremath{\zphot_{\mathrm{EAZY}}}}
\newcommand{\zBEAGLE}{\ensuremath{\zphot_{\mathrm{BEAGLE}}}}
\newcommand{\dzn}{\ensuremath{(\zMUSE-\zphot)/(1+\zMUSE)}}

\newcommand{\lya}{Ly-$\alpha$}
\newcommand{\lyb}{Ly-$\beta$}
\newcommand{\ha}{\ensuremath{\mathrm{H}\alpha}}
\newcommand{\hb}{\ensuremath{\mathrm{H}\beta}}
\newcommand{\oii}[1]{\ensuremath{\mbox{[\textsc{O ii}]}\lambda#1}}
\newcommand{\oiii}[1]{\ensuremath{\mbox{[\textsc{O iii}]}\lambda#1}}
\newcommand{\ciii}[1]{\ensuremath{\mbox{\textsc{C iii}]}\lambda#1}}
\newcommand{\nii}[1]{\ensuremath{\mbox{[\textsc{N ii}]}\lambda#1}}
\newcommand{\udft}{\textsf{udf-10}}
\newcommand{\mosaic}{\textsf{mosaic}}
\newcommand{\mhst}[1]{\ensuremath{\mathrm{m}_{\mathrm{#1}}}}
\newcommand{\sLC}{\ensuremath{s_{\mathrm{LC}}}}
\newcommand{\sLAF}{\ensuremath{s_{\mathrm{LAF}}}}

\newcommand{\median}{\ensuremath{\mathrm{median}}}

\begin{document}

\title{The MUSE Hubble Ultra Deep Field Survey:
III. Testing photometric redshifts to 30th magnitude.}
\author{J. Brinchmann \inst{1,2}\thanks{e-mail: \email{jarle@strw.leidenuniv.nl}} \and H. Inami\inst{3}%
\and R. Bacon \inst{3} \and T. Contini\inst{4} \and M. Maseda%
\inst{1} \and J. Chevallard\inst{5} \and N. Bouch{\'e}\inst{4}%
L. Boogaard\inst{1} \and%
M. Carollo\inst{6} \and S. Charlot\inst{7} \and%
W. Kollatschny\inst{8} \and R. A. Marino\inst{6} \and R.\ Pello\inst{4} \and J.%
Richard\inst{3} \and J. Schaye\inst{1} \and A. Verhamme\inst{3} \and L. Wisotzki\inst{9}}
   \institute{
     Leiden Observatory, Leiden University, P.O. Box 9513, 2300 RA
     Leiden, The Netherlands 
     \and 
     Instituto de Astrof{\'\i}sica e Ci{\^e}ncias do Espaço, Universidade do Porto, CAUP, Rua das Estrelas, PT4150-762 Porto, Portugal
   \and    Univ Lyon, Univ Lyon1, Ens de Lyon, CNRS, Centre de Recherche Astrophysique de Lyon UMR5574, F-69230, Saint-Genis-Laval, France
   \and Institut de Recherche en Astrophysique et Plan{\'e}tologie (IRAP), Universit{\'e} de Toulouse, CNRS, UPS, F-31400 Toulouse, France
   \and Scientific Support Office, Directorate of Science and Robotic
   Exploration, ESA/ESTEC, Keplerlaan 1, NL-2201 AZ Noordwijk, the
   Netherlands 
   \and ETH Zurich, Institute of Astronomy, Wolfgang-Pauli-Str. 27,
   CH-8093 Zurich, Switzerland
   \and Sorbonne Universit{\'e}s, UPMC-CNRS, UMR7095, Institut
   d'Astrophysique de Paris, F-75014 Paris, France
   \and Institut  f{\"u}r Astrophysik, Universit{\"a}t G{\"o}ttingen, Friedrich-Hund-Platz 1, D-37077 G{\"o}ttingen, Germany
   \and Leibniz-Institut f{\"u}r Astrophysik Potsdam (AIP), An der Sternwarte 16, D-14482 Potsdam, Germany
 }
\date{October 13, 2017}
\titlerunning{photo-zs in the UDF}
\abstract{We tested the performance of photometric redshifts for
  galaxies in the Hubble Ultra Deep field down to $30^{\mathrm{th}}$
  magnitude. We compared photometric redshift estimates from three
  spectral fitting codes from the literature (EAZY, BPZ and BEAGLE) to
  high quality redshifts for 1227 galaxies from the MUSE integral
  field spectrograph. All these codes can return photometric redshifts
  with bias $\left|\dzn\right|<0.05$ down to $\mhst{F775W}=30$ and
  spectroscopic incompleteness is unlikely to strongly modify this
  statement. We have, however, identified clear systematic biases in the
  determination of photometric redshifts: in the $0.4<z<1.5$ range,
  photometric redshifts are systematically biased low by as much as
  $\dzn=-0.04$ in the median, and at $z>3$ they are systematically
  biased high by up to $\dzn = 0.05$, an offset that can in part be
  explained by adjusting the amount of intergalactic absorption
  applied. In agreement with previous studies we find little
  difference in the performance of the different codes, but in
  contrast to those we find that adding extensive ground-based and
  IRAC photometry actually can worsen photo-z performance for faint
  gaalxies. We find an outlier fraction, defined through
  $\left|\dzn\right| > 0.15$, of 8\% for BPZ and 10\% for EAZY and
  BEAGLE, and show explicitly that this is a strong function of
  magnitude. While this outlier fraction is high relative to numbers
  presented in the literature for brighter galaxies, they are very
  comparable to literature results when the depth of the data is taken
  into account. Finally, we demonstrate that while a redshift might be
  of high confidence, the association of a spectrum to the photometric
  object can be very uncertain and lead to a contamination of a few
  percent in spectroscopic training samples that do not show up as
  catastrophic outliers, a problem that must be tackled in order to
  have sufficiently accurate photometric redshifts for future
  cosmological surveys.}
\maketitle

\section{Introduction}
\label{sec:introduction}

Deep multi-wavelength imaging of the sky has provided a tremendous
amount of information on galaxies in the distant Universe since the
Hubble Deep Field North was published
\citep{Williams:TheAstronomicalJournal:1996}. The tools to efficiently
and accurately exploit these data by fitting their spectral energy
distributions 
have also evolved in step and have now reached a fairly high level of maturity \citep[see][for
reviews]{Conroy:AnnualReviewOfAstronomyAndAstrophysics:2013,Walcher:AstrophysicsAndSpaceScience:2010}. 

The development of photometric redshift (photo-z) estimation
techniques has been particularly notable. The number of objects with
photometric information is much larger than can be efficiently
followed-up spectroscopically and this means that large-scale
multi-band surveys of the sky have to rely on photo-zs to
determine distances to galaxies. This central role for photo-zs has
also led to the development of a wide range of techniques for photo-z
estimation. These fall basically into two categories: machine learning
techniques which aim to empirically determine the map between colours
and redshift, and the template fitting techniques which take a set of
physically motivated spectral energy distributions (SEDs) and find the
best match of a (combination of) these SEDs to the data. An up-to-date
overview of the photo-z methods can be found in the introduction of 
\citet{Sadeh:PublicationsOfTheAstronomicalSocietyOfThe:2016} and more
extensive comparisons of codes can be found in for instance
\citet{Hildebrandt:AstronomyAndAstrophysics:2010,Abdalla:MonthlyNoticesOfTheRoyalAstronomicalSociety:2011,Acquaviva:TheAstrophysicalJournal:2015}. 

The requirements on photometric redshifts from cosmological weak
lensing survey such as the Kilo Degree Survey (KiDS; Hildebrandt et
al. 2016)\nocite{Hildebrandt:Kids450CosmologicalParameterConstraintsFromTomographicWeak:2016},
the Dark Energy Survey (DES; the Dark Energy Coallaboration
2005)\nocite{Collaboration:EprintArxivAstroPh0510346:2005}, and the
Large Synoptic Survey Telescope \citep[LSST,][]{12288492} ground-based
surveys, and the Euclid
\citep{Laureijs:EuclidDefinitionStudyReport:2011} and Wide-Field
InfraRed Survey Telescope
\citep[WFIRST,][]{Spergel:WideFieldInfrarredSurveyTelescopeAstrophysicsFocusedTelescopeAssets:2015}
space missions, are stringent. The requirement on individual redshift
of $\sigma_{\zphot} < 0.05 (1+z)$ is non-trivial but this requirement is not the
biggest challenge.  The preferential way to carry out the weak lensing
surveys is to do this in redshift bins which leads to strict
constraints on the accuracy of the mean redshift in each bin. In the
case of future surveys, the mean redshift must be constrained to
better than $2\times 10^{-3}\ (1+z)$ which is a very challenging
requirement for future surveys
\citep[e.g.][]{Newman:AstroparticlePhysics:2015}.

As a consequence of these needs, several studies have explored the
performance of different photo-z codes on data appropriate for
cosmological studies
\citep[e.g.][]{Hildebrandt:AstronomyAstrophysics:2008,Hildebrandt:AstronomyAndAstrophysics:2010,Abdalla:MonthlyNoticesOfTheRoyalAstronomicalSociety:2011,Bonnett:PhysicalReviewD:2016,Beck:EprintArxiv170108748:2017}. These
studies typically find that the required constraints on individual
photo-z estimates of $\sigma_{\zphot} < 0.05 (1+z)$ is an achievable
goal for the large missions. The more stringent constraint
\citep[e.g.][]{Zhan:JournalOfCosmologyAndAstroParticlePhysics:2006} is
however on the bias of the mean redshifts in a particular redshift
bin, which must be $<2\times 10^{-3} (1+z)$ to reach the goals of the
upcoming surveys.

The majority of the photometric redshift tests have focused on
relatively bright galaxies ($i_{\mathrm{AB}}$ or $r_{\mathrm{AB}}<24$)
since those are the galaxies targeted by weak lensing surveys, but
also because this is typically the magnitude limit to which most
spectroscopic surveys target galaxies. Among the deepest comparisons
to date is the study by \citet{Dahlen:TheAstrophysicalJournal:2013} of
photo-zs in the Cosmic Assembly Near-infrared Deep Extragalactic
Legacy Survey (CANDELS) which focuses on the GOODS-S field, the PHAT1
photo-z accuracy test
by~\citet{Hildebrandt:AstronomyAndAstrophysics:2010} which uses data
from the GOODS-N field. Both of these studies tested multiple
photometric redshift codes on deep HST data, with spectroscopic
reference samples mostly extending to $\mhst{F160W}=24$ with a small
tail extending to fainter magnitudes. The more recent study by Dahlen
et al finds a mean bias of
$\left\langle
  (z_{\mathrm{spec}}-z_{\mathrm{phot}})/(1+z_{\mathrm{spec}})\right\rangle
= -0.008$ and an outlier fraction of 2.9\% when combining all photo-z
considered, while the rates for the best indicator in the PHAT1 test
gave a bias of $0.009$ and an outlier fraction of 4.5\% for the $R<24$
subsample. While these quantities vary from study to study and between
photometric redshift estimators, outlier fractions well below 10\% and
biases $<0.01$ are frequently seen. In these cases, it is reasonable
to use photometric redshifts to study trends of the galaxy population
as the mean predicted properties are not likely to be strongly
influenced by the errors in the photo-zs. However, it is worth noting
that the results in \citet{Hildebrandt:AstronomyAndAstrophysics:2010}
clearly improve when limiting the study to $R<24$, and
\citet{Dahlen:TheAstrophysicalJournal:2013} show a clearly degraded
performance of photometric redshifts when artifically dimming their
spec-z sample, but as this was not a real test against spectra of
faint objects, its relevance to real data is harder to assess.

The dependence of photo-z performance on magnitude,
  also primarily down to $JH_{\mathrm{IR}}=24$, was explored in detail
  by \citet{Bezanson:TheAstrophysicalJournal:2016}. Those authors
  compared 3D-HST grism redshifts
  \citep{Momcheva:TheAstrophysicalJournalSupplementSeries:2016} to
  photometric redshifts from
  \citet{Skelton:TheAstrophysicalJournalSupplementSeries:2014}. This
  is a different kind of test since grism redshifts can vary
  significantly in accuracy and also depend on the photometric
  information as discussed in
  \citeauthor{Bezanson:TheAstrophysicalJournal:2016}, but a clear
  advantage is that the grism redshifts are available in a fairly
  unbiased way across the galaxy population. They focus on the scatter
  in the redshift comparison as well as the outlier fraction, and find
  comparable outlier fractions (1.9--4.9\%) to the studies cited
  above. They also show that the the scatter between photometric and
  grism redshifts increases strongly towards fainter magnitudes and
  were able to use the test of photo-z accuracy using galaxy pairs
  developed by \citet{Quadri:TheAstrophysicalJournal:2010} to show
  that this extends to $\mhst{F160W}=26$.

The decreasing performance of photo-zs at faint
  magnitudes highlights an important point that is well-known but
  often not stated explicitly: the quality of photometric redshifts
  depends on the quality of the photometry. When it is stated that
  photo-zs are improved by adding filter X, then it is implicitly
  assumed that the photometric quality of that band is high. When this
  is not the case, adding this band might in fact decrease photo-z
  performance. While not a key point of the present paper, we discuss
  this in the context of adding IRAC photometry in
  Appendix~\ref{sec:adding-ground-based} where we will show that for
  faint galaxies adding IRAC photometry worsens photo-z performance.

In contrast to these preceding papers, here we focus on the
performance of photometric redshifts for faint galaxies which have
spectroscopic redshifts from the Multi-Unit Spectroscopic Explorer
(MUSE) instrument \citep{bacon2010muse}. Since we are exploring to
fainter magnitudes than normally used for machine learning techniques
for photo-z estimation we ignore these completely \citep[see][for a
discussion]{Sadeh:PublicationsOfTheAstronomicalSocietyOfThe:2016}. We
also do not explore the full range of template fitting methods, thus
widely used codes, such as Le PHARE
\citep{Arnouts:MonthlyNoticesOfTheRoyalAstronomicalSociety:1999,OIlbert:AstronomyAstrophysics:2006},
ZEBRA
\citet{Feldmann:MonthlyNoticesOfTheRoyalAstronomicalSociety:2006}, and
HyperZ \citep{Bolzonella:AstronomyAndAstrophysics:2000} are not
discussed further. This is because our aim is somewhat different from
the recent tests of photo-zs for cosmological surveys. We wish to
present a first exploration of the performance of the codes at
magnitudes $\mhst{F775W} >24$ so we are limiting our attention to
codes that have already been run on the \citet[][R15
hereafter]{Rafelski:TheAstronomicalJournal:2015} catalogue which was
the starting point for our source extractions. This amounts to two
established template fitting codes: BPZ
\citep{Benitez:TheAstrophysicalJournal:2000} and EAZY
\citep{Brammer:TheAstrophysicalJournal:2008}, as well as the new,
fully Bayesian fitting code BEAGLE
\citep{Chevallard:MonthlyNoticesOfTheRoyalAstronomicalSociety:2016}.
The advantage of template fitting codes over the machine learning
methods that might be preferred for cosmological applications
\citep[e.g.][]{Newman:AstroparticlePhysics:2015}, is that they provide
the user with the possibility to also extract physical parameters for
the galaxies in question. Indeed, the main difference between BEAGLE
relative to BPZ and EAZY is the fact that it was optimised for
physical parameter estimation rather than photometric redshifts ---
making a comparison of the different codes particularly interesting.

In Section~\ref{sec:data} below we provide a brief discussion of the
data used in the study. The reduction and redshift analysis of these
are described in detail in \citet[][, Paper II
hereafter]{Inami2017}. In Appendix~\ref{sec:adding-ground-based} we
justify our focus on the 11 band HST-only catalogue from R15 by
showing that for our faint galaxies, the use of the 11 HST bands only
results in better photo-zs than does the use of 44 bands from
\citet{Skelton:TheAstrophysicalJournalSupplementSeries:2014}.  In
Section~\ref{sec:comp-phot-redsh} we compare our spectroscopic
redshifts to the photometric ones. We will find that the EAZY photo-zs
in the R15 catalogue are surprisingly discrepant and re-calibrate
these in section~\ref{sec:reduc-bias-phot}. The nature of the galaxies
for which the photometric and spectroscopic redshifts are clearly
discrepant is discussed in Section~\ref{sec:redshift-outliers} with
some complementary information in
Appendix~\ref{sec:serial-outliers}. The effect of redshift
incompleteness is discussed in section~\ref{sec:redsh-meas}. We
explore the impact of object superpositions on photometric redshift
estimators, including machine learning ones, in
Section~\ref{sec:impact-galaxy-superp}. We discuss our results and
conclude in Sections~\ref{sec:discussion} and~\ref{sec:conclusions}.

\section{Data}
\label{sec:data}

The data used in the present paper come from MUSE Guaranteed Time
Observing (GTO) observations of a 3'x3' field of the Hubble Ultra Deep
Field (UDF, Beckwith et
al. 2006)\nocite{Beckwith:TheAstronomicalJournal:2006}. This amounts
to a total integration time of 116 hours in the autumns of 2014 and
2015.  The details of the survey strategy and data reduction are given
in \citet[][, Paper I hereafter]{Bacon2017}. For the
  present paper the most imporant aspect of the observations is that
  they were carried out to two different depths: a
  $3\arcmin\times 3\arcmin$ medium deep field, \mosaic, with an
  effective integration time of approximately $10\,$hours, and nested within
  this a $1\arcmin \times 1\arcmin$ ultra-deep field, \udft, which
  contains data with a total of $\approx 31$ hours of integration
  time. The data reduction followed broadly the process used for
the Hubble Deep Field South (HDF-S) observations described in
\citet{Bacon:AstronomyAndAstrophysics:2015} with several
enhancements. The main improvements were to the self-calibration
procedure which now uses a polychromatic correction and works directly
on the pixtables output by the MUSE data reduction pipeline
\citet[DRS,][]{Weilbacher:DesignAndCapabilitiesOfTheMuseData:2012}. Inter-stack
defects were removed by using a bad pixel mask projected onto the 3D
cube, see Paper I for details, and a realistic variance cube was
derived as described there. The resulting data cube is considerably
better calibrated spectrally and spatially than the one produced for
the HDF-S used in \citet{Bacon:AstronomyAndAstrophysics:2015}.

We will also make use of the photometric information from the R15
catalogue. This has photometric measurements in up to 11 bands. F225W,
F275W, and F336W from \citet{Teplitz:TheAstronomicalJournal:2013},
F435W, F606W, F775W, and F850LP from
\citep{Beckwith:TheAstronomicalJournal:2006} and finally F105W, F125W, F140W, and
F160W mostly from the UDF09 and UDF12 programs
\citep{Bouwens:TheAstrophysicalJournal:2011,Oesch:TheAstrophysicalJournal:2010,Koekemoer:TheAstrophysicalJournalSupplementSeries:2013,Ellis:TheAstrophysicalJournalLetters:2013}
with shallower F105W, F125W, and F160W data from
\citet{Koekemoer:TheAstrophysicalJournalSupplementSeries:2011} and
\citet{Grogin:TheAstrophysicalJournalSupplementSeries:2011}. In
addition to providing photometry, R15 also ran the BPZ and EAZY
photo-z codes on their catalogue. We here use the results as reported
by R15, the interested reader can consult their paper for details of
how the codes were run and the templates used. We will refer to the
IDs from this paper as for example RAF 4471, where the number is the ID
number in the R15 catalogue.

We do not seek to add further photometric information
  to the R15 catalogue. The comprehensive compilation of photometry
  for 3D-HST by \citet[][S14
  hereafter]{Skelton:TheAstrophysicalJournalSupplementSeries:2014}
  does provide a total of 44 bands for our field, but it has
  photometry for fewer objects with \zMUSE\ than the R15 catalogue
  (see Figure~\ref{fig:s14_vs_r15_color_comp} in
  Appendix~\ref{sec:adding-ground-based}), and it was not used for the
  object definition and spectrum extraction. This means that the
  association of spectroscopic redshifts with photometric object is
  less secure. Finally, in Appendix~\ref{sec:adding-ground-based}
  we show that although the large number of bands might lead to better
  performing photo-zs at bright magnitudes and an overall smaller
  bias, the S14 catalogue leads to a higher number of outliers at
  faint magnitudes than the R15 catalogue. We therefore do not use
  these data here, although we will return to discuss the performance
  of EAZY run on these data in section~\ref{sec:discussion}.

\subsection{Object definitions and spectrum extraction}
\label{sec:object-defin-spectr}

Paper I explains the method for object detection and the spectrum
extraction in described in detail in Paper II. In short we use two
distinct approaches to define objects. The first takes the existing
segmentation map for the HST catalogue of \citet[][R15
hereafter]{Rafelski:TheAstronomicalJournal:2015} and convolves this
with the MUSE Point Spread Function (PSF) to get a segmentation map
suitable for MUSE. Following Paper II we refer to
  these as continuum selected sources. The second approach uses the
matched filter detection method ORIGIN (Mary et al in prep., see paper
II for an overview of the process) to find emission line sources in
the cube and we will refer to these as emission line
  selected objects. These two approaches provide partially
overlapping object lists and these have been consolidated
by inspection case-by-case.

Both approaches produce a mask defining the spatial
  extent of a source and the resulting object masks have then been
used to extract spectra using both a straight sum and a weighted
extraction.  The redshifts used in the present paper were obtained
from the higher signal-to-noise weighted
extractions. For objects with a full-width
  half-maximum size $>0\farcs 7$ in the HST F775W image the
  white-light image of the object was used as a weight, while for
  smaller objects the estimated PSF as a function of wavelength was
  used as a weight. The process of spectrum extraction is described
  further in Paper II but has no impact on the results presented
  here.

\subsection{Redshift determinations}
\label{sec:z-determinations}

The process of redshift determination is detailed in
  Paper II, but it is important for the present paper to summarise the
  steps. The redshift determination for the continuum selected objects
  was done in a semi-automatic manner using a
  modified version of the MARZ redshift determination software 
  \citep{Hinton:AstronomyAndComputing:2016}. The software provides
  redshift estimates using cross-correlation with a set of
  templates. The redshift solutions are visually inspected by at least
  three researchers. The inspection step looks both at the 1D
  spectrum, as well as narrow-band images over the putative spectral
  features. For the emission line selected objects the main challenge
  is to identify which line is detected and this is done by two
  researchers for both the \udft\ and the \mosaic.

For each redshift determination we assigned a confidence,
  where confidence 3 corresponds to a secure redshift, determined by
  multiple features, confidence 2 to a secure redshift, determined by
  a single feature (frequently \lya\ where the asymmetry of the line
  in obvious), and confidence 1 are considered possible redshifts
  which are determined by a single feature with uncertain
  identification. We will almost exclusively show results based on
  spectra with confidence $\ge 2$.

For the present paper the number of sources is dominated by the 
UDF \mosaic, but we also use the deeper \udft\
data. The process to determine redshifts was slightly
  different in these two fields. In the \udft\ the spectra of all
  objects from the R15 catalogue were inspected visually to attempt to
  determine redshifts, whereas in the UDF \mosaic\ only objects with
  $\mhst{F775W}<27$ were inspected visually. The ORIGIN code was run
  in both fields. This mixture of selection criteria complicates the
  selection function of the sample but this does in general not affect
  our results significantly. One consequence is however the fact that
 at $\mhst{F775W}>27$, we have mostly included objects that have been
detected by blind emission line detection codes directly in the cube
with no recourse to the HST images (see Paper I for details).  The
exception is the \udft\ and a few `split' objects (see Paper II for
details). 

One aspect of this process is important to underline
  for this paper: the redshifts were initially determined without any
  knowledge of the photometric redshift of the objects. The HST images
  of the source were used for instance to help the identification of
  lines but not beyond that. In a later update to the catalogue in
  Paper II photo-zs were occasionally consulted, although they were
  not directly used to determine redshifts. To avoid any possible
  biases caused by this we do not use these updated redshifts
  here. The catalogue used here therefore differs
  slightly from that of Paper II.

In order to compare the results below with other
  results in the literature, it is also important to have an
  understanding of the type of objects for which we have
  redshifts. Since MUSE obtains spectra without photometric
  pre-selection, we are not biased towards continuum bright objects,
  and a consequence of this is that most redshifts are determined on
  the basis of emission lines. This selection leads to a different mix
  of spectral types than commonly seen in magnitude limited
  spectroscopic surveys.

\begin{figure}
  \centering
  \includegraphics[width=84mm]{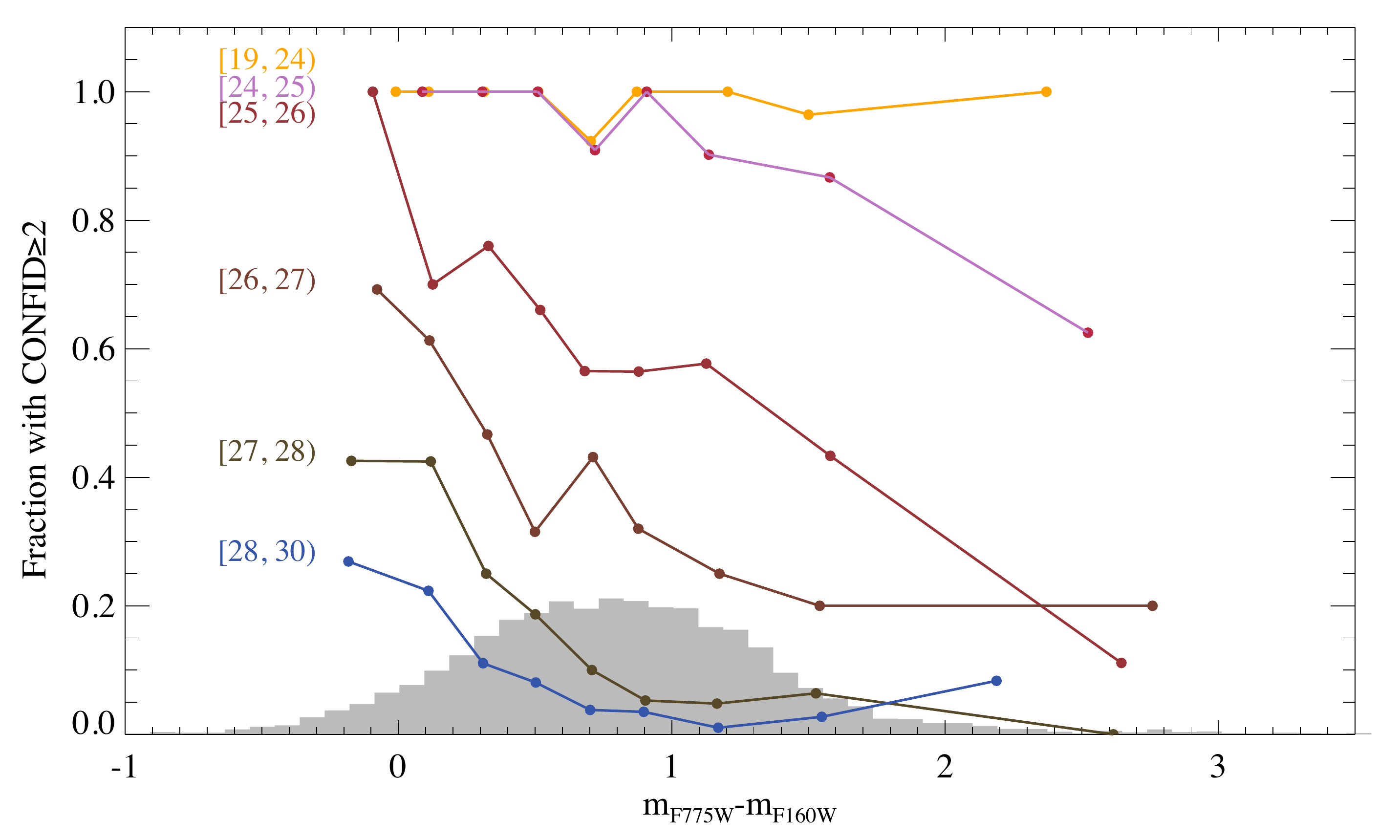}
  \caption{Spectroscopic completeness as a
      function of observed $\mhst{F775W}-\mhst{F160W}$ colour in bins
      of $\mhst{F775W}$ as indicated on the left. We only include
      galaxies with $\mathrm{CONFID}\ge 2$. The bins were chosen to
      span the distribution in colour and the number of objects
      per bin varies strongly. This can be inferred from the
      grey-scale histogram which shows the number of objects as a
      function of colour, summed over all magnitudes down to
      $\mhst{F775W}=30$. A passive galaxy at $z>1.5$ typically has
      $\mhst{F775W}-\mhst{F160W} > 2$.}
  \label{fig:completeness_vs_colour}
\end{figure}

The catalogue discussed in Paper II has 1,329 spectra with redshift
$>0$ and a confidence $\ge 2$. Out of these 1,329 redshifts only 63
(4.5\%) are determined purely on the basis of absorption lines, with
the faintest such galaxy having $\mhst{F775}=26.2$. This is however a
strong function of magnitude, reaching $\approx 20$\% near
$\mhst{F775W}=24$. It should also be noted that most of these
absorption line galaxies are expected to be star-forming as
they are bright in the UV; there are only 11 sources with absorption
line redshifts at $z<1.5$ that are likely to be passive galaxies, most
of which are at $\mhst{F775W}<22$.  The remainder of the spectra have
one or more strong emission lines. The majority of redshifts are
determined on the basis of \lya\ and \oii{3727}, with most
$\mhst{F775W}>27$ sources with secure redshifts being \lya-emitters. A
detailed breakdown of the types of emission lines sources is given in
Table 2 in Paper II but for the present paper this is of little
relevance.

The overall redshift completeness is $>50$\% at $\mhst{F775W}<25.5$
(Paper II) but fainter than this we lose absorption line galaxies. In
Figure~\ref{fig:completeness_vs_colour} we show the spectroscopic
completeness as a function of observed $\mhst{F775W}-\mhst{F160W}$
colour where we have only included galaxies with CONFID$\ge 2$. To put
this in context, a single stellar population from \citet{Bruzual2003}
with $>20$\% solar metallicity forming at $z=10$ will have
$\mhst{F775W}-\mhst{F160W} > 2$ for $z>1.5$. Thus the grey histogram
that is inset in the figure which shows the number of objects in the
R15 catalogue at each colour, shows that very few truly red sources
are in the parent catalogue. Each line shows the completeness in bins
of $\mhst{F775W}$ as indicated by the labels on the left of the
lines. It is clear that down to $\mhst{F775W}=24$ our spectroscopic
sample is complete but going fainter the incompleteness increases
with a clear colour-dependence.  The consequence for the following is
that we will mostly test the performance of photometric redshifts on
star forming galaxies and we will not be able to say anything about
the performance of the codes on passive or very dusty galaxies. We
explore the effects of the spectroscopic incompleteness in
section~\ref{sec:redsh-meas} below.

\subsection{Resolving blends and final sample definition}
\label{sec:resolv-blends-final}

In the present paper our focus is on determining the redshifts of
objects detected in the HST images to be able to compare with the
photometric redshifts of these sources. The procedure outlined above
is not specifically designed for this as the object
  identification step convolves the HST detection mask with the MUSE
  PSF. Hence, although we base ourselves on the catalogue in Paper
II, we spent considerable care on establishing the association with
objects in the HST images defined in the R15 catalogue.

\begin{figure*}
  \centering
  \includegraphics[width=184mm]{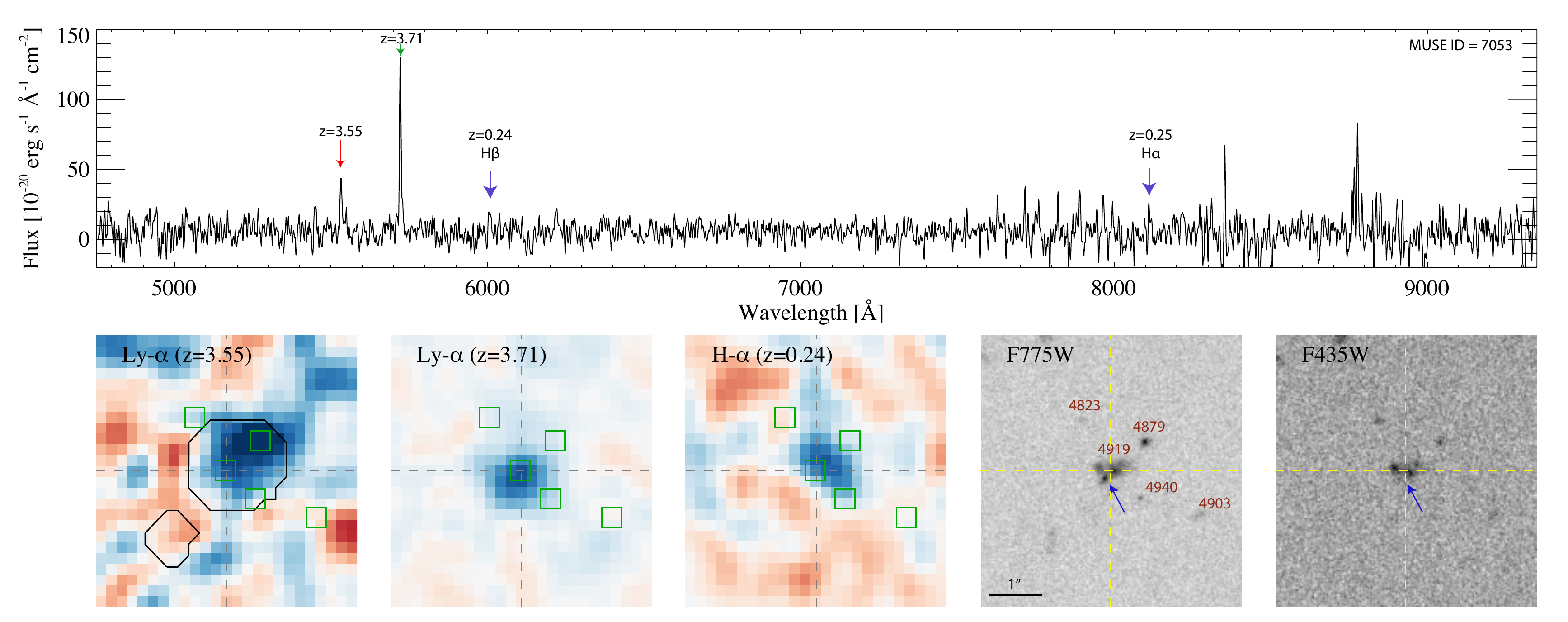}
  \caption{Spectrum and narrow band images of the main spectral
    features of MUSE ID 7053. The top panel shows the spectrum,
    smoothed with a 2.5\AA\ Gaussian. There are spectral features from
    three distinct galaxies in this spectrum, the locations of which
    are indicated by the small arrows. The bluest feature is a \lya\
    line from RAF 4879 (MUSE ID 3052), at z=3.55. The narrow-band
    image over this feature, smoothed with a 3-pixel FWHM Gaussian and
    with side-bands subtracted off, is shown in the lower left
    panel. This panel also shows the segmentation map for this object
    as the black contours. This and the other images are 5\arcsec\ on
    the side. The green squares show the locations of the objects in
    the R15 catalogue that are labelled in the F775W image. The next
    feature is a \lya\ line at $z=3.71$ (MUSE ID 7386) whose
    narrow-band image is shown in the second panel from the left in
    the bottom row. Finally the main bulk galaxy in the image has weak
    \ha\ and \hb\ at $z=0.24$. The combined narrow-band image of \ha\
    and \hb\ is shown in the middle panel on the bottom row. The last
    two panels on the bottom row shows the HST F775W and F435W images
    with an asinh scale. The object most likely associated to the
    $z=3.71$ \lya\ is indicated by the blue arrow. The central four
    objects from the R15 catalogue are labelled with their IDs in the
    R15 catalogue in the F775W image and the cross-hairs in each of
    the lower panels crosses at the same spatial position to help
    compare different panels.}
  \label{fig:7053-example}
\end{figure*}

While the association of spectroscopic features to a particular
photometric object is in most cases straightforward, in a small subset
this is not so. These have been inspected in detail, using the HST
imaging to help resolve the assocation in a number of cases. To do
this, a combination of the centroid of the narrow band images over the
main spectroscopic features in a spectrum and the visibility of the
object in different HST bands was used.

Figure~\ref{fig:7053-example} shows an example of this process, see
also Paper II for more discussion of the redshift determination
process. In a first iteration two \lya\ lines were identified based on
their characteristic line shapes and coherent narrow-band images with
clearly different centroids. The source mask was adjusted to create
two MUSE objects (MUSE IDs 3052 and 7053 at $z=3.71$ and
$z=3.55$). However the association of a $z=3.71$ redshift yet a very
clear detection in F435W was problematic, while for $z=3.55$ it is
still feasible. Subsequent examination then
showed that the \lya\ at $z=3.71$ appears to be associated with a
small point source just below the central galaxy (see the blue arrows
in the HST images in Figure~\ref{fig:7053-example}) which disappears
between F775W and F435W. Re-examination of the spectrum also showed
tentative \ha\ and \hb\ at $z=0.24$. In this case, then, the single
HST object RAF 4919 is a blend of a $z=3.71$ \lya-emitter and a
star-forming $z=0.24$ galaxy. These are not easily separated at
ground-based resolution, although their narrow-band image centroids are
slightly offset (second and third image from the left in the lower
panel). The segmentation map is shown as the black

\begin{figure*}
  \centering
  \includegraphics[width=184mm]{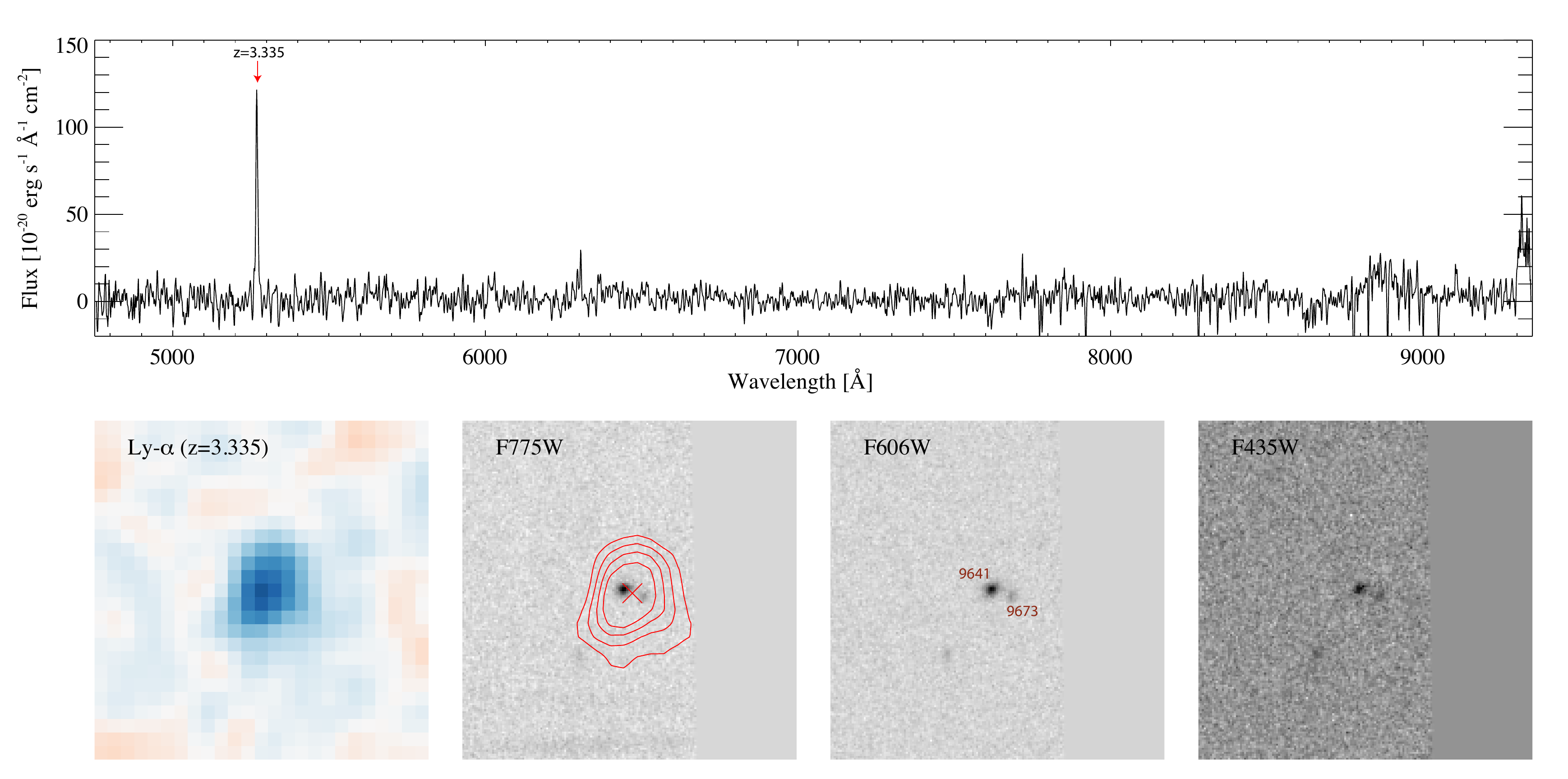}
  \caption{MUSE object \#2277. The top panel shows the spectrum which
    shows a clear \lya\ line at $z=3.335$. The narrow-band image over
    this line is shown in the bottom left. The following three images
    in the bottom row shows the HST images in F775W, F606W and
    F435W. The \lya\ narrow-band image is overlaid as contours in the
    F775W image with contours at approximate S/N per pixel of 2, 3, 4
    and 5 shown and the centroid of the narrow-band image indicated by
    a cross. The two central HST objects from the R15 catalogue are
    labelled in the F606W image and it is clear that the narrow-band
    image is extending across both objects. A comparison of the F775W
    and F435W images also shows that these objects have
    similar colours. }
  \label{fig:2277-example}
\end{figure*}

In most cases this careful examination leads to a unique association
of an emission line to an HST object, but for a total of 58 MUSE
objects this has proven impossible with the existing data so they have
multiple RAF IDs associated to them. For these objects we have a
redshift for the spectrum but we are unable to ascertain which of the
HST objects the spectroscopic feature pertains to. An example of the
latter case is MUSE ID 2277, shown in
Figure~\ref{fig:2277-example}. The MUSE spectrum shows a strong \lya\
line but there is no easy way to determine which object this line
belongs to (indeed it could be associated with both). We refer to
these as blended objects, and they are, unless otherwise stated,
excluded from the plots that follow. However, it is important to
realise that in a study without the exquisite HST images available in
the UDF, these subtleties might not be noticed. This has important
consequences for the accuracy with which we can assign a redshift to
an object and we explore this further in
Section~\ref{sec:impact-galaxy-superp} below.

%
%
The end product of this procedure is a catalogue with 182 objects with
spectroscopic redshift confidence (defined in Paper II) of 1, 602 with
confidence 2 and 557 with confidence 3 (the most secure redshift),
adding up to a total of 1341 HST objects with redshifts. This can be
contrasted with the R15 catalogue in the same spatial region which has
a total of 6362 objects brigther than $\mhst{F775W}=30$ and 1181
objects with $\mhst{F775W}<27$. There are in total 126 HST objects for
where the redshift is clear but the association of the redshift to the
object is impossible due to blending, corresponding to 58 MUSE
sources. This can be compared with the 160 spectroscopic redshifts
known in this area before --- an increase of a factor of 8.2.

\section{Comparison to photometric redshift estimates}
\label{sec:comp-phot-redsh}

We have opted to focus our comparison on photometric redshift
estimates for the UDF data from the recent literature. R15 provide
photometric redshift estimates for their detected objects, one using
the Bayesian Photometric redshift code BPZ
\citep{Benitez:TheAstrophysicalJournal:2000} and one using the EAZY
code described by \citet{Brammer:TheAstrophysicalJournal:2008}, in
addition to these, we will also test the recent photo-z estimates
from the BEAGLE code
\citep{Chevallard:MonthlyNoticesOfTheRoyalAstronomicalSociety:2016}. We
have verified that the results below are also found when using
the photo-zs from S14 which we discuss in more detail
  in Appendix~\ref{sec:adding-ground-based}.

The main difference from previous comparisons in the GOODS-S area is
that we now can test these predictions outside the magnitude range
where they have been validated earlier.  As we will show shortly, the
base photometric redshift estimates in the R15 catalogue and from
BEALGE are systematically biased at faint magnitudes/high redshift,
and the EAZY estimates in particular are significantly off. However we
will also see that we can improve the situation for EAZY considerably
by adopting other spectral templates.

\subsection{Sample definitions}
\label{sec:sample-definitions}

We are interested in characterising how well photometric redshifts
work, including at very faint limits. As outlined above, this presents
some challenges for the association between spectroscopic redshift and
photometric counter-part. Firstly we will only consider galaxies for
which we have a clear HST counterpart, so we require that they are not
flagged as blended.  Our spectroscopic redshift determinations are
considerably less accurate for galaxies with confidence 1. In the
main, therefore, we will limit our studies to galaxies with
spectroscopic redshift confidence $\ge 2$. With these cuts we are
confident that the spectroscopic redshifts are correct and we expect
that any discrepancy with photometric redshifts is due to an incorrect
photometric redshift estimate. We will examine this assumption and the
impact of relaxing the confidence requirement in the next section
when we discuss the results.

We will adopt a notation throughout where we denote photometric
redshifts as \zphot\ and spectroscopic redshifts from MUSE as
\zMUSE. The difference between the spectroscopic and photometric
redshift is denoted $\Delta z$ and is defined as
\begin{equation}
  \label{eq:1}
  \Delta z = \zMUSE-\zphot
\end{equation}
throughout, and we will frequently follow the literature and normalise
this by $1+\zMUSE$.

\begin{figure*}
  \centering
  \includegraphics[width=184mm]{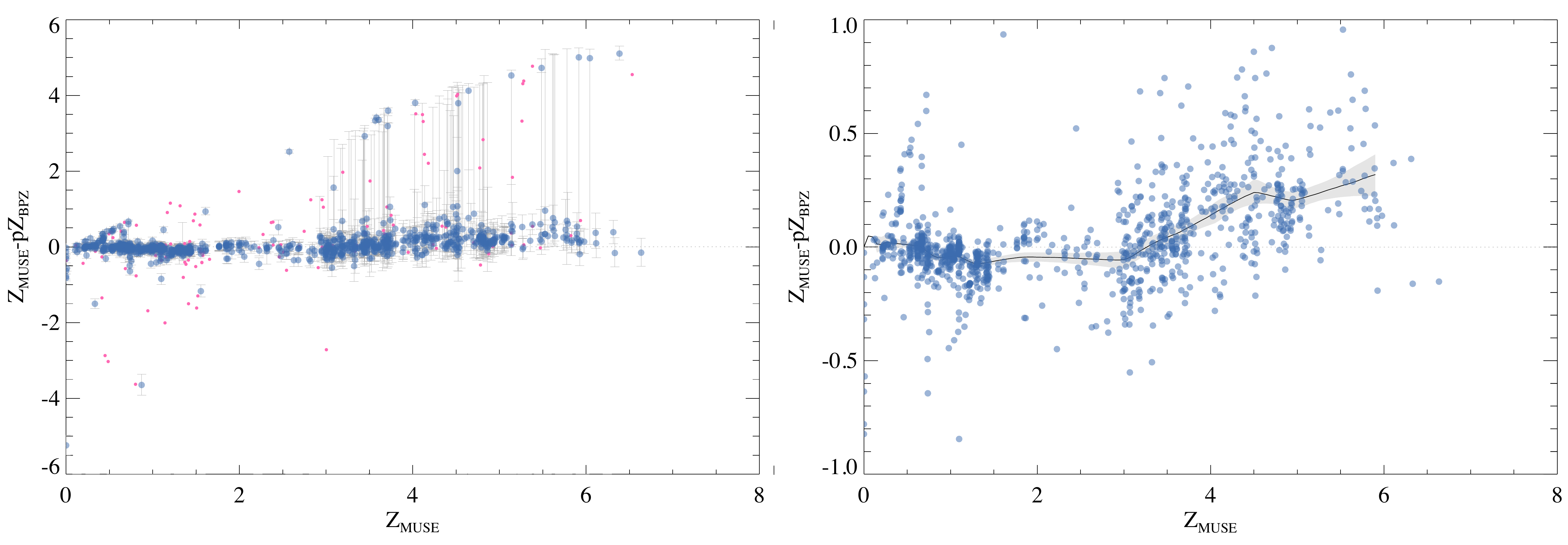}
  \caption{Difference between photometric and spectroscopic
    redshifts as a function of spectroscopic redshifts. The left panel
    shows all galaxies with spectroscopic redshifts. The solid points
    with  error-bars correspond to objects with high confidence
    spectroscopic redshifts (confidence level 2 and 3), while
    confidence 1 objects are plotted as pink points. Galaxies that are
    flagged as blended are not shown. The right panel shows a zoom in
    the y-axis to highlight the systematic offset at high
    redshift. The shaded region shows the median and the 68\%
    posterior region on the median from bootstrap sampling including
    the uncertainties on the photo-zs. } 
  \label{fig:delta_z_bpz}
\end{figure*}

Figure~\ref{fig:delta_z_bpz} shows the difference between the
spectroscopic and photometric redshifts using the BPZ code in R15. The
left panel shows a view of all galaxies that are not blended. The
confidence level 2 and 3 galaxies are shown as black points with
error bars while the confidence level 1 objects are shown as smaller
pink symbols. 

One immediately sees the outliers in this figure. At high redshift
these lie along the degeneracy line between the 4000\AA\ and
Lyman-breaks. This degeneracy ought to be detected by photo-z codes
but for many of the objects that show clear \lya\ lines this is not
the case and the photo-zs are strongly in disagreement with the
spectroscopic redshifts and for most cases the probability
distribution function (PDF) has no second peak at the spectroscopic
redshift.

We will return to the outliers, but first we will turn our attention
to the overall bias between photo-zs and spectroscopic redshifts.
While the agreement at low redshift is reasonable in this difference
metric, there appears to be a significant offset at higher redshift as
previously noted in \citet{Oyarzun:TheAstrophysicalJournal:2016} and
\citet{Herenz:EprintArxiv170508215:2017}. This is made clearer in the
right panel, which zooms the y-axis and suppresses the error-bars to
show the offsets more clearly. The shaded grey region shows the median
trend with the shading corresponding to the 68\% confidence limit on
the median determined by bootstrap resampling plus Monte Carlo
resampling of the photo-zs within their uncertainties. It is
clear there is a weak bias for photometric redshift to
overestimate the true redshift in $0.5<z<1.5$ or even to $z=3$ for BPZ
and BEAGLE, while there is a clear offset towards a systematic
underestimate of the true redshift at $z>3$. It is also worth nothing
that the latter offset is considerably larger than that induced by
the deviation from \lya\ redshifts from the true systemic redshift ---
as an example a 1000 km/s offset between the \lya\ redshift and the
systemic redshift of the galaxy would lead to a normalised difference
$\dzn = -0.0008$ at $z=3$ where the effect is the largest for
MUSE. For cosmological applications this could be crucial but the
amplitude is too small for what is seen here.

\begin{figure*}
  \centering
  \includegraphics[width=184mm]{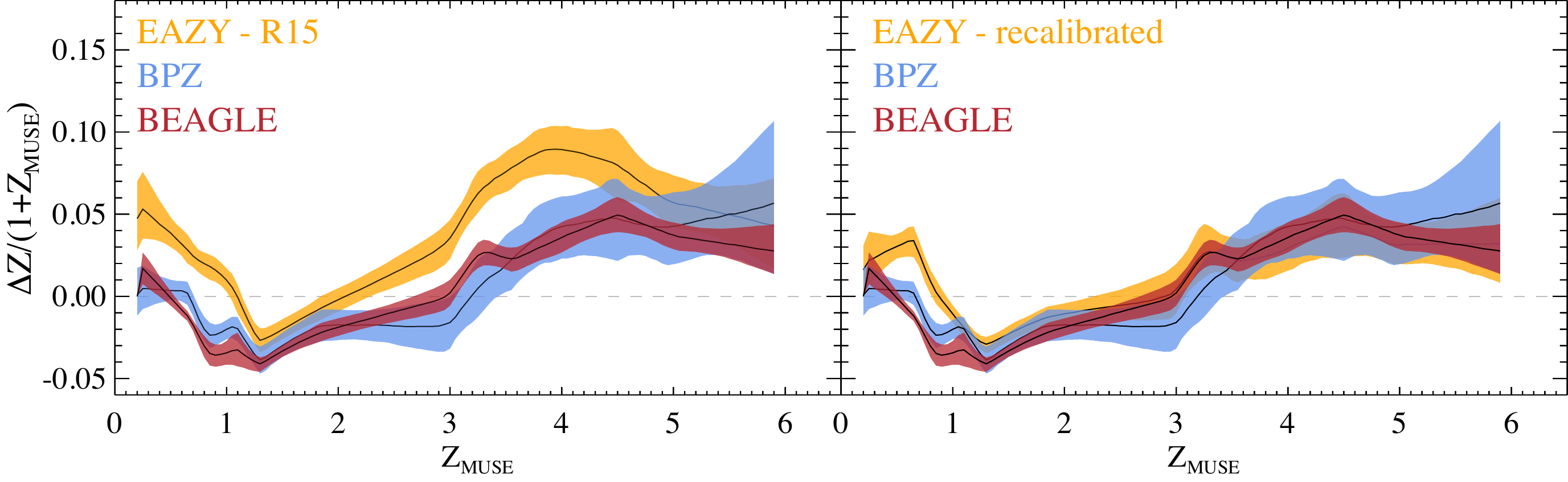}
  \caption{Left: the redshift offset between spectroscopic redshifts
    from MUSE and photometric redshifts from BPZ (blue) and EAZY
    (orange) from the R15 catalogue, and BEAGLE (burgundy) from
    \citet{Chevallard:MonthlyNoticesOfTheRoyalAstronomicalSociety:2016},
    all normalised by $1+\zMUSE$. The solid lines show the median as a
    function of redshift and the uncertainty on the median is shown by
    the shaded areas. Right: the same trends, but now
      using the recalibrated EAZY photo-zs from
      section~\ref{sec:reduc-bias-phot}. The improvment is noticeable
      and all three codes lead to very
      similar trends in the bias as a function of redshift.}
  \label{fig:median_offset}
\end{figure*}

\begin{figure*}
  \centering
  \includegraphics[width=184mm]{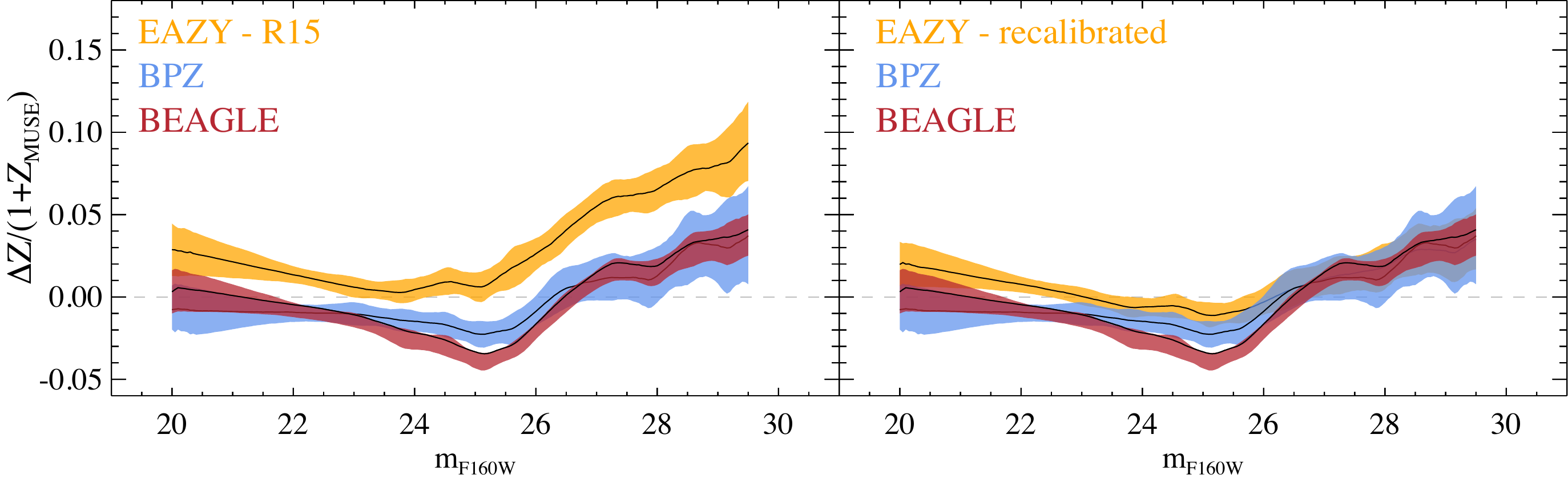}
  \caption{Similar to the previous figure but now showing $\dzn$
    as a function of the apparent F160W magnitude.}
  \label{fig:median_offset_vs_mag}
\end{figure*}

These offsets are not unique to the BPZ photometric redshifts, but are
also seen in the EAZY and BEAGLE photo-zs as shown
in Figure~\ref{fig:median_offset}. The solid lines show the median
offsets as a function of spectroscopic redshift and the shaded regions
show the 68\% uncertainty region on the median from bootstrap
resampling including resampling the photo-zs within their errors. The
systematic offset at high redshift is clear and since we have here
normalised by $1+\zMUSE$, we see the deviations at low redshift also
relatively clearly.  The left panel shows the results
  using the published photo-zs and it can be seen that the EAZY
  predictions have a higher bias than the other two. We return to this
  in the next section.

The bias as a function of redshift is also reflected
in the trend as a function of magnitude, shown in
Figure~\ref{fig:median_offset_vs_mag}. The lines and
  shading show the same as in Figure~\ref{fig:median_offset}, but the
offset between EAZY and the other two methods is even
clearer.

These figures show two separate issues: firstly the
  EAZY predictions from R15 have a clearly different, and larger, bias
  than the other two codes, and secondly, there are systematic biases
  with redshift that are common to all three codes. It is natural to
  ask what the reasons are for these offsets and whether they can be
  reduced. Since the predictions using EAZY are the most strongly
  discrepant, we will focus on these first.

\section{Reducing the bias in the EAZY photometric redshifts}
\label{sec:reduc-bias-phot}

There are at least two immediate possibitilities for the substantial
offsets seen at high redshift: template mismatch and an incorrect
treatment of inter-galactic medium (IGM) absorption. Here we will
first focus on the former because this also has the potential of resolving
the low redshift discrepancies. 

\subsection{Template mismatch}
\label{sec:template-mismatch}

\begin{figure*}
  \centering
  \includegraphics[width=184mm]{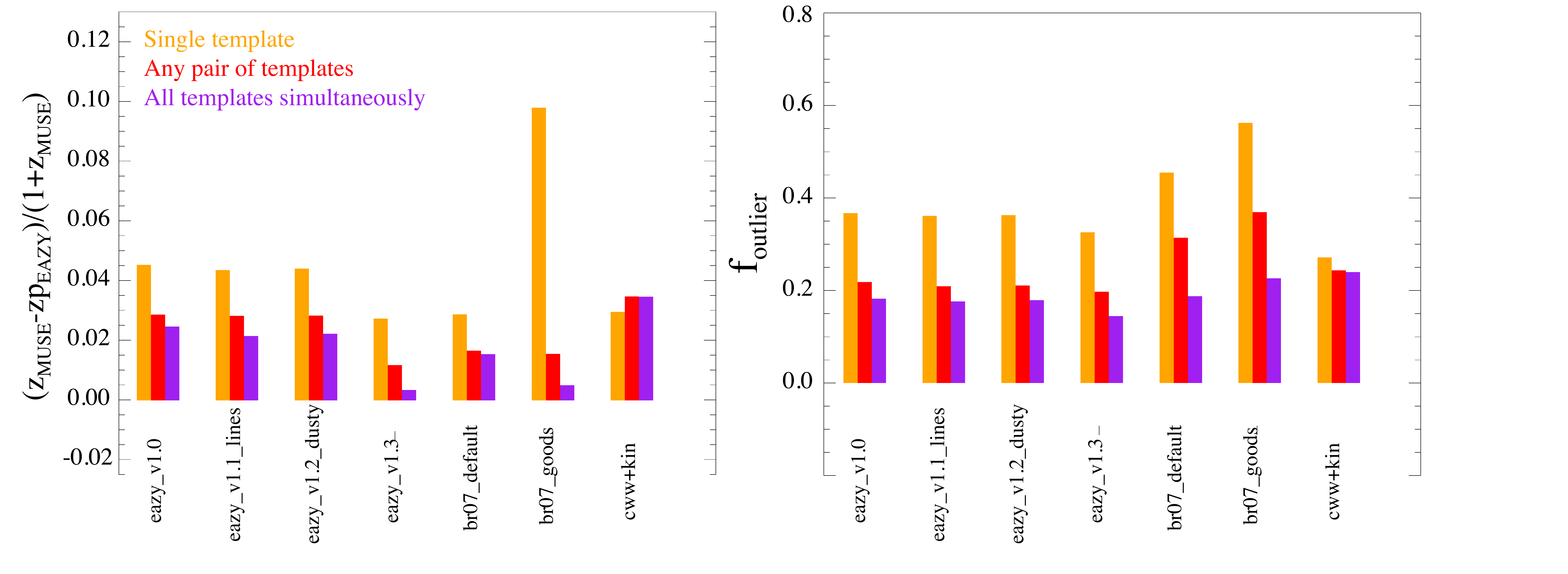}
  \caption{Left: The median bias in photometric redshift as a function of
    template set used and the method used to combine templates. Right:
    The outlier fraction, defined as the fraction of galaxies with a
    absolute normalised redshift difference greater than 15\%. See text for
    details. }
  \label{fig:bias_overall}
\end{figure*}

\begin{table*}
  \centering
  \caption{Template sets used for the runs with EAZY}
  \begin{tabular}{rl}\hline
    \texttt{eazy\_v1.0} & The six templates from
                          \citet{Brammer:TheAstrophysicalJournal:2008}. 
    \\
    \texttt{eazy\_v1.1} & Similar to \texttt{eazy\_v1.0} but with
                          emission lines from \citet{27739469} and \\
                        &  one
                          additional old, red SED\footnote{From the EAZY
                          documentation this is \texttt{ma05\_kr\_z02\_age10.1.dat}} from \citet{Maraston2005}.\\
    \texttt{eazy\_v1.2} & The same as \texttt{eazy\_v1.1} but with
                          one additional dusty \citet{Bruzual2003}
                          SED. \\
    \texttt{eazy\_v1.3} & The same as \texttt{eazy\_v1.2} but
                          including the SED of Q2343-BX418 from
                          \citet{Erb:TheAstrophysicalJournal:2010}. \\
    \texttt{br07\_default} & The default set of templates from
                             \citet[][their appendix B]{Blanton2007}. \\
    \texttt{br07\_goods} & The templates fit to GOODS data from \citet[][their appendix B]{Blanton2007}. \\
    \texttt{cww+kin} &
                       \citet{Coleman:TheAstrophysicalJournalSupplementSeries:1980}
                       combined with
                                                                                                                    \citet{Kinney:TheAstrophysicalJournal:1996} 
                      
    \\
    & as provided with LE PHARE
      \citep{Arnouts:MonthlyNoticesOfTheRoyalAstronomicalSociety:1999,OIlbert:AstronomyAstrophysics:2006}
      \\ 
    & \url{http://www.cfht.hawaii.edu/\~arnouts/lephare.html} \\ \hline
  \end{tabular}
  \label{tab:eazy_templates}
\end{table*}

\begin{table*}
  \centering
  \caption{The median bias and outlier fraction for different templates and template combinations in EAZY}
  \label{tab:mismatch_summary}
    \begin{tabular}{|ll|rrrr|}\hline
    \multicolumn{1}{|c}{Template} & \multicolumn{1}{c}{Combination} & 
    \multicolumn{4}{c|}{$100\times\mathrm{Bias}$}  \\ 
    & &  \multicolumn{1}{r}{$z<1.5$} & \multicolumn{1}{r}{$1.5\le z<2.9$} 
     & \multicolumn{1}{r}{$z\ge2.9$} & \multicolumn{1}{r|}{All $z$} \\ \hline
\texttt{eazy\_v1.0} & Single &   2.8 &  22.6 &   5.2 &   4.7  \\
 & Any two &   1.2 &  -0.0 &   4.7 &   2.9  \\
 & All &   0.7 &  -0.0 &   4.7 &   2.6  \\
    \hline
\texttt{eazy\_v1.1} & Single &   3.2 &  42.8 &   4.9 &   4.6  \\
 & Any two &   1.6 &  -0.3 &   4.2 &   2.8  \\
 & All &   0.6 &  -0.5 &   4.2 &   2.2  \\
    \hline
\texttt{eazy\_v1.2} & Single &   3.3 &  42.8 &   4.9 &   4.7  \\
 & Any two &   1.7 &  -0.3 &   4.2 &   2.9  \\
 & All &   0.6 &  -0.5 &   4.2 &   2.3  \\
    \hline
\texttt{eazy\_v1.3} & Single &   2.2 &  10.1 &   2.8 &   2.9  \\
 & Any two &   1.4 &  -0.6 &   1.4 &   1.2  \\
 & All &  -0.3 &  -0.7 &   1.3 &   0.4  \\
    \hline
\texttt{br07\_default} & Single &  -1.6 &  64.5 &   4.3 &   2.9  \\
 & Any two &  -0.7 &  -2.3 &   3.4 &   1.7  \\
 & All &   0.2 &  -1.0 &   3.4 &   1.6  \\
    \hline
\texttt{br07\_goods} & Single &  16.5 &  65.2 &   5.2 &   9.9  \\
 & Any two &  -3.9 &  -2.3 &   3.7 &   1.6  \\
 & All &  -2.3 &  -1.5 &   3.7 &   0.6  \\
    \hline
\texttt{cww+kin} & Single &  -1.7 &   0.3 &   6.6 &   3.0  \\
 & Any two &  -0.1 &   0.9 &   6.6 &   3.6  \\
 & All &  -0.1 &   0.8 &   6.6 &   3.6  \\
    \hline
  \end{tabular}
\end{table*}

To explore this we have run EAZY\footnote{We used the \texttt{git}
  version from \url{https://github.com/gbrammer/eazy-photoz},
  specifically we cloned the git repository on May 7, 2016 with commit
  id c992854eb9bce2bcf4810ff306d014bda92cdf9c.} with the seven
different template-sets provided with the code detailed in
Table~\ref{tab:eazy_templates}. For each template set we used three of
the combination methods provided by EAZY: we used each template on its
own, any pair of templates and all templates simultaneously. This
provides a total of 21 photo-z runs for the UDF data. We do
not attempt to deal with strong AGNs here, as these present a separate
set of complications
\citep[e.g.][]{Salvato:TheAstrophysicalJournal:2011} and we have very
few strong AGNs in our sample. The version of EAZY
  used by us does not support iterative adjustments of the zeropoints
  but we did this by calling EAZY iteratively. However this did not
  lead to noticeable improvements in photo-z performance, indicating
  that the main problem with the current data is not photometric
  calibration, this also matches the finding by S14 that adjustments
  for HST bands are so small as to be ignored. Given the lack of
  improvement we show the results without the iterative adjustments
  here for simplicity.

Focusing first on the overall performance, we summarise the global
agreement as the median of $\Delta z/(1+\zMUSE)$ over all
redshifts. This is shown in the left panel of
Figure~\ref{fig:bias_overall} and the great variation is clear to
see. It is of course to be expected that the bias will decrease as
more flexibility in the template combination is allowed since there
will be a higher probability to match the actual SED of the galaxy,
in agreement with the arguments in
\citet{Acquaviva:TheAstrophysicalJournal:2015}. This is also borne out
in the figure where the single template fitting results in
significantly higher bias than pairs of templates which again has a
larger bias than fitting all templates together.  This is however not
true for the \texttt{cww+kin} set of templates where the different
combination results are consistent with each other within the
errors. The strong deviation for the \texttt{br07\_goods} template set
when using only one template is likely due to the way it was optimised
for higher redshift work --- Table~\ref{tab:mismatch_summary}
demonstrates that it performs well at high redshift. It is also worth
noting that the bias is positive in all cases, this is caused by the
systematic bias at high redshift which we will discuss further below.

Indeed breaking down the numbers by redshift is also enlightening. We
do this in Table~\ref{tab:mismatch_summary}. This shows (100 times)
the bias for three bins in spectroscopic redshift and the total. What
is striking to see is that in the high redshift case, the ability to
combine multiple templates is of little importance except for the
\texttt{eazy\_1.3} template set, while at low redshift this is
crucial and most template sets perform similarly here. This suggests
that a crucial advantage of the \texttt{eazy\_1.3} template set is the
addition of the one extra SED from
\citet{Erb:TheAstrophysicalJournal:2010} which allows a more flexible
fit to the high-redshift galaxies. The change with redshift is
otherwise as expected: for the low redshift galaxies the photometry
probes the rest optical to near-IR where a mixture of stellar
populations has strong effect on the observed colours, while at high
redshift the photometry mostly probe the shorter wavelength UV where
the age range of stars contributing to the spectrum is fairly narrow
and thus mixtures of SEDs have less effect. The
  addition of high-quality (observed-frame) $K$ and MIR photometry
  would undoubtedly alter the picture, but as we show in
  Appendix~\ref{sec:adding-ground-based}, the addition of ground-based
  $K$-band and Spitzer IRAC fluxes does not currently improve the
  photo-z estimates for the very faint galaxies studied in this
  paper. This will certainly change with the advent of JWST NIRCam
  photometry.

The right panel of figure~\ref{fig:bias_overall} shows the fraction of
significant outliers in the sample. We here define a galaxy to be an
outlier if $\Delta z/(1+\zMUSE) > 0.15$, but the conclusions are not
very sensitive to this choice. What is notable here is that the
absolute outlier fraction is high. We will return to this in
Section~\ref{sec:redshift-outliers} below but for now we merely note that
the relative trends are as expected and similar to the trends for the
bias.  The best-performing setup uses the \texttt{eazy\_1.3} template
set with simultaneous combination of templates.  From now on, when we
refer to the ``best EAZY'' photo-zs, we mean this particular run of
EAZY. This combination has a median $\dzn$ of 0.008, with a median
absolute deviation (MAD) of 0.045 and an outlier fraction of 0.10,
using $z_{\mathrm{peak}}$ as the point estimate of the photo-z and
$\left|\dzn\right| > 0.15$ as the outlier criterion.  This is quite
comparable to the BPZ photo-z estimates from R15 which for the same
sample have a median bias of $0.00$, a MAD of 0.045 and a median
outlier fraction of 0.08, and to the BEAGLE photo-zs which have a
median bias of $0.002$ and a MAD of 0.053 with a median outlier
fraction of 0.10.


\begin{figure}
  \centering
  \includegraphics[width=84mm]{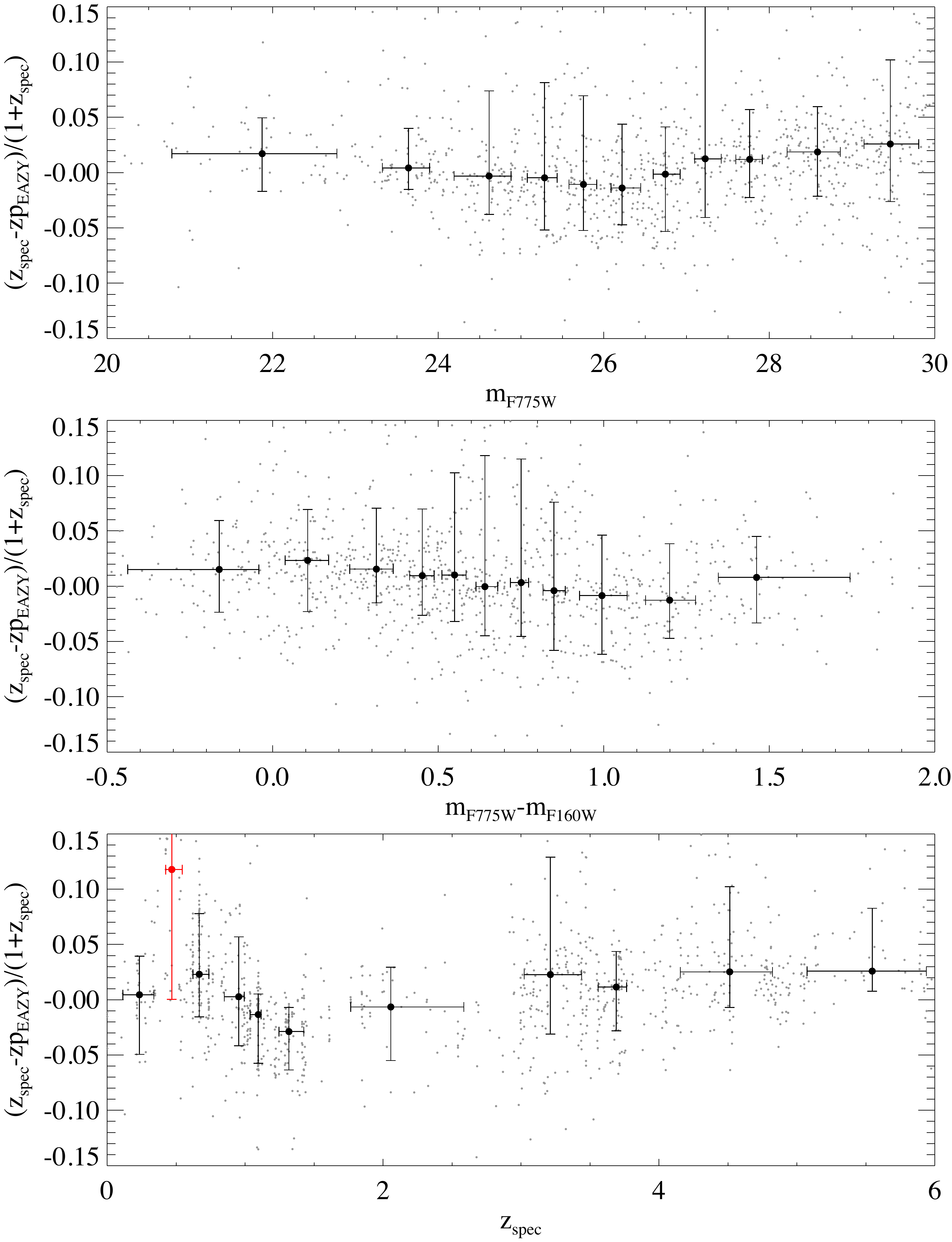}
  \caption{Illustration of the performance of the best-performing
    EAZY template set, namely v1.3 with all possible combinations of
    templates. The top panel shows the normalised redshift difference
    as a function of the F775W magnitude with the points plotted at
    the average magnitude and the median difference in each bin, and
    the error bars indicating the 16th and 84th percentiles of the
    distributions in x and y. The red points with error bars are shown
    where the outlier fraction exceeds 15\%.  The middle panel shows
    the same, but now plotted as a function of the observed
    F775W$-$F160W colour, and the final panel shows the same but 
    against the spectroscopic redshift. Note the one bin near $z=0.4$
    with very discrepany bias.}
  \label{fig:best-eazy}
\end{figure}

\begin{figure}
  \centering
  \includegraphics[width=84mm]{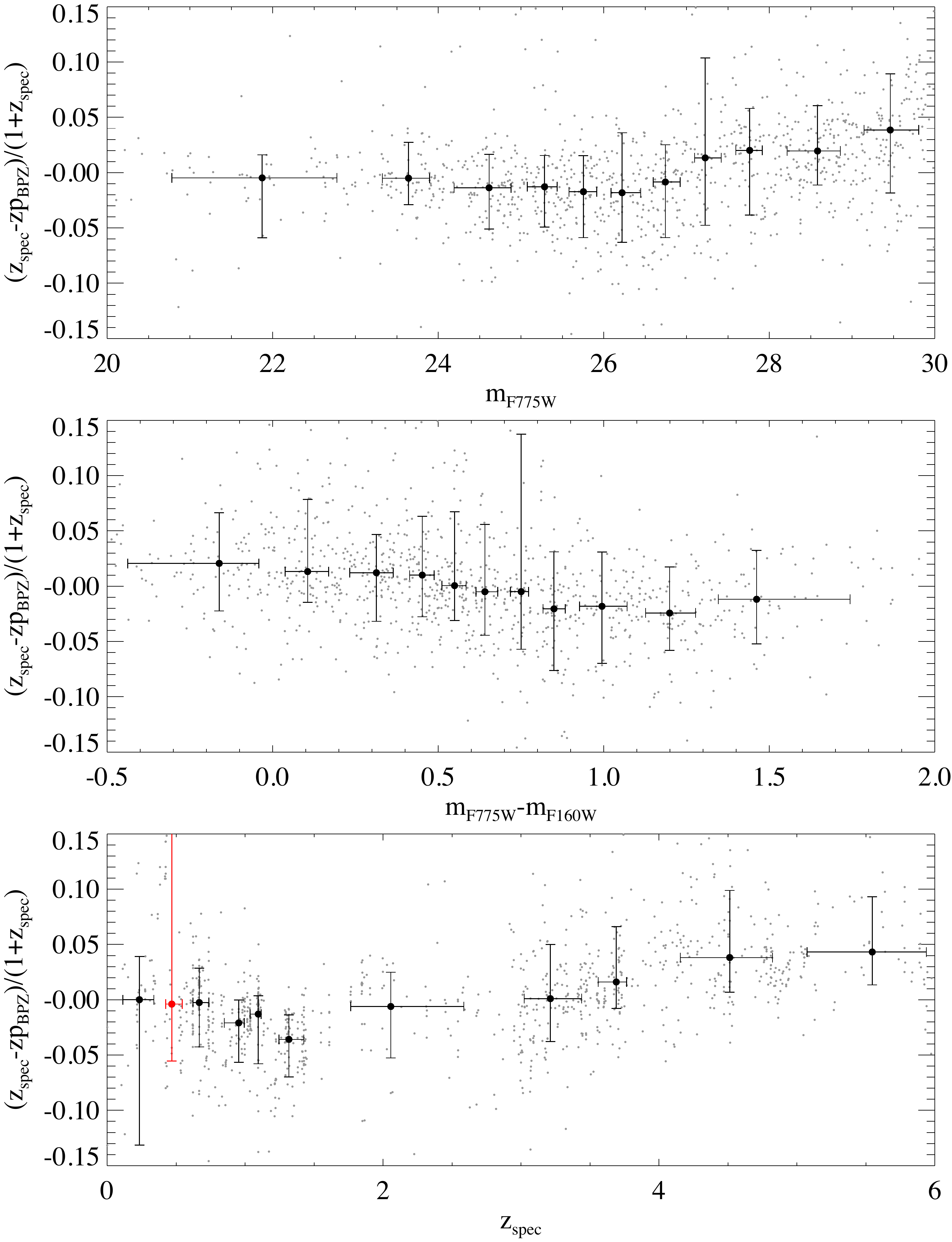}
  \caption{Similar to Figure~\ref{fig:best-eazy} but this time for the
    BPZ photometric redshift estimates in R15. }
  \label{fig:best-bpz}
\end{figure}
\begin{figure}
  \centering
  \includegraphics[width=84mm]{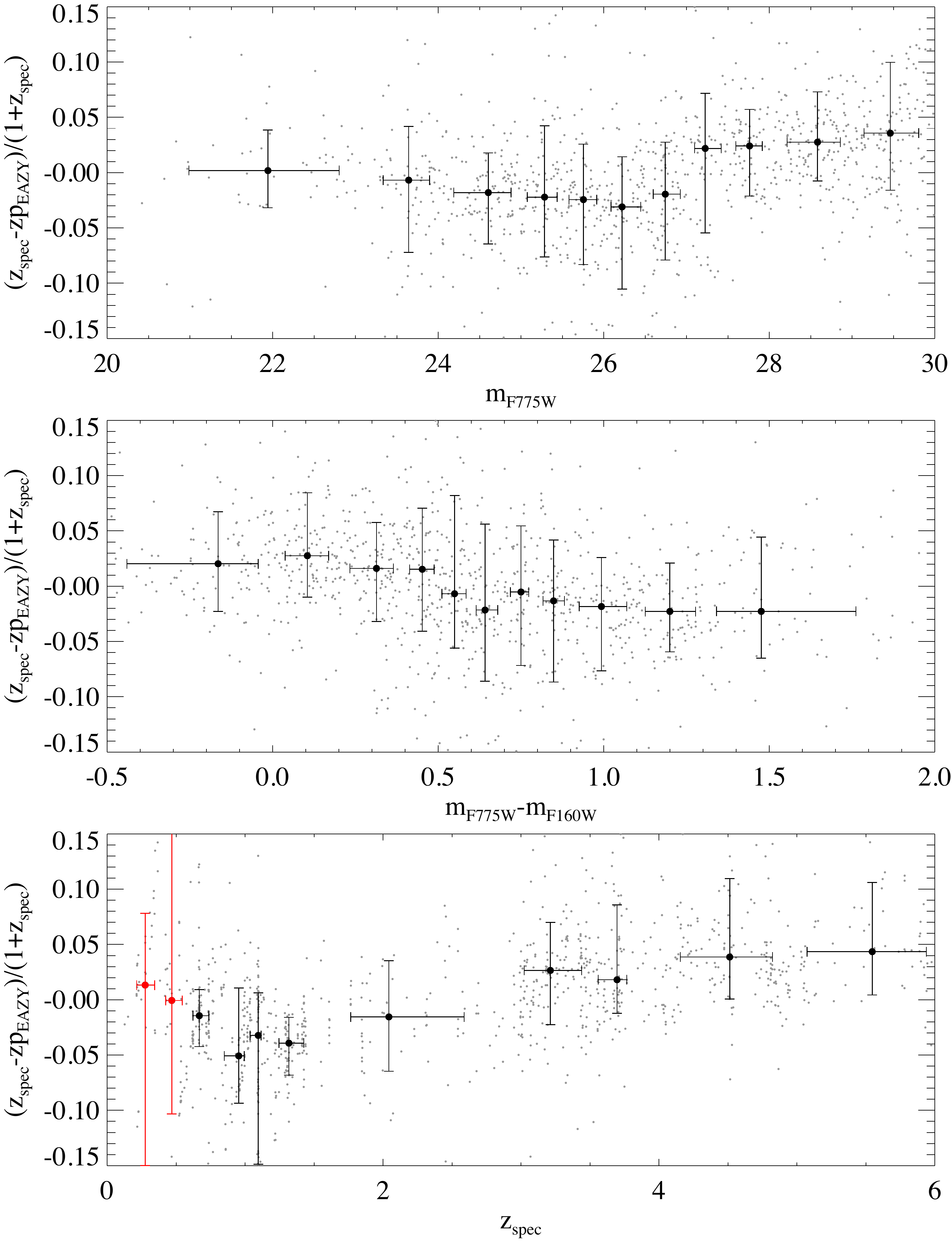}
  \caption{Similar to Figure~\ref{fig:best-eazy} but this time for the
    photometric redshift estimated using BEAGLE. }
  \label{fig:best-beagle}
\end{figure}

The improvement in the EAZY photo-zs is clearly seen when comparing
the left and right panels in Figures~\ref{fig:median_offset} and
\ref{fig:median_offset_vs_mag}. The left-hand panels use the EAZY
photo-zs from R15, while the right-hand panels show the result with
the best EAZY photo-zs. It also clear that after this improvement, the
three photo-z codes examined all show almost identical trends for the
median bias.

Figures~\ref{fig:median_offset} and
  \ref{fig:median_offset_vs_mag} show only the median trends and not
  the individual data points since in that case it would not be
  possible to overplot multiple photo-z codes and still have a
  readable figure. The figures also show the uncertainty on the
  median, and not the scatter in each redshift bin. This information
  is instead shown in
  Figures~\ref{fig:best-eazy}--~\ref{fig:best-beagle}. Each figure
  here shows the trends of the normalised redshift difference against
magnitude (top panel), observed colour (middle panel) and
spectroscopic redshift (bottom panel). The bins along the $x$-axis
were chosen to have similar number of objects per
bin. The error-bars show the range containing 68\% of
  the data points in that bin both in the x \& y direction and bins
  with more than 20\% outliers (defined as having
  $\left|\dzn\right|>0.15$) are plotted in red. These figures show
  that the offset at high redshift persists also for EAZY with the
  \texttt{eazy\_v1.3} template set, and also that there are systematic
  trends with redshift at the few percent level for all three photo-z
  codes.  At magnitudes fainter than $\mhst{F775W}=27$, the scatter
goes down but this is almost certainly a selection effect because we
are only able to securely determine redshifts for the subset of strong
\lya-emitters at those redshifts.  It is also of particular note that
in the $0.4<z<0.6$ bin, which contains 186 galaxies, the bias is
significantly larger. We do not know why this is. Finally, we note that
there is only a very minor residual trend with colour. In contrast,
the EAZY estimates in the R15 catalogue show very significant trends
with colour --- strongly indicating that they suffer from a mismatch
in the template set used.

\subsection{IGM absorption modelling}
\label{sec:igm-absorpt-modell}

The absorption of the neutral medium at high redshift is an important
ingredient in photometric redshift codes for objects with photometry
shortwards of rest-frame \lya. The widely used model of
\citet[][M95 hereafter]{Madau:TheAstrophysicalJournal:1995} was recently revised by 
\citet[][I14 hereafter]{Inoue:MonthlyNoticesOfTheRoyalAstronomicalSociety:2014} who
demonstrated that the updates in the IGM modelling could lead to
modest, but systematic, changes in the photometric redshift estimates
of $\Delta z \sim 0.05$ in the redshift range of interest to us. 

The default EAZY operation uses a hybrid approach where the absorption
longwards of the Lyman limit is treated using the M95 model, while the
Lyman continuum opacitiy is treated using the I14 model. However the
code contains the options to use the I14 model throughut and we have
compared both.

The IGM treatment is statistical in nature as the IGM seen by a given
galaxy will differ depending on the local conditions. Thus ideally a
photo-z code should also marginalise over this unknown but this is not
normally done. To approximately treat this we have modified EAZY to
include a scale factor for the attenuation shortwards of the Lyman
limit, \sLC, and another for the Lyman-forest
attenuation, \sLAF. We then ran EASY varying \sLC\ and \sLAF\ between
0.1 and 2.5 in steps of 0.1 for both the default IGM treatment using a
mixture of M95 and I14, and only using the I14 IGM treatment. 

\begin{figure}
  \centering
  \includegraphics[width=84mm]{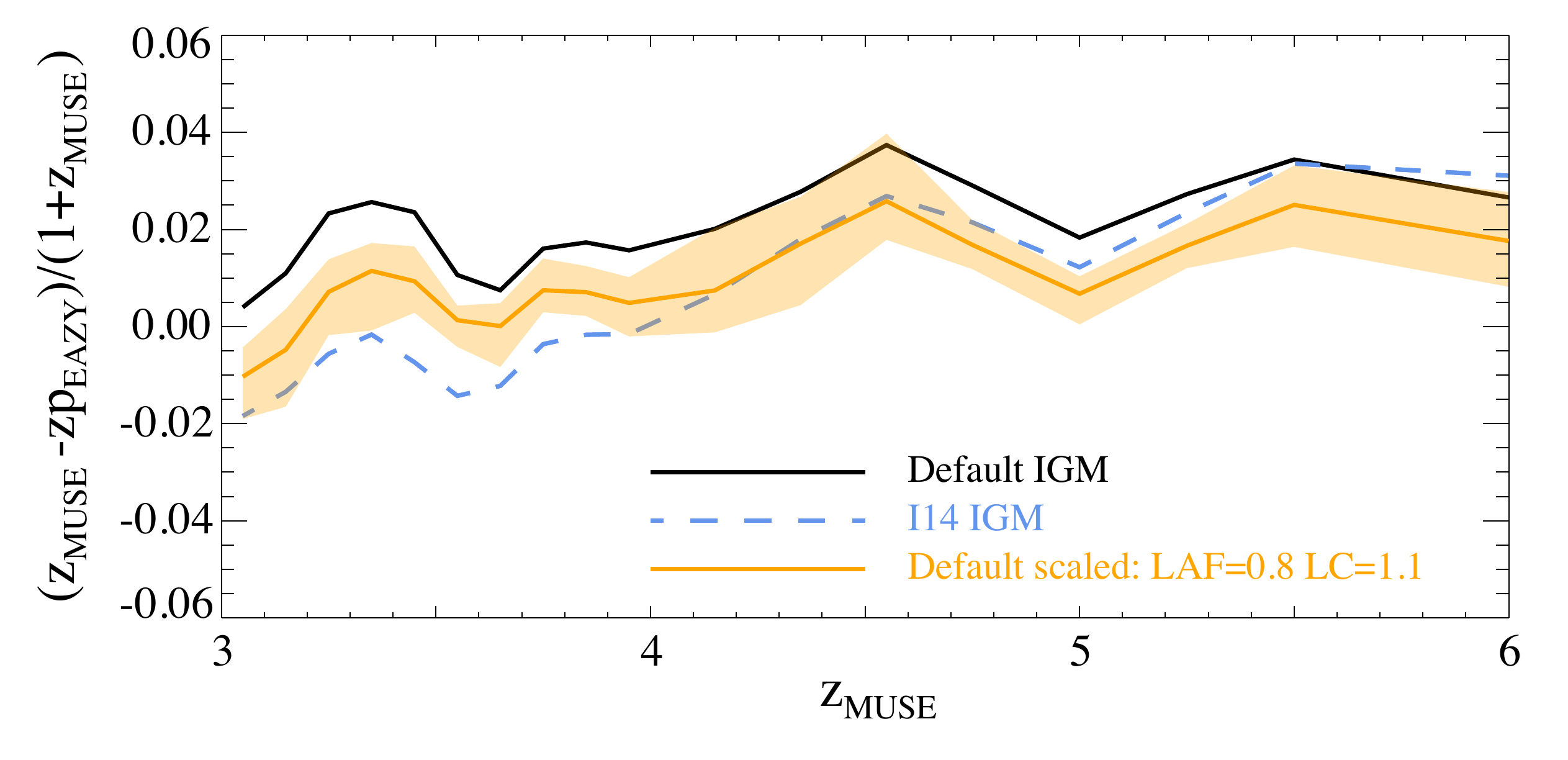}
  \caption{Median normalised bias between $\zMUSE$ and $\zEAZY$
    for three different IGM models. The solid black and dashed blue
    lines show the default EAZY IGM model and the IGM model using the
    I14 prescription respectively. The orange line with shading shows
    the adjusted IGM model with the jointly smallest bias at $z=3$ and
    slope (see text for details). The shaded region shows the
    uncertainty on the median and this is suppressed for the blue and
    black lines as these are near identical to that shown for the
    orange line.}
  \label{fig:igm_dependence_of_bias}
\end{figure}

We found that the dependence on $\sLC$ is signficantly weaker than
that of $\sLAF$. This is natural since the attenuation bluewards of
the Lyman limit is very large and any small change in its value
has little effect. To assess the effect of the scale factors we
calculate a running median of $\dzn$ versus \zMUSE\ between $z=3.0$
and $z=6.5$. We calculate the median with 31 galaxies per bin and
calculate the uncertainty on the median using 999 bootstrap
repetitions. We then fit a linear function $\dzn = a + b(\zMUSE-3)$
and minimise $\left|a\right| + \left|b\right|$, that is, we attempt to
jointly minimise the bias at $z=3$ and the slope. This gave us a
minimum for $\sLAF=0.8$ and $\sLC=1.1$ for the default IGM treatment
in EAZY.

The results are summarised in
Figure~\ref{fig:igm_dependence_of_bias}. The black solid line shows
the normalised bias as a function of $\zMUSE$ for the optimised EAZY
model, similar to the bottom panel of Figures~\ref{fig:best-eazy} but
with finer sampling in $\zMUSE$ and a zoomed $y$-axis. The dashed blue
line shows the bias for the I14 IGM model and this shows a slightly
stronger redshift trend but overall a slightly smaller bias. The
orange solid line shows the best adjusted IGM model with the shading
indicating the 68\% confidence region around the median.

The main point to note here is that the optimised IGM model works
somewhat better but still fails to correct all the bias at high
redshift. This should not be taken as a failure of the IGM model,
however. As mentioned above this is statistical in nature and the
scaling factors we have introduced are also mean quantities for the
whole redshift range. By letting the scale factors vary with redshift,
we can remove the trend also at high redshift but we do not have
enough data points to carry out this minimisation in a statistically
meaningful way.

It is also interesting to ask whether changes in the IGM model is
sufficient to explain all scatter in the normalised bias. The answer
to this is a qualified ``yes''. For objects for which
$\left|\dzn\right|>0.2$, the effect of the IGM is insufficient to
change the photometric redshifts to agree with the
spectroscopic. However for galaxies with $\left|\dzn\right|<0.1$, we
can find values for $\sLAF$ and $\sLC$ that bring photometric and
spectroscopic redshifts in agreement.  However, these solutions show
two clear problems: the scatter in the implied \lya\ to \lyb\ flux
depression is much larger than that inferred from \lya-forest studies
\citep[e.g.][]{fan2006constraining,FaucherGiguere:TheAstrophysicalJournal:2008},
and secondly, the fits tend to prefer the extremes of the scales
($\sLAF=0.1$ or $\sLAF=2.5$), rather than something in the
middle. This strongly hints that while the IGM treatment might be part
of the problem, the template spectra are also not fully covering the
range of real galaxy spectra at these redshifts.

To summarise this section, we wanted to compare the three photo-z codes
used here and found, in good agreement with previous work
\citep[e.g.][]{Hildebrandt:AstronomyAndAstrophysics:2010}, that these all
agree well. Overall the BPZ estimates performs very slightly better
than the re-calibrated EAZY estimates, except perhaps at high
redshift, and show less scatter than BEAGLE although the median trend
is very similar. In the following we will therefore mostly use the BPZ
photo-zs but occasionally also show the results of using EAZY or
BEAGLE, however we stress that for practical use these can be used
interchangeably --- although we do not recommend using the EAZY
estimates from the R15 catalogue.

\section{Redshift outliers}
\label{sec:redshift-outliers}

\begin{figure}
  \centering
  \includegraphics[width=84mm]{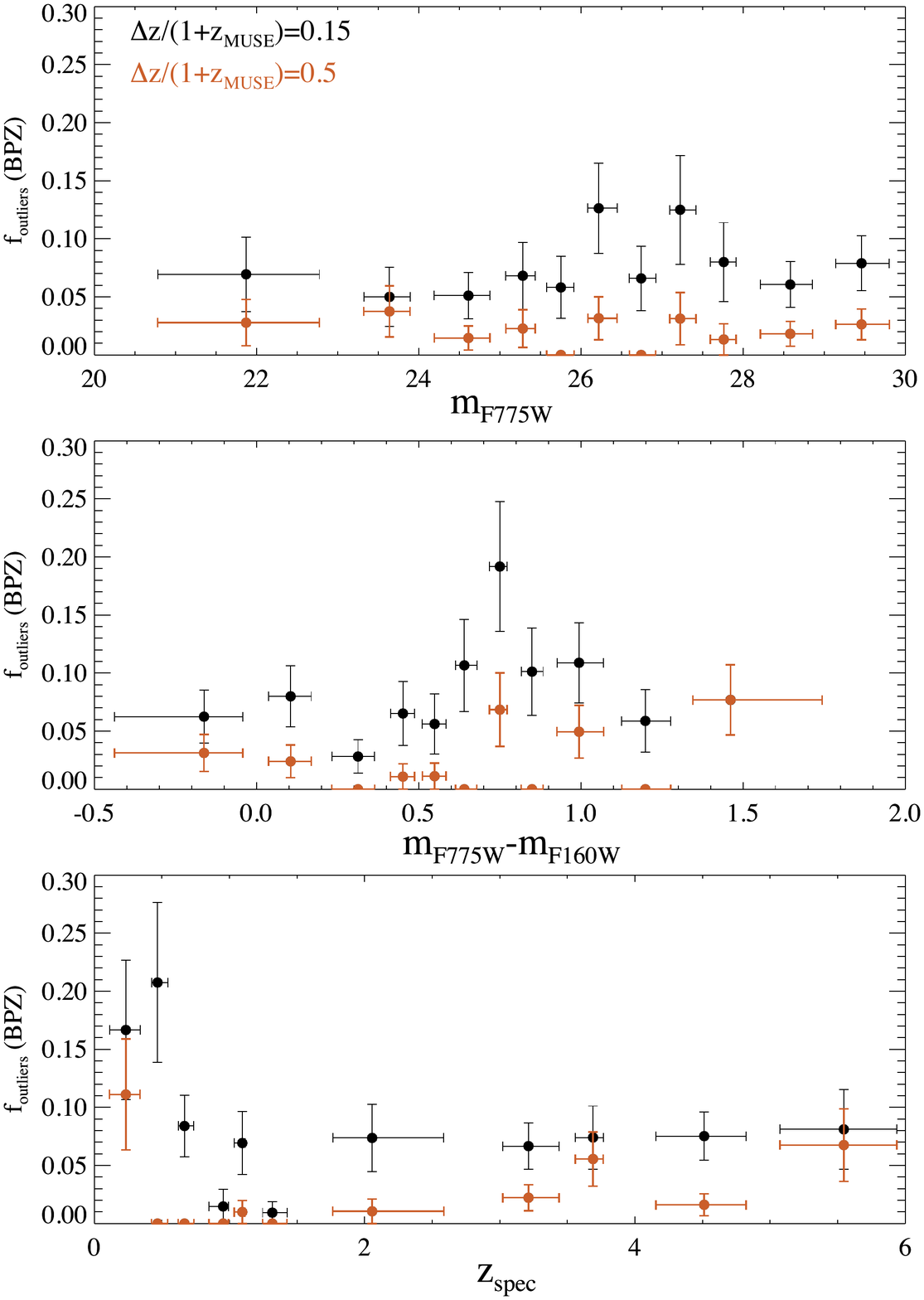}
  \caption{Fraction of objects with
    $\left|\Delta z\right|/(1+\zMUSE) >0.15$ and 0.5, in black and red
    respectively, as a function of magnitude (top), colour (middle),
    and spectroscopic redshift (bottom). The redshifts
      used are BPZ but the figure looks similar using BEAGLE photo-zs
      and also with the improved
      EAZY photo-zs from section~\ref{sec:reduc-bias-phot}. The error bars assume
    $\sqrt{N}$ uncertainty and standard propagation of errors and
    should be considered as indicative only. }
  \label{fig:outlier_fraction_vs_many}
\end{figure}

\begin{figure}
  \centering
  \includegraphics[width=84mm]{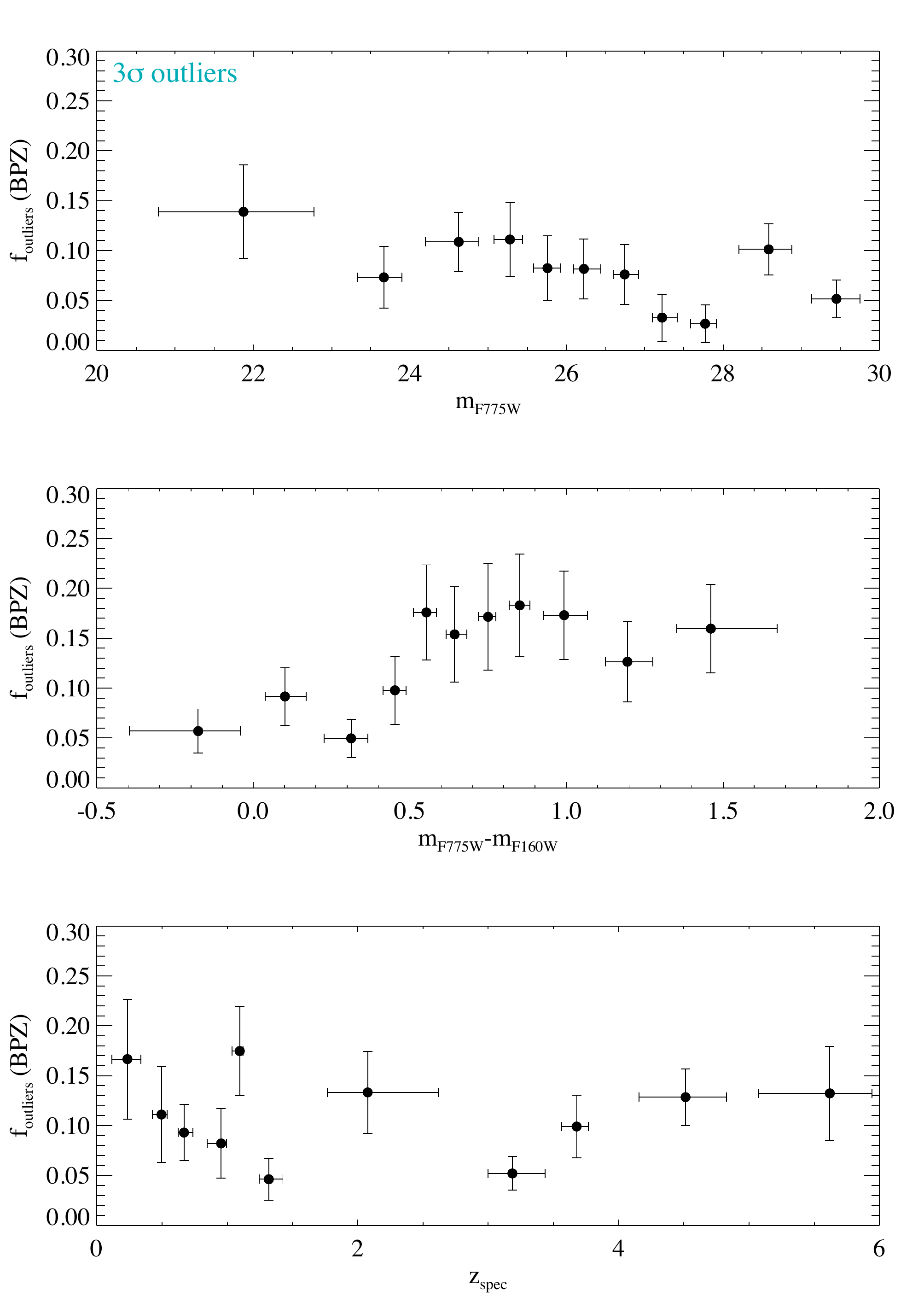}
  \caption{Fraction of objects with redshifts
    $\left|\dzn\right|>3\sigma_{\mathrm{MAD}}$ away from zero, where
      $\sigma_{\mathrm{MAD}}$ is calculated for each bin. The fraction
      is shown as a function of magnitude in the top panel,
      $\mhst{F775W}-\mhst{F160W}$ colour in the middle panel, and
      spectroscopic redshift in the bottom panel. As in
      Figure~\ref{fig:outlier_fraction_vs_many}, the error bars assume
      $\sqrt{N}$ ucertainty and standard propagation of errors and
      should be considered indicative only. }
  \label{fig:rel_outlier_fraction_vs_many}
\end{figure}

It is customary, albeit arbitrary, to define outliers as those
galaxies for which $\Delta z/(1+\zMUSE) > x$, where $x$ is often
taken to be 0.1 or 0.15, or as $\Delta z/(1+\zMUSE) >
n\sigma_{\mathrm{MAD}}$, where $\sigma_{\mathrm{MAD}}$ is the median
absolute deviation around the median, adjusted to correspond to a
standard deviation for a Gaussian distribution,
\begin{equation}
  \label{eq:2}
  \sigma_{\mathrm{MAD}}(x) = 1.48 \median\left(\left|x -
      \median(x)\right|\right). 
\end{equation}
We take $n=3$ in this section since this gives a reasonable number of
galaxies and hence avoid the problems of small-number statistics. The
statistics of truly significant outliers, with $x=0.5$ or $n=5$, is
also interesting and we discuss these catastrophic outliers in some
detail in Appendix~\ref{sec:serial-outliers}. We show there that
BEAGLE and BPZ have the lowest fraction of catastrophic outliers, with
EAZY having a factor of 2--3 more catastrophic outliers, depending on
the definition.

In Figure~\ref{fig:outlier_fraction_vs_many} we show the outlier
fraction for the BPZ photo-zs against \mhst{F775W} magnitude (top),
$\mhst{F775W}-\mhst{F160W}$ colour (middle) and spectroscopic redshift
(bottom). 

\begin{figure}
  \centering
  \includegraphics[width=84mm]{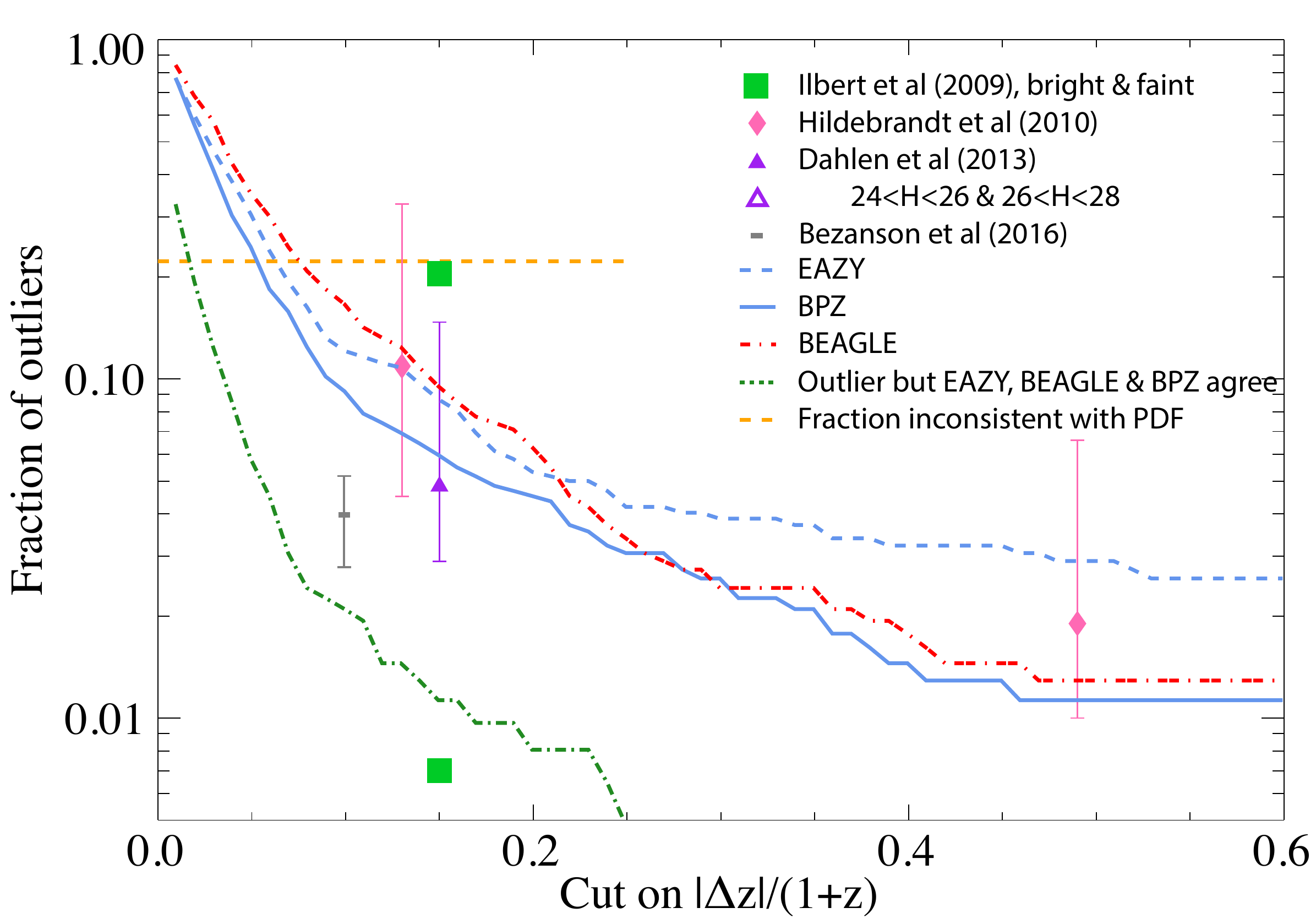}
  \caption{Outlier fraction as a function of the
    $\left|\dzn\right|$ threshold adopted. The blue solid line shows
    the trend for BPZ photo-zs and the dashed blue line that using
    the best EAZY photo-zs as derived in
    Section~\ref{sec:reduc-bias-phot}. The dotted green line shows the
    fraction of galaxies that are outliers both relative to EAZY and
    BPZ but where EAZY and BPZ agree. The fraction of objects that
    have redshifts inconsistent with the PDF of the photo-z are shown
    by the dashed orange horizontal line. The outlier fractions
    measured by \citet{27739469} are shown by the green squares, with
    the one at a fraction of 0.007 corresponding to the $i^{+}<22.5$
    sample in COSMOS and the higher value at 0.2, for a faint
    subsample $24<i^+<25$. The outlier fraction measured by
    \citet{Hildebrandt:AstronomyAndAstrophysics:2010} is shown by the
    pink lozenges with the lozenge arbitrarily placed at the median
    outlier fraction and the error spanning the range of outlier
    fractions found in their study. The measurement at a
    threshold of 0.15 has been shifted to 0.14 to avoid overlap with
    other data. The outlier fractions found
    in~\citet{Dahlen:TheAstrophysicalJournal:2013} are shown by the
    purple triangles. The solid triangle shows the median outlier
    fraction and the full range found, while the open triangles show
    the results for their $24<\mhst{F160W}<26$ (at 0.16) and
    $26<\mhst{F160W}<28$ (at 0.28) subsamples. The outlier fractions
    reported for GOODS-S in
    \citet{Bezanson:TheAstrophysicalJournal:2016} is shown as the grey
    enlongated rectangle with error bars covering the range from comparison
    to grism redshifts (lowest) to spectroscopic redshifts
    (highest).}
  \label{fig:olf}
\end{figure}

There are some points worth noting in
Figure~\ref{fig:outlier_fraction_vs_many}. The first is that the
outlier fractions are relatively high compared with what is normally
reported in the literature. In the most similar studies to our work,
\citet{Dahlen:TheAstrophysicalJournal:2013},
\citet{Hildebrandt:AstronomyAndAstrophysics:2010} and R15 all found
outlier fractions that were $<5\%$ in the best cases with a cut-off of
0.15 (see Figure~\ref{fig:olf} below for a quantitative
illustration). We are generally above this level at all
magnitudes. The second point of note is that for the more stringent
0.15 case (black symbols), the outlier fraction rises somewhat towards
fainter magnitudes and lower redshifts. For the catastrophic outliers
(orange symbols) the trend is much less pronounced and more a tendency
for catastrophic outliers to be more prevalent at high redshift (see
Appendix~\ref{sec:serial-outliers} for more details on catastrophic
outliers). 

The increase in outlier fraction with fainter magnitude is real, but
it is in part a reflection of an increased scatter in the relationship
between photometric and spectroscopic redshifts and in part a
reflection of the systematic bias in the photometric redshifts,
particularly at high redshift, that we saw above. This is made
explicit in Figure~\ref{fig:rel_outlier_fraction_vs_many} where we
have defined outliers to be those that have
$\left|\Delta z/(1+\zMUSE)\right| > 3\sigma_{\mathrm{MAD}}$, where
$\sigma_{\mathrm{MAD}}$ was calculated locally in each bin, choosing
$5\sigma_{\mathrm{MAD}}$ would lead to very similar trends but the
points would be shifted down by on average a factor of 1.7-1.8. The
panels are the same as in Figure~\ref{fig:outlier_fraction_vs_many}
and the contrast is clear: there is no increase in the number of
outliers with magnitude or redshift, however the trend with colour is
approximately maintained.

To put these results in context, Figure~\ref{fig:olf} shows the
outlier fraction in our sample down to $\mhst{F775W}=30$, as a
function of the \dzn\ cut-off. 
The strong trend with the cut-off value is clear and it is also clear
that the outlier fraction for EAZY (dashed line) is consistently
somewhat higher than for BPZ (solid blue line) and BEAGLE (red
dot-dashed line) but recall that the points here are not
independent. We can also estimate a limit to how well we can reduce
the outlier fraction by combining photo-z estimates. We do this by
calculating the fraction of objects that are outliers for EAZY, BPZ
and BEAGLE while at the same time all three photo-z codes give
redshift estimates in agreement with each other. The resulting values
(green dashed line) lie below the other curves but remains above
1\% for cut-off values $<0.15$.

A comparison of these results to other studies is not
  entirely straightforward. Firstly, the photometry might be of
  different quality and not all objects might be included. To reduce
  the effect of this we will compare to studies that use well-studied
  photometric data sets, including HST data. Secondly, different
  studies have different limiting magnitudes and this has significant
  effects on outliers, both due to the photometric quality and because
  the mean redshift will differ. And finally, the methods used are
  different in different surveys which might make comparisons
  hard. With those caveats we have chosen four different surveys to
  compare against.

We show the results from the comprehensive set of photo-zs in the
COSMOS field by \citet{27739469} as the green
squares. This survey is considerably brighter on
  average than ours but their photo-zs use 30 bands in contrast to
  our 11. They report outlier fractions for various subsamples. The
fraction found for their bright sample, $i^+<22.5$, is very low
($<1$\%) while that for the faintest sample $24<i^+<25$, and also for
their $1<z<3$ sample, reaches 20\% --- considerably higher than our
estimates at the same threshold but of course for less deep
photometry. We plot the outlier fraction reported for
  the GOODS-S field by \citet{Bezanson:TheAstrophysicalJournal:2016}
  as a grey rectangle with error-bars extending from outliers in the
  grism-z vs photo-z comparison (lowest) to the spec-z vs photo-z
  comparison (highest). We also report results for
  \citet{Hildebrandt:AstronomyAndAstrophysics:2010} and
  \citet{Dahlen:TheAstrophysicalJournal:2013} who both use a mixture
  of space- and ground-based photometry which is somewhat shallower
  than the photometry in the UDF. They present results for a range of
codes and we show this as an error bar with a symbol placed at the
median outlier fraction. It is reasonable to compare to the best results
from their studies which is shown as the lower end of the error
bars. 
It is obvious that the best-performing codes in these
studies give lower outlier fractions than we are finding, by a factor
of approximately two. However,
\citeauthor{Dahlen:TheAstrophysicalJournal:2013} also report the
performance of the best codes in two faint magnitude bins
($24<\mhst{F160W}<26$ and $26<\mhst{F160W}<28$). This is shown by the
open triangles and it is clear that the outlier fractions for these
fainter subsamples are higher than what we find for our sample down to
$\mhst{F775W}=30$.

The CANDELS data are of course shallower on average than the HUDF data
used here. The median $\mhst{F160W}$ for the galaxies with redshift
confidence level $\ge 2$ in our sample is 25.89, while for the
\citet{Dahlen:TheAstrophysicalJournal:2013} it is close to 22 and for
\citet{Hildebrandt:AstronomyAndAstrophysics:2010} is it somewhat
fainter. Thus we argue that our outlier fraction estimates are as good
or lower than what has been reported in the literature at comparable
depth.

That said, however, it is important to realise when using photo-zs,
that the outlier fraction at faint magnitudes, even with ultra-deep
photometry, is going to be considerably higher than what is found for
bright galaxy samples such as the $i^{+}<22.5$ COSMOS sample.

\section{The impact of redshift completeness}
\label{sec:redsh-meas}

Our statements above about photo-z accuracy are, by necessity, only
made about galaxies for which we have spectroscopic
redshifts. However, it is also of major interest to know how well the
photometric redshifts work also for galaxies for which we have not
managed to secure a redshift. An accurate assessment of this will of
course have to wait until a deeper data set becomes available, but we
can already make some progress with the data in hand.

\begin{figure}
  \centering
  \includegraphics[width=84mm]{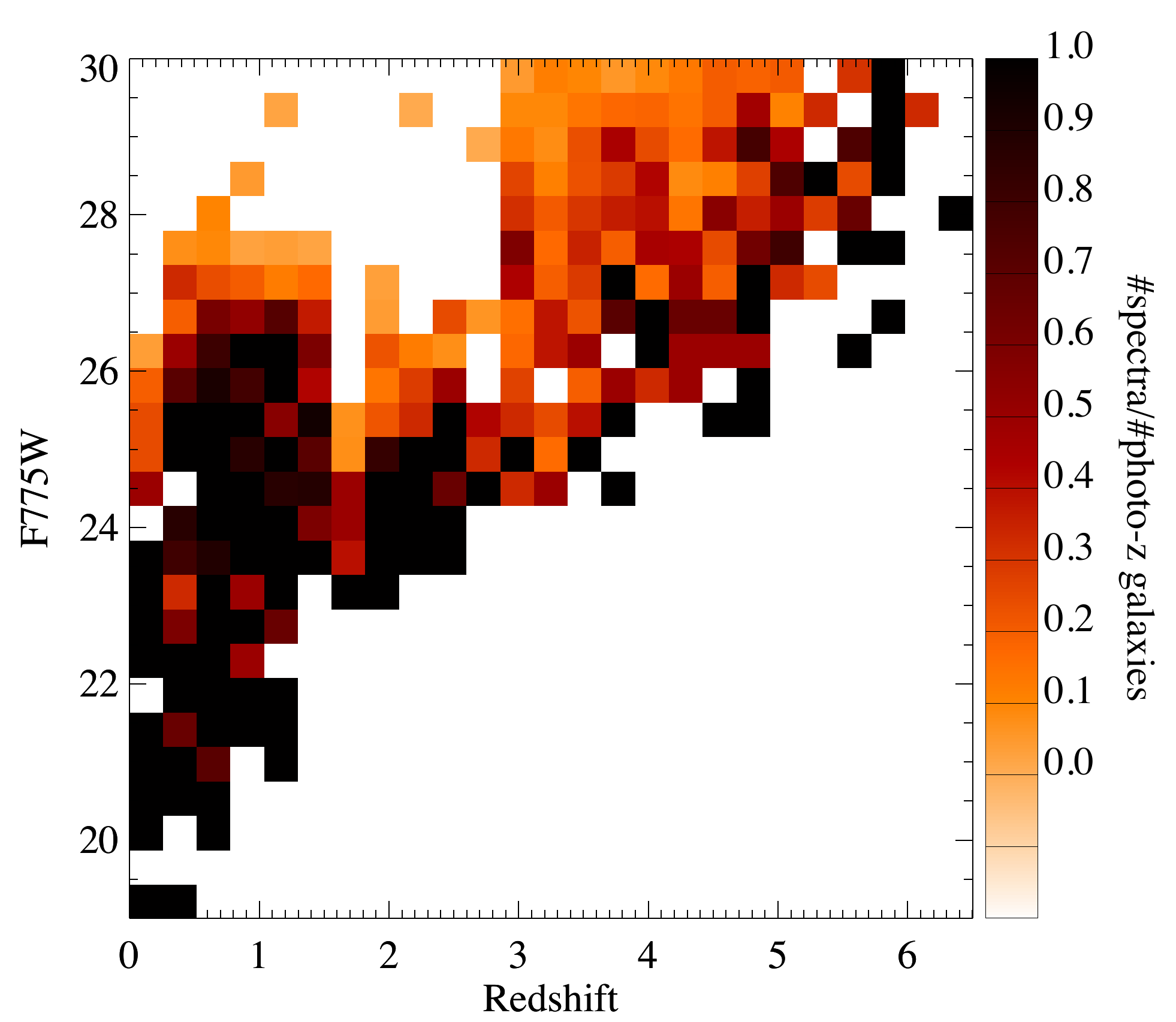}
  \caption{Redshift incompleteness in the redshift-$m_{\mathrm F775W}$
    plane. The fractions have been calculated by binning the objects
    with high confidence ($\ge 2$) MUSE redshifts in a $25\times 25$
    grid and dividing the resulting map by one created by binning
    using the BPZ photometric redshifts.}
  \label{fig:z_mag_completeness}
\end{figure}

Figure~\ref{fig:z_mag_completeness} shows the redshift incompleteness
in the redshift-$m_{\mathrm F775W}$ plane, assuming perfect photometric
redshifts (the $\zphot_{\mathrm{BPZ}}$ were used). It shows very
clearly the preponderance of \lya-emitters at faint magnitudes and the
lack of redshifts in the $1.5<z<2.9$ MUSE redshift desert. While the
completeness at bright magnitudes is close to 1, it is not everywhere
at that level. The first thing to ask is therefore whether this
incompleteness can affect our conclusions about the performance of the
photometric redshifts at bright magnitudes. 

There are only two galaxies with $m_{\mathrm{F775W}} \le 24$ for which
we have no redshift in the catalogue. These are RAF 2231 and
24499. The first of these does in fact have a clear absorption line
redshift ($z=0.66708$) but was missed in the first iteration of
redshifts (c.f. section~\ref{sec:z-determinations})
because of the strong \oii{3727} line from RAF 2236 next to
it. However it is clear from narrow-band images over Ca K that the
absorption line redshift is associated to RAF
2231. Since we recognised this discrepant redshift
  during the photo-z comparison, we have not added these updated
  redshifts here, nor are they in the catalogue in Paper II but will
  be in future updates.

RAF 24499 is very close to a much brighter star (RAF 24467)
which would require a more sophisticated spectrum extraction than we
have adopted here to have a chance to determine its redshift (it
clearly has no very strong emission lines). Thus, the failure to
determine a redshift for this object has nothing to do with its
intrinsic properties. 

Faintwards of $\mhst{F775W}=24$ however, a few objects have no
redshift at present purely due to their intrinsic properties.  There
are 10 objects with $\mhst{F775W}\le 25$ (out of 310 or 3.2\%) that
have no spectroscopic redshift estimate. While we have been unable to
determine a precise redshift, the sources all have clearly detected
continua that are consistent with their photo-z estimates, which fall
mostly in the $1.5<z<2.8$ redshift range where the main diagnostic
feature in the MUSE spectra is the relatively weak \ciii{1909}
doublet \citep{Maseda2017}.

Turning next to the fainter galaxies, we would like to ask whether the
galaxies for which we have no spectroscopic redshift fall in distinct
regions in the space of spectral energy distributions (SEDs). To do
this, we then need to partition this space or project it down to a
smaller-dimensional space. The latter approach has been taken by
\citet{Masters:TheAstrophysicalJournal:2015} who use a self-organising
map to project photometry from the COSMOS survey on to a 2D
plane. While this approach shows significant promise for photo-z
calibration, given our significantly smaller sample, we here adopt the
first option and use k-means clustering to partition the space spanned
by $\mhst{F435W}-\mhst{F606W}$, $\mhst{F606W}-\mhst{F775W}$,
$\mhst{F775W}-\mhst{F850LP}$, and $\mhst{F606W}-\mhst{F125W}$
colours. This is very similar to, and was inspired by, the approach
taken by \citet{Bonnett:PhysicalReviewD:2016} for their investigation
of the photo-zs in the DES shear catalogue.

\begin{figure}
  \centering
  \includegraphics[width=84mm]{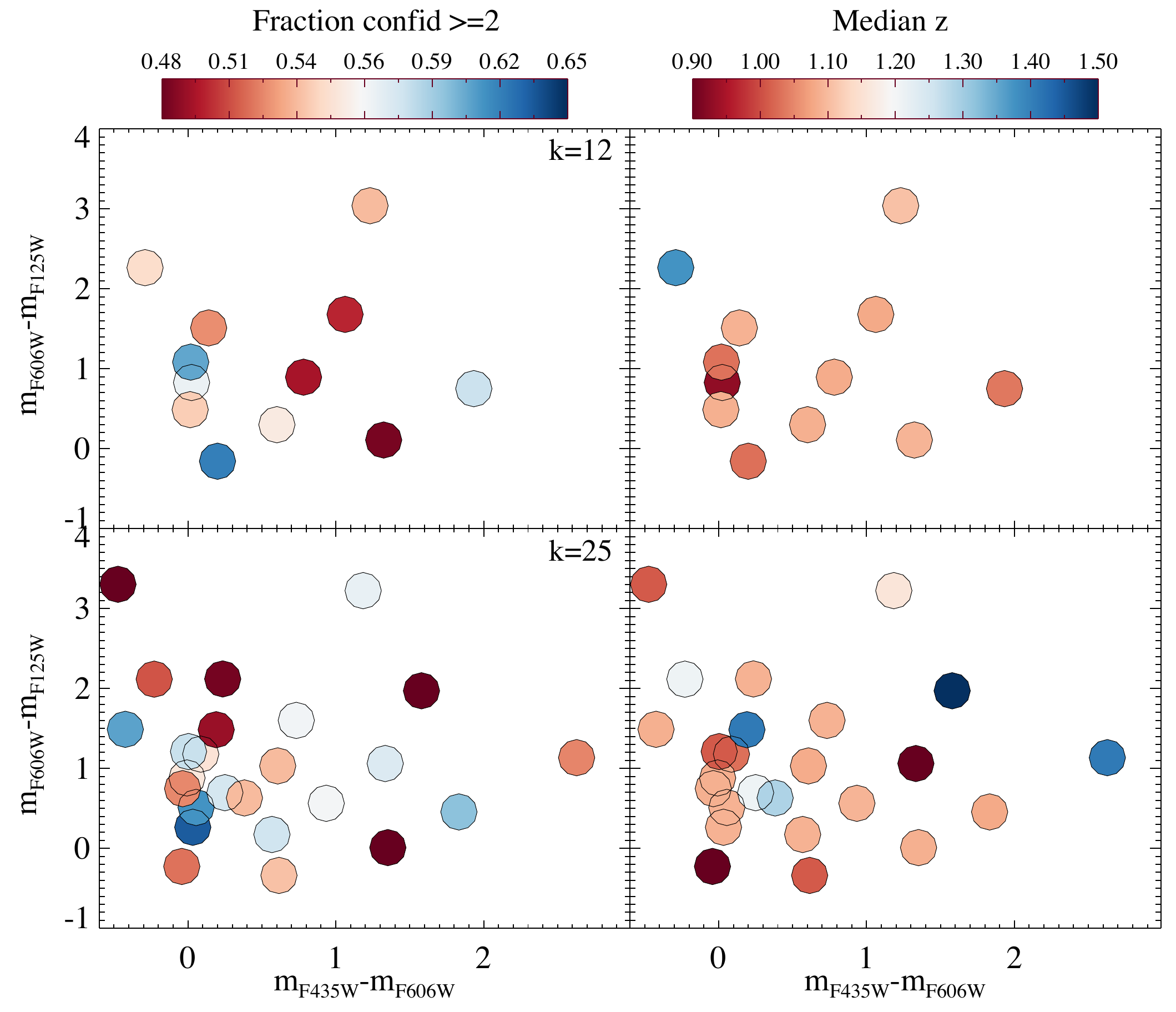}
  \caption{The top left panel shows the fraction of galaxies with
    secure (Confid $>=2$) redshifts in 12 k-means bins (see text for
    details) projected onto the $\mhst{F435W}-\mhst{F606W}$ vs $\mhst{F606W}-\mhst{F125W}$
    plane. The top right panel shows the median redshift in each
    bin. The bottom row shows the same, but this time using 25 k-means
    bins to segment four-colour space. }
  \label{fig:kmeans-illustration}
\end{figure}

The k-means algorithm assigns points in space to k clusters with the
clusters located so that the distances to the cluster centres is
minimal. In two dimensions this is more commonly called a Voronoi
binning of the data and the clusters are called Voronoi cells. Here we
divide our 4-dimensional colour-space in to 12 and 25 bins
respectively. For $k=12$ this leads to bins with
  334--1176 galaxies per bin, with an average of 650, while for $k=25$
  the bins have from 82 to 591 galaxies with a mean of 312. The bins
are chosen by the k-means algorithm and each galaxy is assigned to one
of these bins. The process creates bins that are adapted to the
distribution of data which is very convenient for small data sets such
as ours. To illustrate the properties of the sample,
Figure~\ref{fig:kmeans-illustration} shows the median redshift (right)
and redshift completeness (left) for $k=12$ (top) and
$k=25$ (bottom) bin segmentation of four-colour space. For these
figures we only include galaxies with $\mhst{F775W} \le 27$, but the
qualitative appearance remains when including all galaxies to
$\mhst{F775W}=30$ although the reshift completeness goes down and the
mean redshift goes up.

One main point to note is that while there is variation in completeness
between the different bins, the quantitative variation is modest. It
is also true that with 12 bins, the median redshift in each bin is
fairly close to the overall mean of the sample ($\mathrm{median} (z)=
1.091$), while 25 bins clearly segregate the galaxies somewhat
more but at the obvious cost of fewer galaxies per bin.

\begin{figure}
  \centering
  \includegraphics[width=84mm]{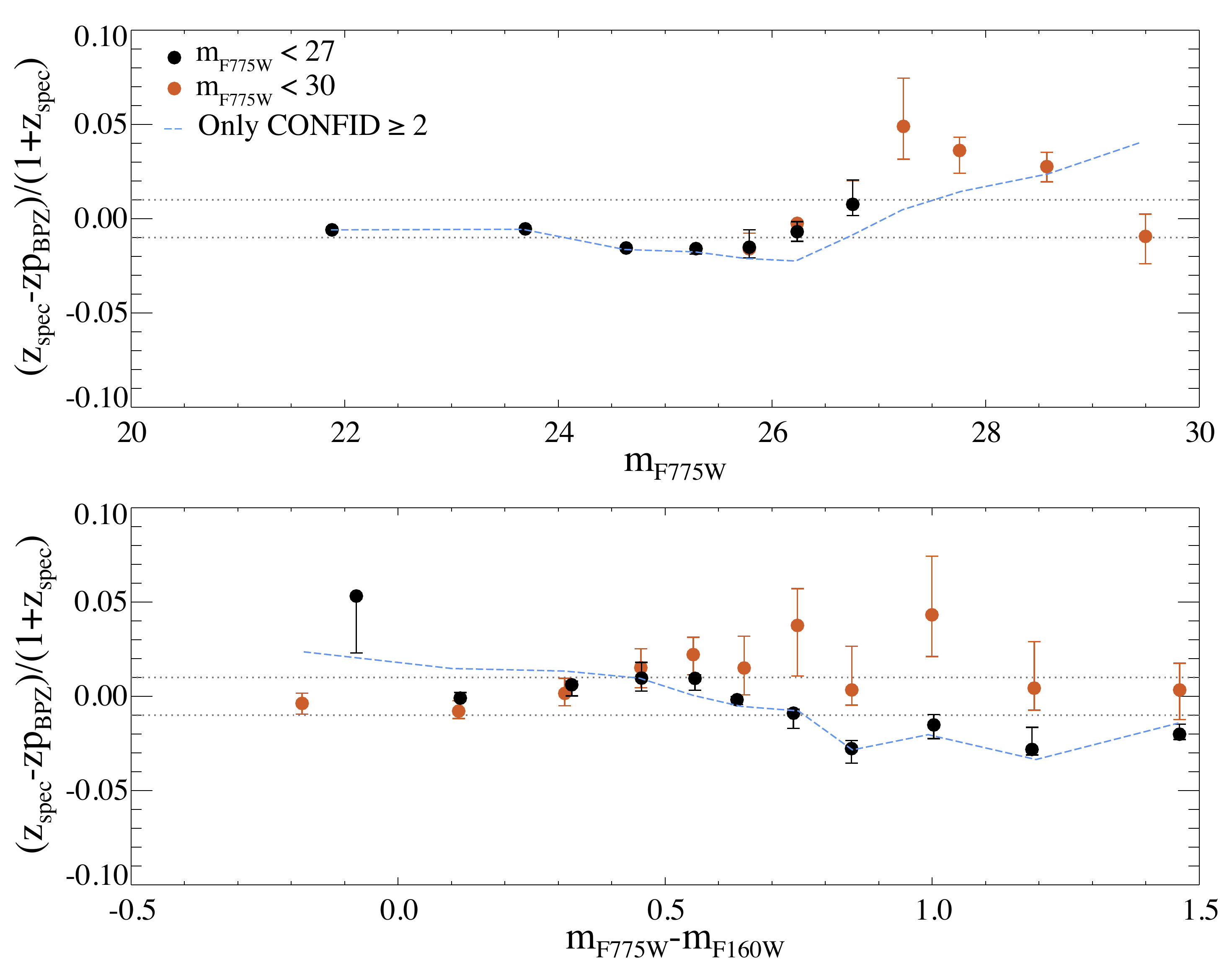}
  \caption{Top panel: The normalised mean bias in photometric redshift
    estimates as a function of apparent magnitude of the galaxy. Bottom
    panel: The same, but now against the $\mhst{F775W}-\mhst{F160W}$
    colour. The dashed black lines show the level corresponding to a
    bias of $\pm  10^{-2}$. The blue dashed line shows the trend found
  when considering galaxies with secure (confidence $\ge 2$) MUSE
  redshifts.}
  \label{fig:redshift_trends_with_incompleteness}
\end{figure}

Moving now to the overall trends of bias in redshift determinations,
it is natural to ask whether the bias we found in
Section~\ref{sec:comp-phot-redsh} is artificially large or small given
that we only considered galaxies with secure spectroscopic
redshifts. To tackle this question, we need to have a way to assign
likely redshifts to the objects for which no spectroscopic redshift
was possible, and we would need to do this independent of our
photometric redshift codes. For this we will use our k-means bins. The
galaxies in each of these bins will have very similar colours so our
assumption is that the distribution of redshifts for the sample
with MUSE redshifts is similar to the distribution of redshifts
for all galaxies in that bin. Given this assumption, we assign each
galaxy without a secure spectroscopic redshift a redshift at random
from the galaxies with a spectroscopic redshift in the k-means bin in which
the galaxy falls. This is of course to some extent a very simple
machine-learning method for photometric redshifts, but we choose to
use the k-means clusters rather than a more sophisticated algorithm as
the method is simple and its limitations are therefore easier to
identify. This approach is rather conservative as the k-means bins are
not particularly small and any residual degeneracy in colour-redshift
space is therefore not resolved.

Figure~\ref{fig:redshift_trends_with_incompleteness} shows the
resulting trend in normalised photo-z bias as a function of magnitude
(top) and $\mhst{F775W}-\mhst{F160W}$ colour (bottom). The black points show the
result when only galaxies with $\mhst{F775W} < 27$ are used, while the orange
points result when the full sample of galaxies with $\mhst{F775W} < 30$ is
used. The photometric redshifts used here are BPZ ones, but the
conclusions are not dependent on this choice. 

The blue line in each panel is the trend found when only considering
galaxies with $\mathrm{Confid}\ge 2$ which we showed in
Figure~\ref{fig:best-bpz}. At bright magnitudes we have very low
spectroscopic incompleteness and as expected the blue line and the
black points fall on top of each other in the top panel. At fainter
magnitudes we predict comparable, but not identical biases in the
redshift determination. To guide the eye a pair of dotted lines at a
bias of $\pm 10^{-2}$ are included --- recall that for cosmological
weak lensing applications a bias $<2\times 10^{-3}$ is desired. It is
clear that for most redshift bins the photo-z codes considered here
are not able to determine the mean redshift to that level, as
commented above, but they are able to constrain it to within
5\%. From this analysis we can tentatively conclude that spectroscopic
incompleteness is unlikely to change this conclusion significantly.

The assignment of redshifts is done in bins of colour so it is not
surprising that the predicted bias shows a dependence on colour,
but we note that the colour used here is not used in the k-means cluster
definition. The bias changes clearly in the redder bins when including
faint galaxies. This is expected and reflects the mean colour with
redshift, but the important point is that at no colour do we see a
mean bias above 6\%, although that should be tempered by noting that
in many bins the bias is larger than 1\%.

We do not show the comparison as a function of redshift because the
method is inherently biased in this case. Each k-means bin has an
upper and lower redshift cut-off given by the sample we have redshifts
for. This is not corrected for in the Monte Carlo approach used above
and as a consequence the mean redshift at high redshift is biased low
and the mean redshift at low redshift is biased high.

\section{The impact of galaxy superpositions and blends}
\label{sec:impact-galaxy-superp}

As discussed in the introduction, for cosmological uses of photometric
redshifts the requirements on biases are now very strict. A frequently
considered approach is therefore to calibrate photometric redshifts
using an existing spectroscopic redshift catalogue. A comprehensive
overview of the different methods is given in
\citet{Newman:AstroparticlePhysics:2015} and references therein, while
\citet{Hildebrandt:Kids450CosmologicalParameterConstraintsFromTomographicWeak:2016}
show the impact of different calibration methods on derived
cosmological parameters.

This approach places considerable constraints on the properties of the
spectroscopic training/calibration sample, and this has been discussed
extensively in the literature as reviewed by
\citeauthor{Newman:AstroparticlePhysics:2015} While the size of the
spectroscopic sample has got much attention,
\citet{Cunha:MonthlyNoticesOfTheRoyalAstronomicalSociety:2014} showed
convincingly that the fraction of incorrect redshifts might have a
severe effect on cosmological parameter determinations and that it is
necessary to limit this to $\sim 1$\%. 

In \citeauthor{Cunha:MonthlyNoticesOfTheRoyalAstronomicalSociety:2014}
and other work in the literature, the focus is on redshifts that are
incorrectly determined from the spectrum but this is only one half of
the problem. To carry out such an endeavour, it is also necessary to be
able to map reliably between spectroscopic redshift and photometric
object. At bright magnitudes this presents few challenges as
comparatively bright objects are well separated on the sky, but at 
fainter magnitudes this is not the case. Thus we might have cases
where the redshift is very certain, but that the assignment of the
redshift to a photometric object might be uncertain --- or indeed in
many cases it might appear certain when in fact it is wrong.  We have
already discussed this fact above in
Section~\ref{sec:sample-definitions}, but we
did not quantify the effect.

The challenge when using a fibre or long-slit spectrograph is that
other, fainter, sources within the aperture sampled by the
spectrograph can contribute emission lines that lead to incorrect
redshift determinations. In the MUSE data this is frequently seen at
faint magnitudes and requires careful inspection of narrow band images
to ensure correct assignment of redshifts to objects. 

\begin{figure}
  \centering
  \includegraphics[width=84mm]{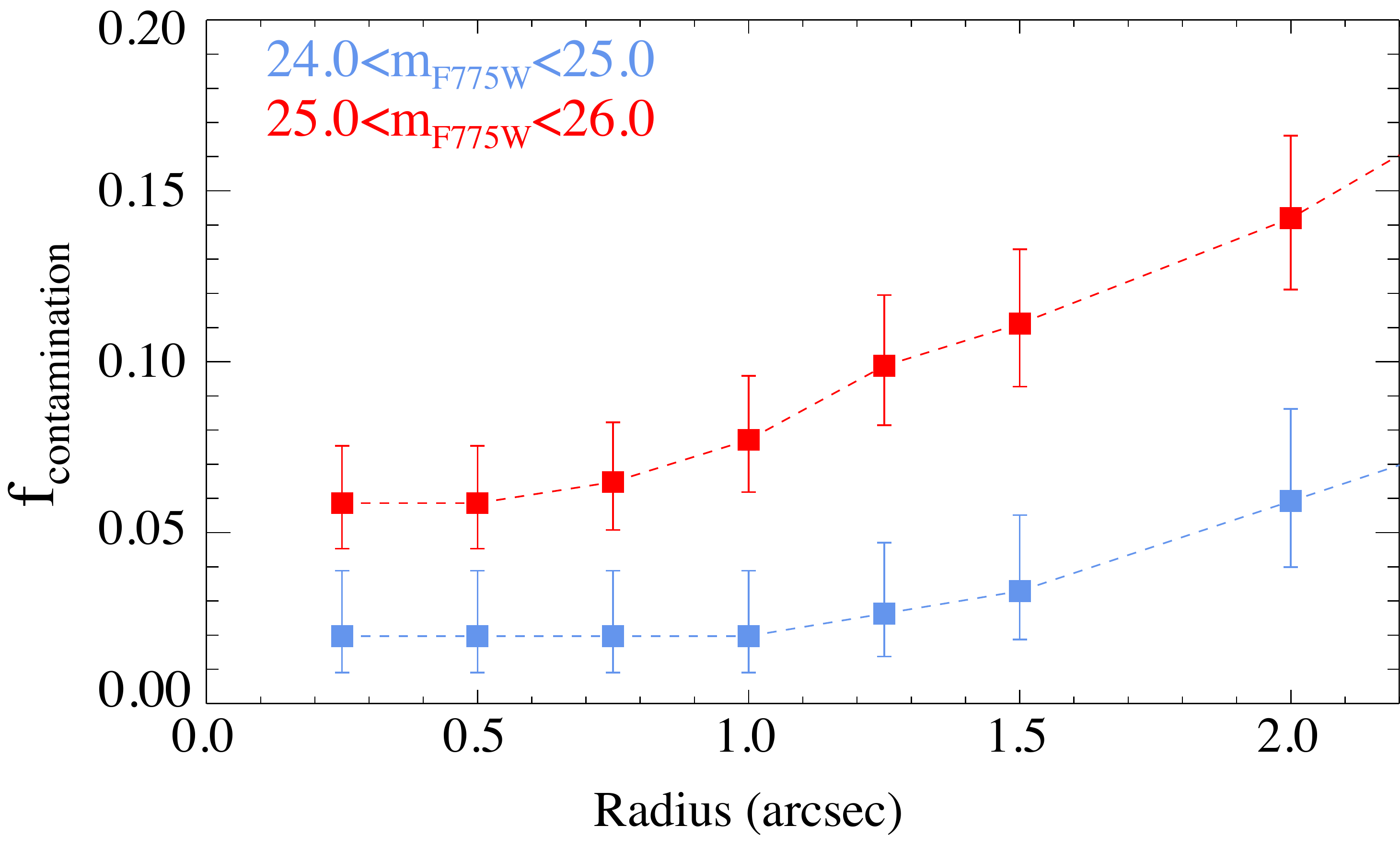}
  \caption{Fraction of potentially contaminated objects as a
    function of search radius. For very small separations our data are
    insufficient to resolve galaxies. Even at the smallest
    separations there is a small level of contamination, amounting to
    a couple of percent. Note also the clear dependence on apparent
    magnitude.}
  \label{fig:f_cont_vs_aperture}
\end{figure}

\begin{figure}
  \centering
  \includegraphics[width=84mm]{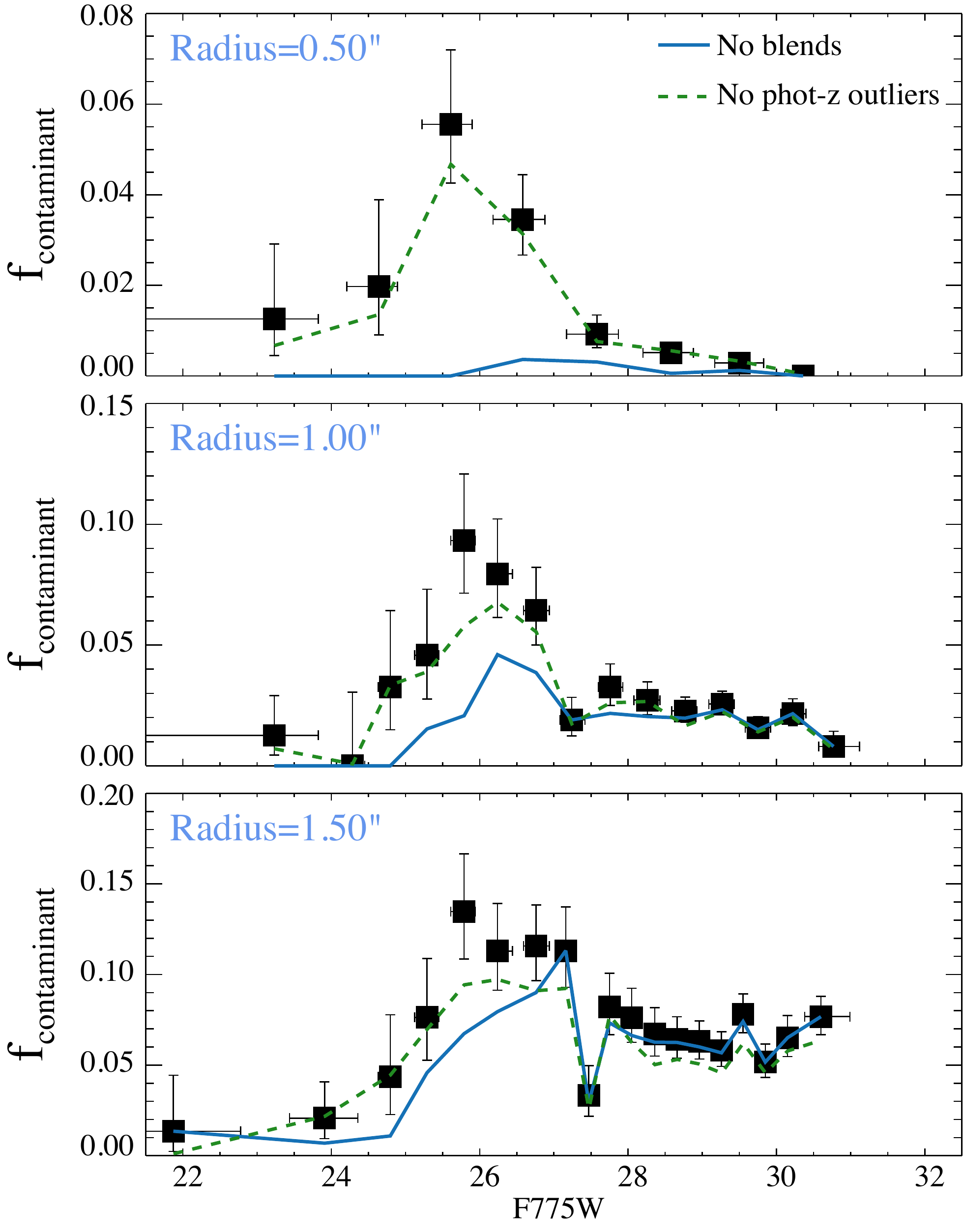}
  \caption{Trend of the fraction of potential contaminants as a
    function of the magnitude of the galaxy. The top panel uses a
    radius of 0\farcs 5 to search for contaminants, the middle panel
    1\farcs 0, and the bottom panel 1\farcs 5. The downturn at faint
    magnitudes with the smallest search radii is a reflection of the
    lower redshift completeness. The solid blue line shows the level
    when we consider only objects we were able to deblend with
    MUSE. The dashed green line shows the result if only galaxies for
    which $\left|\dzn\right|<0.15$ are kept. }
  \label{fig:f_cont_vs_magnitude}
\end{figure}

Here we seek to quantify this effect in an approximate way for the
sample as a whole, we will discuss the shortcomings of this approach
below. Specifically we ask, for each galaxy, whether there is a galaxy
with a fainter $\mhst{F775W}$ within a circular region of radius $r$
that has a confidence $\ge 2$ emission line redshift, and where the
strongest line in the fainter galaxy has a higher flux than the
strongest line in the galaxy considered.  If this is the case, we
consider that this is a possible contamination source for an
observation taken with an aperture or seeing comparable to the size
$r$.  This effectively assumes that the contamination source is a
point source whereas in truth they are extended. In addition we also
keep the sources that we are unable to resolve, even with MUSE, as a
possible source of contamination. For these we lack the neccessary
information to treat these on the same footing, so we assume the line
detected comes from the faintest object but would be assigned to the
brightest. We also do not know the spatial separation for these and
set it to zero --- this will then give a minimum level of
contamination at all resolutions which is clearly unphysical but with
the present data we cannot do better.

Figure~\ref{fig:f_cont_vs_aperture} shows the fraction of galaxies
with a potential contaminant as a function of separation for two
magnitude ranges. The contamination potential for objects with
magnitude $24<m_{F775W}<25$ (blue line) is lower than that for objects
one magnitude fainter (red line). The general shape in both cases is
as expected, with the larger search area increasing the chance of
finding a contaminant. The zero level at small radii is caused by the
blended objects as discussed above.

The magnitude dependence clearly indicates that we should also inspect
this as a function of magnitude. This is shown in
Figure~\ref{fig:f_cont_vs_magnitude} which shows the trends as a
function of magnitude in three different aperture sizes. There are two
noticeable trends: a drop at bright magnitudes and another at faint
magnitudes. The drop at faint magnitudes is a consequence of the
declining redshift completeness at fainter magnitudes. The blue solid
line shows the effect of ignoring all blended objects --- since these
truly are confused this is of course an over-simplification but it
shows that the majority of the blended objects at bright magnitudes
are impossible to deblend even with MUSE.

For a spectroscopic observation with a spatial resolution of 0\farcs
5, one can expect significant contamination in $\approx 1$\% of all
spectra, while for a larger aperture, say a fibre spectrograph with
2\arcsec aperture, typically $\approx 3$\% of all spectra will have
stronger emission lines from a galaxy that is not the brightest source
in the aperture, at least for galaxies fainter than
$\mathrm{F775W}=24$ which is where the UDF sample has significant
number of galaxies. That this is a real concern can also be seen in
the fact reported in Paper II, that 2 of 12 redshifts in this area
from the VIMOS Ultra Deep Survey (VUDS) data release 1
\citep{Tasca:AstronomyAstrophysics:2017}, are incorrectly associated
with HST objects due to contamination. VUDS primarily selects objects
with $i_{\mathrm{AB}}\ge 25$ so this rate of 17\% can be contrasted
with our estimate of $\sim 5$\% at a similar depth for a $1\arcsec$
radius.

That the VUDS contamination is larger is not entirely surprising as
our our estimate is in fact likely to be an understimate at radii less
than a couple of arcseconds. This is because the contaminating sources
are typically extended, as spectacularly seen for \lya\ emitters in
MUSE data
\citep[e.g.][]{Wisotzki:AstronomyAndAstrophysics:2016,Leclercq2017}. Thus
in reality the contaminants will not be delta functions in spatial
extent and this will boost the number of contaminants at small
separations. However as this depends on the line flux profiles and
might be mitigated by examining spatial line profiles in slit spectra,
a precise quantification of this effect is beyond the present work.

The importance of this contamination depends of course on whether it
leads to incorrect redshift determinations for the galaxy. This will
clearly depend on the depth of spectroscopy. With a well-detected
continuum it is more likely that one can recognise that the emission
line does not belong to that galaxy or not, but for very faint
galaxies it is probable that this contamination will lead to incorrect
redshift determinations and hence biases in the calibration of
photometric redshifts.

One might also imagine that the blends might be identified as outliers
in photo-z versus spectroscopic redshift comparisons. To test the
effect of this, we assigned each blended object the redshift implied
by the strongest line, but in the case of multiple contaminating
sources we chose the one that would minimise the discrepancy with the
photometric redshift. We then calculated the fraction of contaminants
only counting those objects with $\left|\dzn\right| < 0.15$. The
result of this test is shown by the dashed green line in
Figure~\ref{fig:f_cont_vs_magnitude} and as can be seen this reduces
the contamination problem somewhat but does not remove it, the effect
is similar if blends are removed. At faint magnitudes the overall
effect of the $\dzn$ cut is very small because almost all sources
involved are \lya\ emitters.

\section{Discussion}
\label{sec:discussion}

The requirements on photometric redshift accuracy naturally depend on
the scientific question that they are needed for. We have found that
the mean bias in redshift determination is always below $0.05(1+z)$
even down to $\mhst{F775W}=30$ with a $\sigma_{\mathrm{MAD}}<0.06$ in
$\dzn$ for all but the $0.4<z<0.6$ bin. While this might appear to be
a small bias, it is clearly too high for cosmological needs. For many
galaxy evolution studies, however, it is an acceptable bias. That
said, the clear systematic trends with both magnitude, redshifts and
colour are important to note and also be aware that the scatter varies
somewhat less with these properties.

\begin{figure*}
  \centering
  \includegraphics[width=184mm]{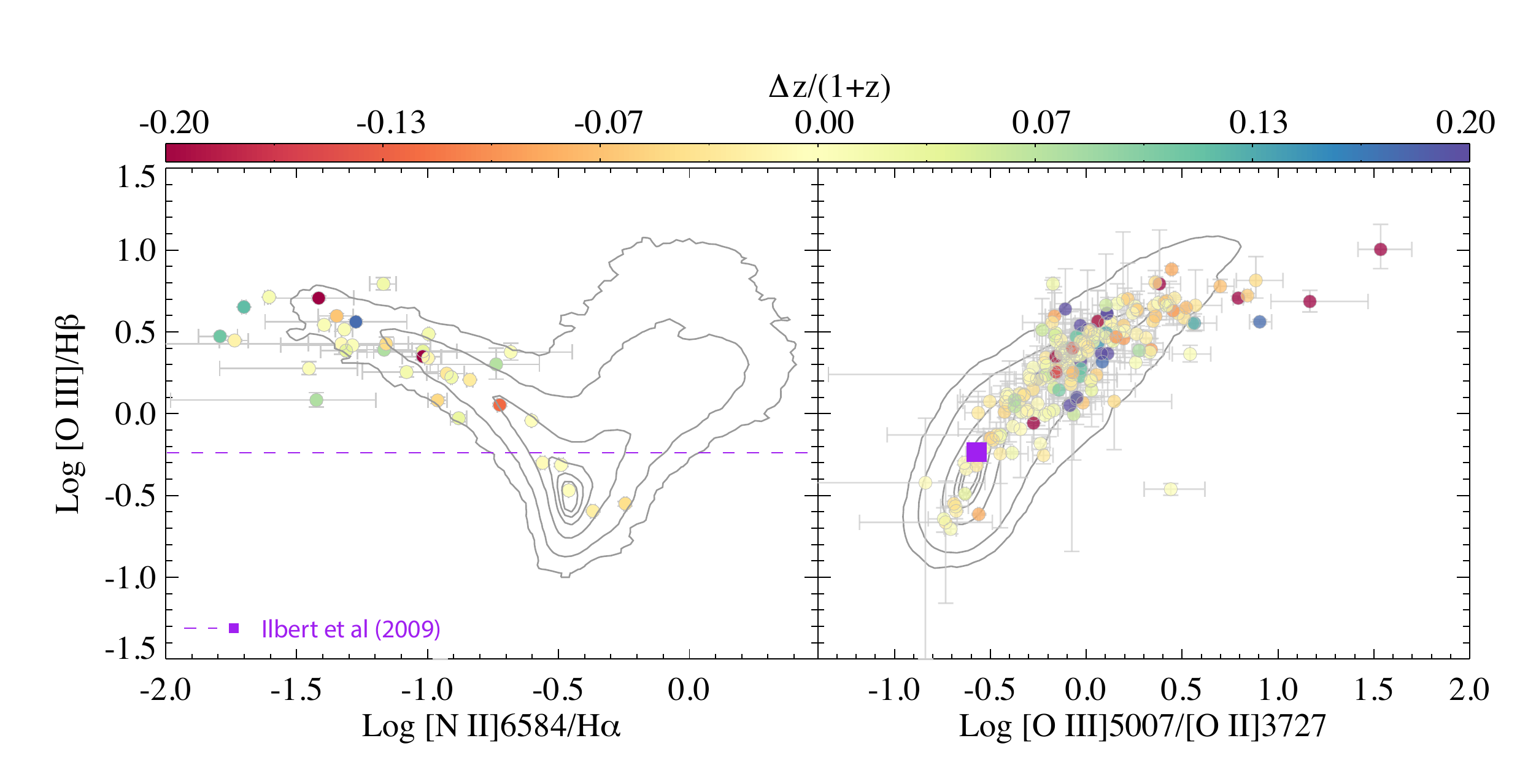}
  \caption{Left: the BPT \nii{6584}/\ha\ vs \oiii{5007}/\hb\
    diagram. The location of SDSS galaxies is shown by the grey
    contours with the outermost contour encolosing 99\% of the
    galaxies. The MUSE UDF sources are plotted in top with colour
    encoding the discrepancy between BPZ photometric and spectroscopic
    redshift as given by the colour key on top. On the right the
    \oiii{5007}/\oii{3727} vs \oiii{5007}/\hb\ diagram (see also
    Paalvast et al 2017, in prep.) showing the same. The purple line and square
    shows the line ratios proposed for photo-z work by \citet{27739469}.}
  \label{fig:line_ratios_vs_dz}
\end{figure*}

We find two regions in redshift space where spectroscopic and
photometric redshifts differ significantly. These are at low redshift,
$0.5<z<1.5$, and at higher redshift $z>3.5$. Armed with the MUSE data
it is reasonable to ask whether the discrepancies we see at low
redshift can be due to the spectral templates being systematically
unrepresentative. To explore this, Figure~\ref{fig:line_ratios_vs_dz}
shows the location of our MUSE sources in two different line ratio
diagrams where we have used the reference (weighted) line measurements
although this particular choice has no significant influence on the
results shown here. The figure on the left shows the classical BPT
\nii{6584}/\ha\ vs \oiii{5007}/\hb\ diagram. This is appropriate for
the lowest redshift galaxies where \ha\ is still available. To give
context to the MUSE galaxies, we make use of line flux measurements
for galaxies from the Sloan Digital Sky Survey (SDSS) Data Release 7
\citep{Abazajian2009} from the MPA-JHU
compilation\footnote{\url{http://www.mpa-garching.mpg.de/SDSS/DR7}}
which is an update of that discussed in \citet{Brinchmann2004}. The
distribution of the SDSS galaxies with S/N in each line $>3$ are shown
by the grey contours where the outermost contour encloses 99\% of
galaxies. It is immediately obvious that our objects fall mostly
within the locus of SDSS galaxies albeit with a relative overabundance
of lower metallicity (low \nii{6584}/\ha\ ratio) galaxies. The colour
gives the relative discrepancy between BPZ photometric and
spectroscopic redshift, $\Delta z/(1+z)$. There is no evidence of an
increased discrepancy away from the bulk of the SDSS galaxies in this
figure.

The right-hand panel shows the same but now plotting
\oiii{5007}/\oii{3727} on the x-axis. This line ratio is more
comprehensively explored, for a larger sample, in Paalvast et al (in
prep), here we simply use it as a way to explore the spectroscopic
properties of the higher redshift part of the sample. The meaning of
the symbols and lines are the same and there is perhaps some evidence
that the most extreme \oiii{5007}/\oii{3727} galaxies show increased
photo-z discrepancy and this might reflect an incomplete template
sample. However, the bulk does not show such a correlation so this is
unlikely to explain the offset in the median between photometric and
spectroscopic redshifts.

In the literature, line ratios are either predicted by the code
providing the underlying SEDs for photo-z fitting
\citep[e.g.][]{Brammer:TheAstrophysicalJournal:2008}, but more
commonly line ratios are taken from some compilation and the flux of
the lines is tied to the star formation rate of the SED template. The
most widely used line ratios are those provided by
\citet{27739469}. Those authors compiled observed line
  ratios (relative to \oii{3727}) provided in the literature for a
  mixture of galaxies at $z<0.1$. The resulting
line ratios are plotted as the dashed purple line in the left-hand
panel and as the purple square in the right-hand panel. It is clear
that these deviate from the bulk of the galaxies in the
MUSE sample. This is not entirely surprising given that the
  line ratios used by Ilbert et al are more representative for low-z
  massive galaxies than our sample of faint, star-forming
  galaxies. For application to fainter sources it would therefore be
desirable to have a more comprehensive treatment of emission lines
such as that taken by \citet{Salmon:TheAstrophysicalJournal:2015} who
use a library of simulated emission line spectra for metal-poor
galaxies from
\citet{Inoue:MonthlyNoticesOfTheRoyalAstronomicalSociety:2011}, or the
BEAGLE
\citep{Chevallard:MonthlyNoticesOfTheRoyalAstronomicalSociety:2016}
which exploits a large library of photoionisation models from
\citet{Gutkin:MonthlyNoticesOfTheRoyalAstronomicalSociety:2016}.

The importance of emission line ratios does however appear to be
relatively minor given the lack of clear correlations seen for our
sample. One might instead worry that the offsets would correlate with
the relative strength of the emission lines. We have explored this by
looking for correlations between $\Delta z$ and the equivalent width
of the strongest lines. Our assumption here is that this line is
representative for all emission lines in the spectrum. Except for
tentative evidence of the 2--3 systems with the highest \lya\
equivalent width to have slightly higher $\Delta z$, we have not found
any robust evidence for the equivalent width to have an impact on the
photo-z estimates. Since the strengths of lines are tied to the star
formation rate of each template in the template fitting codes, this is
not entirely unexpected.

Taken together with the lack of clear correlation between bias and
location in the k-means bins in Section~\ref{sec:redsh-meas}, we find
no obvious ``second-parameter'' that can be used to reduce the
photometric redshifts errors. A logical next step would be to combine
multiple photometric redshift estimators to create a consensus
photo-z
\citep[e.g.][]{CarrascoKind:MonthlyNoticesOfTheRoyalAstronomicalSociety:2014,Cavuoti:MonthlyNoticesOfTheRoyalAstronomicalSociety:2017,Suveges:LearnFromEveryMistakeHierarchicalInformationCombination:2017},
but we postpone this to a future work.

The addition of more photometric bands might offer another way to
reduce bias and indeed \citet{Bezanson:TheAstrophysicalJournal:2016}
showed convincingly that increasing the number of bands, and in
particular adding IRAC photometry does improve photo-z performance
significantly for brighter galaxies. In
Appendix~\ref{sec:adding-ground-based} we explore the effect of using
the 44 bands in the S14 photometric catalogue for the fainter galaxies
considered here. We find there that indeed the bias is lower when all
44 bands are included, but at a cost of a significiantly increased
scatter, particularly at faint magnitudes. Thus for faint
$\mhst{F775W}>25$ galaxies the current photometric catalogues from the
ground and Spitzer are not of sufficient quality to help improve
photo-z determinations.

While reducing the bias for individual galaxies is important,
re-calibration methods either using spectroscopic redshifts directly
\citep[e.g.][]{Lima:MonthlyNoticesOfTheRoyalAstronomicalSociety:2008}
or using spatial cross-correlations
\citep{Newman:TheAstrophysicalJournal:2008} show great potential to
reduce the bias in the estimated mean redshift of a sample of
galaxies.  However these all need correct redshifts, and we have
demonstrated above that for a few percent of galaxies the redshifts
might be of high confidence, but may be assigned to the wrong
photometric object. This presents significant challenges for future
surveys \citep{Cunha:MonthlyNoticesOfTheRoyalAstronomicalSociety:2014}
and ways must be found to counter this --- particularly for studies
that go fainter than $24^{\mathrm{th}}$ magnitude, such as the LSST
weak lensing survey.  The most obvious, but less practical, way to
reduce this problem is to use spatially resolved spectroscopy for the
fainter galaxy population. The various data sets obtained by MUSE are
particularly well suited for this. While not quite as powerful because
of the loss of information, slit spectra also provide a way to at
least address some of these problems and spectra should ideally have
not just a quality flag for the redshift, but also one for the
association to photometric objects. There is no easy way to tackle
this problem in surveys using fibre-based spectrographs, and the
residual uncertainy will remain and must be included as a possible
source of error in subsequent analysis.

\section{Conclusions}
\label{sec:conclusions}

Our aim with this study was to quantify the performance of photometric
redshifts in galaxies down to $\mhst{F775W}=30$ with predominantly
emission line galaxies at $\mhst{F775W} > 25.5$. We have extended the
validation of photometric redshifts to fainter magnitude limits than
previous studies and found that they do in general perform well
throughout. The median redshift in bins of redshifts, colour and
magnitude is always determined to $\left|\dzn\right| <0.05$ for the
galaxies for which we have spectroscopic redshifts from MUSE as shown
in Figures~\ref{fig:best-eazy}--\ref{fig:best-beagle}. This conclusion
appears insensitive to the choice of photo-z estimation code with
EAZY, BPZ and BEAGLE all performing comparably well. This does give us
some confidence in results based on photo-zs at faint magnitudes. We
do, however find systematic trends with redshift and colour: the
photometric redshifts being systematically biased high for $0.4<z<1.5$
by $\dzn=-0.02$ to $-0.04$, and biased low by a similar to slightly
higher amount at $z>3$.

While the cause of the deviation at low redshift is unclear, we showed
in Figure~\ref{fig:igm_dependence_of_bias} that allowing for free
scaling of the IGM model for \lya-forest and Lyman continuum
absorption does reduce the discrepancy and that it is possible to
remove all bias from galaxies with $\left|\dzn\right|<0.1$ by
adjusting the amount of IGM absorption for each galaxy
individually, although the amount required appear to be inconsistent
with the scatter seen in \lya-forest absorption data indicating that
the changing the IGM modelling is unlikely to be the full
explanation. 

The number of objects with strongly discrepant photometric redshifts
is considerably higher than that found in brighter subsamples of
galaxies in the literature. However we also demonstrate in
Figure~\ref{fig:rel_outlier_fraction_vs_many} that the fraction is a
strong function of how an outlier is defined, but one should at least
expect an outlier fraction of 10--20\% for these faint galaxies, where
an outlier is defined as having $\left|\dzn\right|>0.15$. The sample
is too small to make strong statements on catastrophic outliers,
defined as those satisfying $\left|\dzn\right|>0.5$ or
$\left|\dzn\right|>5\sigma_{\mathrm{MAD}}$, but we did show that EAZY
has 2--3 times more catastrophic outliers than BEAGLE or
BPZ. Encouragingly, only $<0.6$\% of the galaxies are catastrophic
outliers for all codes simultaneously (Figure~\ref{fig:olf} and
Table~\ref{tab:significant_outliers}), indicating that comparisons of
codes might help identify possible problem cases.

Finally, we found that the fraction of spatially overlapping galaxies
is sufficiently high to be of concern for cosmological surveys using
faint galaxies for weak lensing studies. We showed in
Figure~\ref{fig:f_cont_vs_magnitude} that at $\mhst{F775W}=25$, 1--2\%
of all galaxies will have a fainter source within 0.5'' that has a
stronger emission line than the strongest line in the galaxy in
question. While the redshift determined from the emission line(s)
might be a very secure redshift, the association of the redshift to
the photometric object will be incorrect, leading to incorrect
redshifts in photo-z calibration samples at this percentile
level. This problem is expected to become considerably more severe for
spectroscopic surveys of faint galaxies using fibres, where $>5$\% of
the galaxies might be affected by this, depending on the fibre
diameter, and it is clearly necessary to develop mitigating strategies
for this problem. Unfortunately a simple constraint on the difference
between photometric and spectroscopic redshift is not enough as we
showed in the same figure, thus a more sophisticated approach is
called for and ultimately a statistical approach to correct for the
effect might be needed.

\begin{acknowledgements}
  We thank the referee for a constructive and helpful report that
  helped improve the paper. JB acknowledges useful discussions with
  Peter Capak. This work was in part supported by Funda{\c c}{\~a}o
  para a Ci{\^e}ncia e a Tecnologia (FCT) through national funds
  (UID/FIS/04434/2013) and by FEDER through COMPETE2020
  (POCI-01-0145-FEDER-007672). During part of this work, JB was
  supported by FCT through Investigador FCT contract
  IF/01654/2014/CP1215/CT0003.  This work is supported by the ERC
  advanced grant 339659-MUSICOS (R. Bacon). TC acknowledges support of
  the ANR FOGHAR (ANR-13-BS05-0010-02), the OCEVU Labex (ANR-11-
  LABX-0060) and the A*MIDEX project (ANR-11-IDEX-0001-02) funded by
  the “Investissements d’avenir” French government program managed by
  the ANR. JR acknowledges support from the ERC starting grant
  336736-CALENDS. JS acknowledges support from ERC Grant
  278594-GasAroundGalaxies.  LW acknowledges funding by the
  Competitive Fund of the Leibniz Association through grant
  SAW-2015-AIP-2. RAM acknowledges support by the Swiss National
  Science Foundation. SC acknowledges support from the ERC advanced
  grant 321323-NEOGAL.
\end{acknowledgements}


\begin{thebibliography}{74}
\expandafter\ifx\csname natexlab\endcsname\relax\def\natexlab#1{#1}\fi

\bibitem[{Abazajian {et~al.}(2009)Abazajian, Adelman-McCarthy, Ag\"{u}eros,
  Allam, Prieto, An, Anderson, Anderson, Annis, Bahcall, Bailer-Jones,
  Barentine, Bassett, Becker, Beers, Bell, Belokurov, Berlind, Berman,
  Bernardi, Bickerton, Bizyaev, Blakeslee, Blanton, Bochanski, Boroski,
  Brewington, Brinchmann, Brinkmann, Brunner, Budav\'{a}ri, Carey, Carliles,
  Carr, Castander, Cinabro, Connolly, Csabai, Cunha, Czarapata, Davenport,
  de~Haas, Dilday, Doi, Eisenstein, Evans, Evans, Fan, Friedman, Frieman,
  Fukugita, G\"{a}nsicke, Gates, Gillespie, Gilmore, Gonzalez, Gonzalez,
  Grebel, Gunn, Gy\"{o}ry, Hall, Harding, Harris, Harvanek, Hawley, Hayes,
  Heckman, Hendry, Hennessy, Hindsley, Hoblitt, Hogan, Hogg, Holtzman, Hyde,
  Ichikawa, Ichikawa, Im, Ivezi\'{c}, Jester, Jiang, Johnson, Jorgensen,
  Juri\'{c}, Kent, Kessler, Kleinman, Knapp, Konishi, Kron, Krzesinski,
  Kuropatkin, Lampeitl, Lebedeva, Lee, Lee, Leger, L\'{e}pine, Li, Lima, Lin,
  Long, Loomis, Loveday, Lupton, Magnier, Malanushenko, Malanushenko,
  Mandelbaum, Margon, Marriner, Mart\'{i}nez-Delgado, Matsubara, McGehee,
  McKay, Meiksin, Morrison, Mullally, Munn, Murphy, Nash, Nebot, Neilsen,
  Newberg, Newman, Nichol, Nicinski, Nieto-Santisteban, Nitta, Okamura,
  Oravetz, Ostriker, Owen, Padmanabhan, Pan, Park, Pauls, Peoples, Percival,
  Pier, Pope, Pourbaix, Price, Purger, Quinn, Raddick, Fiorentin, Richards,
  Richmond, Riess, Rix, Rockosi, Sako, Schlegel, Schneider, Scholz, Schreiber,
  Schwope, Seljak, Sesar, Sheldon, Shimasaku, Sibley, Simmons, Sivarani, Smith,
  Smith, Smol\v{c}i\'{c}, Snedden, Stebbins, Steinmetz, Stoughton, Strauss,
  SubbaRao, Suto, Szalay, Szapudi, Szkody, Tanaka, Tegmark, Teodoro, Thakar,
  Tremonti, Tucker, Uomoto, Vanden~Berk, Vandenberg, Vidrih, Vogeley, Voges,
  Vogt, Wadadekar, Watters, Weinberg, West, White, Wilhite, Wonders, Yanny,
  Yocum, York, Zehavi, Zibetti, \& Zucker}]{Abazajian2009}
Abazajian, K.~N., Adelman-McCarthy, J.~K., Ag\"{u}eros, M.~a., {et~al.} 2009,
  The Astrophysical Journal Supplement Series, 182, 543

\bibitem[{Abdalla {et~al.}(2011)Abdalla, Banerji, Lahav, \&
  Rashkov}]{Abdalla:MonthlyNoticesOfTheRoyalAstronomicalSociety:2011}
Abdalla, F.~B., Banerji, M., Lahav, O., \& Rashkov, V. 2011, Monthly Notices of
  the Royal Astronomical Society, 417, 1891

\bibitem[{Acquaviva {et~al.}(2015)Acquaviva, Raichoor, \&
  Gawiser}]{Acquaviva:TheAstrophysicalJournal:2015}
Acquaviva, V., Raichoor, A., \& Gawiser, E. 2015, The Astrophysical Journal,
  804

\bibitem[{Arnouts {et~al.}(1999)Arnouts, Cristiani, Moscardini, Matarrese,
  Lucchin, Fontana, \&
  Giallongo}]{Arnouts:MonthlyNoticesOfTheRoyalAstronomicalSociety:1999}
Arnouts, S., Cristiani, S., Moscardini, L., {et~al.} 1999, Monthly Notices of
  the Royal Astronomical Society, 310, 540

\bibitem[{Bacon {et~al.}(2010)Bacon, Accardo, Adjali, Anwand, Bauer, Biswas,
  Blaizot, Boudon, Brau-Nogue, \& Brinchmann}]{bacon2010muse}
Bacon, R., Accardo, M., Adjali, L., {et~al.} 2010, in SPIE Astronomical
  Telescopes+ Instrumentation, International Society for Optics and Photonics,
  773508--773508

\bibitem[{Bacon {et~al.}(2015)Bacon, Brinchmann, Richard, Contini, Drake,
  Franx, Tacchella, Vernet, Wisotzki, Blaizot, Bouch\'{e}, Bouwens, Cantalupo,
  Carollo, Carton, Caruana, Cl\'{e}ment, Dreizler, Epinat, Guiderdoni, Herenz,
  Husser, Kamann, Kerutt, Kollatschny, Krajnovic, Lilly, Martinsson,
  Michel-Dansac, Patricio, Schaye, Shirazi, Soto, Soucail, Steinmetz, Urrutia,
  Weilbacher, \& de~Zeeuw}]{Bacon:AstronomyAndAstrophysics:2015}
Bacon, R., Brinchmann, J., Richard, J., {et~al.} 2015, Astronomy and
  Astrophysics, 575, 75

\bibitem[{{Bacon} {et~al.}(2017){Bacon}, {Conseil}, {Mary},
  {et~al.}}]{Bacon2017}
{Bacon}, R., {Conseil}, D., {Mary}, D., {et~al.} 2017, \aap, in press (MUSE UDF
  SI paper I)

\bibitem[{Beck {et~al.}(2017)Beck, Lin, Ishida, Gieseke, de~Souza,
  Costa-Duarte, Hattab, Krone-Martins, \&
  Collaboration}]{Beck:EprintArxiv170108748:2017}
Beck, R., Lin, C.~A., Ishida, E. E.~O., {et~al.} 2017, eprint arXiv:1701.08748

\bibitem[{Beckwith {et~al.}(2006)Beckwith, Stiavelli, Koekemoer, Caldwell,
  Ferguson, Hook, Lucas, Bergeron, Corbin, Jogee, Panagia, Robberto, Royle,
  Somerville, \& Sosey}]{Beckwith:TheAstronomicalJournal:2006}
Beckwith, S. V.~W., Stiavelli, M., Koekemoer, A.~M., {et~al.} 2006, The
  Astronomical Journal, 132, 1729

\bibitem[{Ben\'{i}tez(2000)}]{Benitez:TheAstrophysicalJournal:2000}
Ben\'{i}tez, N. 2000, The Astrophysical Journal, 536, 571

\bibitem[{Bezanson {et~al.}(2016)Bezanson, Wake, Brammer, van Dokkum, Franx,
  Labb\'{e}, Leja, Momcheva, Nelson, Quadri, Skelton, Weiner, \&
  Whitaker}]{Bezanson:TheAstrophysicalJournal:2016}
Bezanson, R., Wake, D.~A., Brammer, G.~B., {et~al.} 2016, The Astrophysical
  Journal, 822

\bibitem[{Blanton \& Roweis(2007)}]{Blanton2007}
Blanton, M.~R. \& Roweis, S. 2007, The Astronomical Journal, 133, 734

\bibitem[{Bolzonella {et~al.}(2000)Bolzonella, Miralles, Pell\'{o}, Pell\'{o},
  \& R.}]{Bolzonella:AstronomyAndAstrophysics:2000}
Bolzonella, M., Miralles, J.~M., Pell\'{o}, R., Pell\'{o}, \& R. 2000,
  Astronomy and Astrophysics, 363, 476

\bibitem[{Bonnett {et~al.}(2016)Bonnett, Troxel, Hartley, Amara, Leistedt,
  Becker, Bernstein, Bridle, Bruderer, Busha, Carrasco~Kind, Childress,
  Castander, Chang, Crocce, Davis, Eifler, Frieman, Gangkofner, Gaztanaga,
  Glazebrook, Gruen, Kacprzak, King, Kwan, Lahav, Lewis, Lidman, Lin, MacCrann,
  Miquel, ONeill, Palmese, Peiris, Refregier, Rozo, Rykoff, Sadeh, S\'{a}nchez,
  Sheldon, Uddin, Wechsler, Zuntz, Abbott, Abdalla, Allam, Armstrong, Banerji,
  Bauer, Benoit-L\'{e}vy, Bertin, Brooks, Buckley-Geer, Burke, Capozzi,
  Carnero~Rosell, Carretero, Cunha, DAndrea, da~Costa, DePoy, Desai, Diehl,
  Dietrich, Doel, Fausti~Neto, Fernandez, Flaugher, Fosalba, Gerdes, Gruendl,
  Honscheid, Jain, James, Jarvis, Kim, Kuehn, Kuropatkin, Li, Lima, Maia,
  March, Marshall, Martini, Melchior, Miller, Neilsen, Nichol, Nord, Ogando,
  Plazas, Reil, Romer, Roodman, Sako, Sanchez, Santiago, Smith, Soares-Santos,
  Sobreira, Suchyta, Swanson, Tarle, Thaler, Thomas, Vikram, \&
  Walker}]{Bonnett:PhysicalReviewD:2016}
Bonnett, C., Troxel, M.~., Hartley, W., {et~al.} 2016, Physical Review D, 94

\bibitem[{Bouwens {et~al.}(2011)Bouwens, Illingworth, Oesch, Labb\'{e}, Trenti,
  van Dokkum, Franx, Stiavelli, Carollo, Magee, \&
  Gonzalez}]{Bouwens:TheAstrophysicalJournal:2011}
Bouwens, R.~J., Illingworth, G.~D., Oesch, P.~A., {et~al.} 2011, The
  Astrophysical Journal, 737

\bibitem[{Brammer {et~al.}(2008)Brammer, van Dokkum, \&
  Coppi}]{Brammer:TheAstrophysicalJournal:2008}
Brammer, G.~B., van Dokkum, P.~G., \& Coppi, P. 2008, The Astrophysical
  Journal, 686

\bibitem[{Brinchmann {et~al.}(2004)Brinchmann, Charlot, White, Tremonti,
  Kauffmann, Heckman, \& Brinkmann}]{Brinchmann2004}
Brinchmann, J., Charlot, S., White, S. D.~M., {et~al.} 2004, Monthly Notices of
  the Royal Astronomical Society, 351, 1151

\bibitem[{Bruzual \& Charlot(2003)}]{Bruzual2003}
Bruzual, G. \& Charlot, S. 2003, Monthly Notices of the Royal Astronomical
  Society, 344, 1000

\bibitem[{Carrasco~Kind \&
  Brunner(2014)}]{CarrascoKind:MonthlyNoticesOfTheRoyalAstronomicalSociety:2014}
Carrasco~Kind, M. \& Brunner, R.~J. 2014, Monthly Notices of the Royal
  Astronomical Society, 442, 3380

\bibitem[{Cavuoti {et~al.}(2017)Cavuoti, Tortora, Brescia, Longo, Radovich,
  Napolitano, Amaro, Vellucci, La~Barbera, Getman, \&
  Grado}]{Cavuoti:MonthlyNoticesOfTheRoyalAstronomicalSociety:2017}
Cavuoti, S., Tortora, C., Brescia, M., {et~al.} 2017, Monthly Notices of the
  Royal Astronomical Society, 466, 2039

\bibitem[{Chevallard \&
  Charlot(2016)}]{Chevallard:MonthlyNoticesOfTheRoyalAstronomicalSociety:2016}
Chevallard, J. \& Charlot, S. 2016, Monthly Notices of the Royal Astronomical
  Society, 462, 1415

\bibitem[{Coleman {et~al.}(1980)Coleman, Wu, \&
  Weedman}]{Coleman:TheAstrophysicalJournalSupplementSeries:1980}
Coleman, G.~D., Wu, C.~C., \& Weedman, D.~W. 1980, The Astrophysical Journal
  Supplement Series, 43, 393

\bibitem[{Collaboration(2005)}]{Collaboration:EprintArxivAstroPh0510346:2005}
Collaboration, T. D. E.~S. 2005, eprint arXiv:astro-ph/0510346

\bibitem[{Conroy(2013)}]{Conroy:AnnualReviewOfAstronomyAndAstrophysics:2013}
Conroy, C. 2013, Annual Review of Astronomy and Astrophysics, 51, 393

\bibitem[{Cunha {et~al.}(2014)Cunha, Huterer, Lin, Busha, \&
  Wechsler}]{Cunha:MonthlyNoticesOfTheRoyalAstronomicalSociety:2014}
Cunha, C.~E., Huterer, D., Lin, H., Busha, M.~T., \& Wechsler, R.~H. 2014,
  Monthly Notices of the Royal Astronomical Society, 444, 129

\bibitem[{Dahlen {et~al.}(2013)Dahlen, Mobasher, Faber, Ferguson, Barro,
  Finkelstein, Finlator, Fontana, Gruetzbauch, Johnson, Pforr, Salvato,
  Wiklind, Wuyts, Acquaviva, Dickinson, Guo, Huang, Huang, Newman, Bell,
  Conselice, Galametz, Gawiser, Giavalisco, Grogin, Hathi, Kocevski, Koekemoer,
  Koo, Lee, McGrath, Papovich, Peth, Ryan, Somerville, Weiner, \&
  Wilson}]{Dahlen:TheAstrophysicalJournal:2013}
Dahlen, T., Mobasher, B., Faber, S.~M., {et~al.} 2013, The Astrophysical
  Journal, 775, 93

\bibitem[{Ellis {et~al.}(2013)Ellis, McLure, Dunlop, Robertson, Ono, Schenker,
  Koekemoer, Bowler, Ouchi, Rogers, Curtis-Lake, Schneider, Charlot, Stark,
  Furlanetto, \& Cirasuolo}]{Ellis:TheAstrophysicalJournalLetters:2013}
Ellis, R.~S., McLure, R.~J., Dunlop, J.~S., {et~al.} 2013, The Astrophysical
  Journal Letters, 763, L7

\bibitem[{Erb {et~al.}(2010)Erb, Pettini, Shapley, Steidel, Law, \&
  Reddy}]{Erb:TheAstrophysicalJournal:2010}
Erb, D.~K., Pettini, M., Shapley, A.~E., {et~al.} 2010, The Astrophysical
  Journal, 719, 1168

\bibitem[{Fan {et~al.}(2006)Fan, Strauss, Becker, White, Gunn, Knapp, Richards,
  Schneider, Brinkmann, \& Fukugita}]{fan2006constraining}
Fan, X., Strauss, M.~A., Becker, R.~H., {et~al.} 2006, The Astronomical
  Journal, 132, 117

\bibitem[{Faucher-Gigu\`{e}re {et~al.}(2008)Faucher-Gigu\`{e}re, Prochaska,
  Lidz, Hernquist, \&
  Zaldarriaga}]{FaucherGiguere:TheAstrophysicalJournal:2008}
Faucher-Gigu\`{e}re, C.-A. .~A., Prochaska, J.~X., Lidz, A., Hernquist, L., \&
  Zaldarriaga, M. 2008, The Astrophysical Journal, 681, 831

\bibitem[{Feldmann {et~al.}(2006)Feldmann, Carollo, Porciani, Lilly, Capak,
  Taniguchi, Le~F\`{e}vre, Renzini, Scoville, Ajiki, Aussel, Contini,
  McCracken, Mobasher, Murayama, Sanders, Sasaki, Scarlata, Scodeggio, Shioya,
  Silverman, Takahashi, Thompson, \&
  Zamorani}]{Feldmann:MonthlyNoticesOfTheRoyalAstronomicalSociety:2006}
Feldmann, R., Carollo, C.~M., Porciani, C., {et~al.} 2006, Monthly Notices of
  the Royal Astronomical Society, 372, 565

\bibitem[{Grogin {et~al.}(2011)Grogin, Kocevski, Faber, Ferguson, Koekemoer,
  Riess, Acquaviva, Alexander, Almaini, Ashby, Barden, Bell, Bournaud, Brown,
  Caputi, Casertano, Cassata, Castellano, Challis, Chary, Cheung, Cirasuolo,
  Conselice, Roshan~Cooray, Croton, Daddi, Dahlen, Dav\'{e}, de~Mello, Dekel,
  Dickinson, Dolch, Donley, Dunlop, Dutton, Elbaz, Fazio, Filippenko,
  Finkelstein, Fontana, Gardner, Garnavich, Gawiser, Giavalisco, Grazian, Guo,
  Hathi, H\"{a}ussler, Hopkins, Huang, Huang, Jha, Kartaltepe, Kirshner, Koo,
  Lai, Lee, Li, Lotz, Lucas, Madau, McCarthy, McGrath, McIntosh, McLure,
  Mobasher, Moustakas, Mozena, Nandra, Newman, Niemi, Noeske, Papovich,
  Pentericci, Pope, Primack, Rajan, Ravindranath, Reddy, Renzini, Rix, Robaina,
  Rodney, Rosario, Rosati, Salimbeni, Scarlata, Siana, Simard, Smidt,
  Somerville, Spinrad, Straughn, Strolger, Telford, Teplitz, Trump, van~der
  Wel, Villforth, Wechsler, Weiner, Wiklind, Wild, Wilson, Wuyts, Yan, \&
  Yun}]{Grogin:TheAstrophysicalJournalSupplementSeries:2011}
Grogin, N.~A., Kocevski, D.~D., Faber, S.~M., {et~al.} 2011, The Astrophysical
  Journal Supplement Series, 197

\bibitem[{Gutkin {et~al.}(2016)Gutkin, Charlot, \&
  Bruzual}]{Gutkin:MonthlyNoticesOfTheRoyalAstronomicalSociety:2016}
Gutkin, J., Charlot, S., \& Bruzual, G. 2016, Monthly Notices of the Royal
  Astronomical Society, 462, 1757

\bibitem[{Herenz {et~al.}(2017)Herenz, Urrutia, Wisotzki, Kerutt, Saust,
  Werhahn, Schmidt, Caruana, Diener, Bacon, Brinchman, Schaye, Maseda, \&
  Weilbacher}]{Herenz:EprintArxiv170508215:2017}
Herenz, E.~C., Urrutia, T., Wisotzki, L., {et~al.} 2017, Astronomy and
Astrophysics, 606

\bibitem[{Hildebrandt {et~al.}(2010)Hildebrandt, Arnouts, Capak, Moustakas,
  Wolf, Abdalla, Assef, Banerji, Ben\'{i}tez, Brammer, Budav\'{a}ri, Carliles,
  Coe, Dahlen, Feldmann, Gerdes, Gillis, Ilbert, Kotulla, Lahav, Li, Miralles,
  Purger, Schmidt, \& Singal}]{Hildebrandt:AstronomyAndAstrophysics:2010}
Hildebrandt, H., Arnouts, S., Capak, P., {et~al.} 2010, Astronomy and
  Astrophysics, 523

\bibitem[{Hildebrandt {et~al.}(2016)Hildebrandt, Viola, Heymans, Joudaki,
  Kuijken, Blake, Erben, Joachimi, Klaes, Miller, Morrison, Nakajima,
  Verdoes~Kleijn, Amon, Choi, Covone, de~Jong, Dvornik, Fenech~Conti, Grado,
  Harnois-D\'{e}raps, Herbonnet, Hoekstra, K\"{o}hlinger, McFarland, Mead,
  Merten, Napolitano, Peacock, Radovich, Schneider, Simon, Valentijn, van~den
  Busch, van Uitert, \&
  Van~Waerbeke}]{Hildebrandt:Kids450CosmologicalParameterConstraintsFromTomographicWeak:2016}
Hildebrandt, H., Viola, M., Heymans, C., {et~al.} 2016, KiDS-450: Cosmological
  parameter constraints from tomographic weak gravitational lensing

\bibitem[{Hildebrandt {et~al.}(2008)Hildebrandt, Wolf, \&
  Ben\'{i}tez}]{Hildebrandt:AstronomyAstrophysics:2008}
Hildebrandt, H., Wolf, C., \& Ben\'{i}tez, N. 2008, Astronomy \& Astrophysics,
  480, 703

\bibitem[{Hinton {et~al.}(2016)Hinton, Davis, Lidman, Glazebrook, \&
  Lewis}]{Hinton:AstronomyAndComputing:2016}
Hinton, S.~R., Davis, T.~M., Lidman, C., Glazebrook, K., \& Lewis, G.~F. 2016,
  Astronomy and Computing, 15, 61

\bibitem[{Ilbert {et~al.}(2006)Ilbert, Arnouts, McCracken, Bolzonella, Bertin,
  F\`{e}vre, Mellier, Zamorani, Pell\`{o}, Iovino, Tresse, Brun, Bottini,
  Garilli, Maccagni, Picat, Scaramella, Scodeggio, Vettolani, Zanichelli,
  Adami, Bardelli, Cappi, Charlot, Ciliegi, Contini, Cucciati, Foucaud,
  Franzetti, Gavignaud, Guzzo, Marano, Marinoni, Mazure, Meneux, Merighi,
  Paltani, Pollo, Pozzetti, Radovich, Zucca, Bondi, Bongiorno, Busarello,
  Torre, Gregorini, Lamareille, Mathez, Merluzzi, Ripepi, Rizzo, \&
  Vergani}]{OIlbert:AstronomyAstrophysics:2006}
Ilbert, O., Arnouts, S., McCracken, H., {et~al.} 2006, Astronomy \&
  Astrophysics, 457, 841

\bibitem[{Ilbert {et~al.}(2009)Ilbert, Capak, Salvato, Aussel, McCracken,
  Sanders, Scoville, Kartaltepe, Arnouts, Floc'h, Mobasher, Taniguchi,
  Lamareille, Leauthaud, Sasaki, Thompson, Zamojski, Zamorani, Bardelli,
  Bolzonella, Bongiorno, Brusa, Caputi, Carollo, Contini, Cook, Coppa,
  Cucciati, de~Torre, de~Ravel, Franzetti, Garilli, Hasinger, Iovino, Kampczyk,
  Kneib, Knobel, Kovac, Borgne, Brun, F\`{e}vre, Lilly, Looper, Maier,
  Mainieri, Mellier, Mignoli, Murayama, Pell\`{o}, Peng, P\'{e}rez-Montero,
  Renzini, Ricciardelli, Schiminovich, Scodeggio, Shioya, Silverman, Surace,
  Tanaka, Tasca, Tresse, Vergani, \& Zucca}]{27739469}
Ilbert, O., Capak, P., Salvato, M., {et~al.} 2009, Astrophysical Journal, 690,
  1236

\bibitem[{{Inami} {et~al.}(2017){Inami}, {Bacon}, {Brinchmann},
  {et~al.}}]{Inami2017}
{Inami}, H., {Bacon}, R., {Brinchmann}, J., {et~al.} 2017, \aap, submitted
  (MUSE UDF SI paper II)

\bibitem[{Inoue(2011)}]{Inoue:MonthlyNoticesOfTheRoyalAstronomicalSociety:2011}
Inoue, A.~K. 2011, Monthly Notices of the Royal Astronomical Society, 415, 2920

\bibitem[{Inoue {et~al.}(2014)Inoue, Shimizu, Iwata, \&
  Tanaka}]{Inoue:MonthlyNoticesOfTheRoyalAstronomicalSociety:2014}
Inoue, A.~K., Shimizu, I., Iwata, I., \& Tanaka, M. 2014, Monthly Notices of
  the Royal Astronomical Society, 442, 1805

\bibitem[{Ivezic {et~al.}(2008)Ivezic, Tyson, Acosta, Allsman, Anderson,
  Andrew, Angel, Axelrod, Barr, Becker, Becla, Beldica, Blandford, Bloom,
  Borne, Brandt, Brown, Bullock, Burke, Chandrasekharan, Chesley, Claver,
  Connolly, Cook, Cooray, Covey, Cribbs, Cutri, Daues, Delgado, Ferguson,
  Gawiser, Geary, Gee, Geha, Gibson, Gilmore, Gressler, Hogan, Huffer, Jacoby,
  Jain, Jernigan, Jones, Juric, Kahn, Kalirai, Kantor, Kessler, Kirkby, Knox,
  Krabbendam, Krughoff, Kulkarni, Lambert, Levine, Liang, Lim, Lupton,
  Marshall, Marshall, May, Miller, Mills, Monet, Neill, Nordby, O'Connor,
  Oliver, Olivier, Olsen, Owen, Peterson, Petry, Pierfederici, Pietrowicz,
  Pike, Pinto, Plante, Radeka, Rasmussen, Ridgway, Rosing, Saha, Schalk,
  Schindler, Schneider, Schumacher, Sebag, Seppala, Shipsey, Silvestri, Smith,
  Smith, Strauss, Stubbs, Sweeney, Szalay, Thaler, Berk, Walkowicz, Warner,
  Willman, Wittman, Wolff, Wood-Vasey, Yoachim, \& Zhan}]{12288492}
Ivezic, Z., Tyson, J.~A., Acosta, E., {et~al.} 2008

\bibitem[{Kinney {et~al.}(1996)Kinney, Calzetti, Bohlin, McQuade,
  Storchi-Bergmann, \& Schmitt}]{Kinney:TheAstrophysicalJournal:1996}
Kinney, A.~L., Calzetti, D., Bohlin, R.~C., {et~al.} 1996, The Astrophysical
  Journal, 467, 38

\bibitem[{Koekemoer {et~al.}(2013)Koekemoer, Ellis, McLure, Dunlop, Robertson,
  Ono, Schenker, Ouchi, Bowler, Rogers, Curtis-Lake, Schneider, Charlot, Stark,
  Furlanetto, Cirasuolo, Wild, \&
  Targett}]{Koekemoer:TheAstrophysicalJournalSupplementSeries:2013}
Koekemoer, A.~M., Ellis, R.~S., McLure, R.~J., {et~al.} 2013, The Astrophysical
  Journal Supplement Series, 209, 3

\bibitem[{Koekemoer {et~al.}(2011)Koekemoer, Faber, Ferguson, Grogin, Kocevski,
  Koo, Lai, Lotz, Lucas, McGrath, Ogaz, Rajan, Riess, Rodney, Strolger,
  Casertano, Castellano, Dahlen, Dickinson, Dolch, Fontana, Giavalisco,
  Grazian, Guo, Hathi, Huang, van~der Wel, Yan, Acquaviva, Alexander, Almaini,
  Ashby, Barden, Bell, Bournaud, Brown, Caputi, Cassata, Challis, Chary,
  Cheung, Cirasuolo, Conselice, Roshan~Cooray, Croton, Daddi, Dav\'{e},
  de~Mello, de~Ravel, Dekel, Donley, Dunlop, Dutton, Elbaz, Fazio, Filippenko,
  Finkelstein, Frazer, Gardner, Garnavich, Gawiser, Gruetzbauch, Hartley,
  H\"{a}ussler, Herrington, Hopkins, Huang, Jha, Johnson, Kartaltepe,
  Khostovan, Kirshner, Lani, Lee, Li, Madau, McCarthy, McIntosh, McLure,
  McPartland, Mobasher, Moreira, Mortlock, Moustakas, Mozena, Nandra, Newman,
  Nielsen, Niemi, Noeske, Papovich, Pentericci, Pope, Primack, Ravindranath,
  Reddy, Renzini, Rix, Robaina, Rosario, Rosati, Salimbeni, Scarlata, Siana,
  Simard, Smidt, Snyder, Somerville, Spinrad, Straughn, Telford, Teplitz,
  Trump, Vargas, Villforth, Wagner, Wandro, Wechsler, Weiner, Wiklind, Wild,
  Wilson, Wuyts, \&
  Yun}]{Koekemoer:TheAstrophysicalJournalSupplementSeries:2011}
Koekemoer, A.~M., Faber, S.~M., Ferguson, H.~C., {et~al.} 2011, The
  Astrophysical Journal Supplement Series, 197

\bibitem[{Laureijs {et~al.}(2011)Laureijs, Amiaux, Arduini, Augu\`{e}res,
  Brinchmann, Cole, Cropper, Dabin, Duvet, Ealet, Garilli, Gondoin, Guzzo,
  Hoar, Hoekstra, Holmes, Kitching, Maciaszek, Mellier, Pasian, Percival,
  Rhodes, Saavedra~Criado, Sauvage, Scaramella, Valenziano, Warren, Bender,
  Castander, Cimatti, Le~F\`{e}vre, Kurki-Suonio, Levi, Lilje, Meylan, Nichol,
  Pedersen, Popa, Rebolo~Lopez, Rix, Rottgering, Zeilinger, Grupp, Hudelot,
  Massey, Meneghetti, Miller, Paltani, Paulin-Henriksson, Pires, Saxton,
  Schrabback, Seidel, Walsh, Aghanim, Amendola, Bartlett, Baccigalupi,
  Beaulieu, Benabed, Cuby, Elbaz, Fosalba, Gavazzi, Helmi, Hook, Irwin, Kneib,
  Kunz, Mannucci, Moscardini, Tao, Teyssier, Weller, Zamorani, Zapatero~Osorio,
  Boulade, Foumond, Di~Giorgio, Guttridge, James, Kemp, Martignac, Spencer,
  Walton, Bl\"{u}mchen, Bonoli, Bortoletto, Cerna, Corcione, Fabron, Jahnke,
  Ligori, Madrid, Martin, Morgante, Pamplona, Prieto, Riva, Toledo, Trifoglio,
  Zerbi, Abdalla, Douspis, Grenet, Borgani, Bouwens, Courbin, Delouis, Dubath,
  Fontana, Frailis, Grazian, Koppenh\"{o}fer, Mansutti, Melchior, Mignoli,
  Mohr, Neissner, Noddle, Poncet, Scodeggio, Serrano, Shane, Starck, Surace,
  Taylor, Verdoes-Kleijn, Vuerli, Williams, Zacchei, Altieri, Escudero~Sanz,
  Kohley, Oosterbroek, Astier, Bacon, Bardelli, Baugh, Bellagamba, Benoist,
  Bianchi, Biviano, Branchini, Carbone, Cardone, Clements, Colombi, Conselice,
  Cresci, Deacon, Dunlop, Fedeli, Fontanot, Franzetti, Giocoli, Garcia-Bellido,
  Gow, Heavens, Hewett, Heymans, Holland, Huang, Ilbert, Joachimi, Jennins,
  Kerins, Kiessling, Kirk, Kotak, Krause, Lahav, van Leeuwen, Lesgourgues,
  Lombardi, Magliocchetti, Maguire, Majerotto, Maoli, Marulli, Maurogordato,
  McCracken, McLure, Melchiorri, Merson, Moresco, Nonino, Norberg, Peacock,
  Pello, Penny, Pettorino, Di~Porto, Pozzetti, Quercellini, Radovich, Rassat,
  Roche, Ronayette, Rossetti, Sartoris, Schneider, Semboloni, Serjeant,
  Simpson, Skordis, Smadja, Smartt, Spano, Spiro, Sullivan, Tilquin, Trotta,
  Verde, Wang, Williger, Zhao, Zoubian, \&
  Zucca}]{Laureijs:EuclidDefinitionStudyReport:2011}
Laureijs, R., Amiaux, J., Arduini, S., {et~al.} 2011, Euclid Definition Study
  Report

\bibitem[{{Leclercq} {et~al.}(2017){Leclercq}, {Bacon}, {Wisotzki},
  {et~al.}}]{Leclercq2017}
{Leclercq}, F., {Bacon}, R., {Wisotzki}, L., {et~al.} 2017, \aap, submitted
  (MUSE UDF SI paper VIII)

\bibitem[{Lima {et~al.}(2008)Lima, Cunha, Oyaizu, Frieman, Lin, \&
  Sheldon}]{Lima:MonthlyNoticesOfTheRoyalAstronomicalSociety:2008}
Lima, M., Cunha, C.~E., Oyaizu, H., {et~al.} 2008, Monthly Notices of the Royal
  Astronomical Society, 390, 118

\bibitem[{Madau(1995)}]{Madau:TheAstrophysicalJournal:1995}
Madau, P. 1995, The Astrophysical Journal, 441, 18

\bibitem[{Maraston(2005)}]{Maraston2005}
Maraston, C. 2005, Monthly Notices of the Royal Astronomical Society, 362, 799

\bibitem[{{Maseda} {et~al.}(2017){Maseda}, {Brinchmann}, {Franx},
  {et~al.}}]{Maseda2017}
{Maseda}, M., {Brinchmann}, J., {Franx}, M., {et~al.} 2017, \aap, in press
  (MUSE UDF SI paper IV)

\bibitem[{Masters {et~al.}(2015)Masters, Capak, Stern, Ilbert, Salvato,
  Schmidt, Longo, Rhodes, Paltani, Mobasher, Hoekstra, Hildebrandt, Coupon,
  Steinhardt, Speagle, Faisst, Kalinich, Brodwin, Brescia, \&
  Cavuoti}]{Masters:TheAstrophysicalJournal:2015}
Masters, D., Capak, P., Stern, D., {et~al.} 2015, The Astrophysical Journal,
  813, 53

\bibitem[{Momcheva {et~al.}(2016)Momcheva, Brammer, van Dokkum, Skelton,
  Whitaker, Nelson, Fumagalli, Maseda, Leja, Franx, Rix, Bezanson, Da~Cunha,
  Dickey, F\"{o}rster~Schreiber, Illingworth, Kriek, Labb\'{e}, Ulf~Lange,
  Lundgren, Magee, Marchesini, Oesch, Pacifici, Patel, Price, Tal, Wake,
  van~der Wel, \&
  Wuyts}]{Momcheva:TheAstrophysicalJournalSupplementSeries:2016}
Momcheva, I.~G., Brammer, G.~B., van Dokkum, P.~G., {et~al.} 2016, The
  Astrophysical Journal Supplement Series, 225

\bibitem[{Newman(2008)}]{Newman:TheAstrophysicalJournal:2008}
Newman, J.~A. 2008, The Astrophysical Journal, 684

\bibitem[{Newman {et~al.}(2015)Newman, Abate, Abdalla, Allam, Allen, Ansari,
  Bailey, Barkhouse, Beers, Blanton, Brodwin, Brownstein, Brunner,
  Carrasco~Kind, Cervantes-Cota, Cheu, Chisari, Colless, Comparat, Coupon,
  Cunha, de~la Macorra, DellAntonio, Frye, Gawiser, Gehrels, Grady, Hagen,
  Hall, Hearin, Hildebrandt, Hirata, Ho, Honscheid, Huterer, Ivezi\'{c}, Kneib,
  Kruk, Lahav, Mandelbaum, Marshall, Matthews, M\'{e}nard, Miquel, Moniez,
  Moos, Moustakas, Myers, Papovich, Peacock, Park, Rahman, Rhodes, Ricol,
  Sadeh, Slozar, Schmidt, Stern, Anthony~Tyson, von~der Linden, Wechsler,
  Wood-Vasey, \& Zentner}]{Newman:AstroparticlePhysics:2015}
Newman, J.~A., Abate, A., Abdalla, F.~B., {et~al.} 2015, Astroparticle Physics,
  63, 81

\bibitem[{Oesch {et~al.}(2010)Oesch, Bouwens, Carollo, Illingworth, Magee,
  Trenti, Stiavelli, Franx, Labb\'{e}, \& van
  Dokkum}]{Oesch:TheAstrophysicalJournal:2010}
Oesch, P.~A., Bouwens, R.~J., Carollo, C.~M., {et~al.} 2010, The Astrophysical
  Journal, 725, L150

\bibitem[{Oyarz\'{u}n {et~al.}(2016)Oyarz\'{u}n, Blanc, Gonz\'{a}lez, Mateo,
  Bailey, Finkelstein, Lira, Crane, \&
  Olszewski}]{Oyarzun:TheAstrophysicalJournal:2016}
Oyarz\'{u}n, G.~A., Blanc, G.~A., Gonz\'{a}lez, V., {et~al.} 2016, The
  Astrophysical Journal, 821

\bibitem[{Quadri \& Williams(2010)}]{Quadri:TheAstrophysicalJournal:2010}
Quadri, R.~F. \& Williams, R.~J. 2010, The Astrophysical Journal, 725, 794

\bibitem[{Rafelski {et~al.}(2015)Rafelski, Teplitz, Gardner, Coe, Bond,
  Koekemoer, Grogin, Kurczynski, McGrath, Bourque, Atek, Brown, Colbert,
  Codoreanu, Ferguson, Finkelstein, Gawiser, Giavalisco, Gronwall, Hanish, Lee,
  Mehta, de~Mello, Ravindranath, Ryan, Scarlata, Siana, Soto, \&
  Voyer}]{Rafelski:TheAstronomicalJournal:2015}
Rafelski, M., Teplitz, H.~I., Gardner, J.~P., {et~al.} 2015, The Astronomical
  Journal, 150

\bibitem[{Sadeh {et~al.}(2016)Sadeh, Abdalla, \&
  Lahav}]{Sadeh:PublicationsOfTheAstronomicalSocietyOfThe:2016}
Sadeh, I., Abdalla, F.~B., \& Lahav, O. 2016, Publications of the Astronomical
  Society of the Pacific, 128, 104502

\bibitem[{Salmon {et~al.}(2015)Salmon, Papovich, Finkelstein, Tilvi, Finlator,
  Behroozi, Dahlen, Dav\'{e}, Dekel, Dickinson, Ferguson, Giavalisco, Long, Lu,
  Mobasher, Reddy, Somerville, \&
  Wechsler}]{Salmon:TheAstrophysicalJournal:2015}
Salmon, B., Papovich, C., Finkelstein, S.~L., {et~al.} 2015, The Astrophysical
  Journal, 799, 183

\bibitem[{Salvato {et~al.}(2011)Salvato, Ilbert, Hasinger, Rau, Civano,
  Zamorani, Brusa, Elvis, Vignali, Aussel, Comastri, Fiore, Floc'h, Mainieri,
  Bardelli, Bolzonella, Bongiorno, Capak, Caputi, Cappelluti, Carollo, Contini,
  Garilli, Iovino, Fotopoulou, Fruscione, Gilli, Halliday, Kneib, Kakazu,
  Kartaltepe, Koekemoer, Kovac, Ideue, Ikeda, Impey, Fevre, Lamareille,
  Lanzuisi, Borgne, Brun, Lilly, Maier, Manohar, Masters, McCracken, Messias,
  Mignoli, Mobasher, Nagao, Pello, Puccetti, Perez-Montero, Renzini, Sargent,
  Sanders, Scodeggio, Scoville, Shopbell, Silvermann, Taniguchi, Tasca, Tresse,
  Trump, \& Zucca}]{Salvato:TheAstrophysicalJournal:2011}
Salvato, M., Ilbert, O., Hasinger, G., {et~al.} 2011, The Astrophysical
  Journal, 742, 61

\bibitem[{Skelton {et~al.}(2014)Skelton, Whitaker, Momcheva, Brammer, van
  Dokkum, Labb\'{e}, Franx, van~der Wel, Bezanson, Da~Cunha, Fumagalli,
  F\"{o}rster~Schreiber, Kriek, Leja, Lundgren, Magee, Marchesini, Maseda,
  Nelson, Oesch, Pacifici, Patel, Price, Rix, Tal, Wake, \&
  Wuyts}]{Skelton:TheAstrophysicalJournalSupplementSeries:2014}
Skelton, R.~E., Whitaker, K.~E., Momcheva, I.~G., {et~al.} 2014, The
  Astrophysical Journal Supplement Series, 214

\bibitem[{Spergel {et~al.}(2015)Spergel, Gehrels, Baltay, Bennett,
  Breckinridge, Donahue, Dressler, Gaudi, Greene, Guyon, Hirata, Kalirai,
  Kasdin, Macintosh, Moos, Perlmutter, Postman, Rauscher, Rhodes, Wang,
  Weinberg, Benford, Hudson, Jeong, Mellier, Traub, Yamada, Capak, Colbert,
  Masters, Penny, Savransky, Stern, Zimmerman, Barry, Bartusek, Carpenter,
  Cheng, Content, Dekens, Demers, Grady, Jackson, Kuan, Kruk, Melton, Nemati,
  Parvin, Poberezhskiy, Peddie, Ruffa, Wallace, Whipple, Wollack, \&
  Zhao}]{Spergel:WideFieldInfrarredSurveyTelescopeAstrophysicsFocusedTelescopeAssets:2015}
Spergel, D., Gehrels, N., Baltay, C., {et~al.} 2015, Wide-Field InfrarRed
  Survey Telescope-Astrophysics Focused Telescope Assets WFIRST-AFTA 2015
  Report

\bibitem[{S\"{u}veges {et~al.}(2017)S\"{u}veges, Fotopoulou, Coupon, Paltani,
  Eyer, \&
  Rimoldini}]{Suveges:LearnFromEveryMistakeHierarchicalInformationCombination:2017}
S\"{u}veges, M., Fotopoulou, S., Coupon, J., {et~al.} 2017, Learn from every
  mistake! Hierarchical information combination in astronomy

\bibitem[{Tasca {et~al.}(2017)Tasca, F\`{e}vre, Ribeiro, Thomas, Moreau,
  Cassata, Garilli, Brun, Lemaux, Maccagni, Pentericci, Schaerer, Vanzella,
  Zamorani, Zucca, Amorin, Bardelli, Cassar\`{a}, Castellano, Cimatti,
  Cucciati, Durkalec, Fontana, Giavalisco, Grazian, Hathi, Ilbert, Paltani,
  Pforr, Scodeggio, Sommariva, Talia, Tresse, Vergani, Capak, Charlot, Contini,
  de~Torre, Dunlop, Fotopoulou, Guaita, Koekemoer, L\'{o}pez-Sanjuan, Mellier,
  Salvato, Scoville, Taniguchi, \& Wang}]{Tasca:AstronomyAstrophysics:2017}
Tasca, L. A.~M., F\`{e}vre, O.~L., Ribeiro, B., {et~al.} 2017, Astronomy \&
  Astrophysics, 600, A110

\bibitem[{Teplitz {et~al.}(2013)Teplitz, Rafelski, Kurczynski, Bond, Grogin,
  Koekemoer, Atek, Brown, Coe, Colbert, Ferguson, Finkelstein, Gardner,
  Gawiser, Giavalisco, Gronwall, Hanish, Lee, de~Mello, Ravindranath, Ryan,
  Siana, Scarlata, Soto, Voyer, \& Wolfe}]{Teplitz:TheAstronomicalJournal:2013}
Teplitz, H.~I., Rafelski, M., Kurczynski, P., {et~al.} 2013, The Astronomical
  Journal, 146

\bibitem[{Walcher {et~al.}(2010)Walcher, Groves, Budav\'{a}ri, \&
  Dale}]{Walcher:AstrophysicsAndSpaceScience:2010}
Walcher, J., Groves, B., Budav\'{a}ri, T., \& Dale, D. 2010, Astrophysics and
  Space Science, 331, 1

\bibitem[{Weilbacher {et~al.}(2012)Weilbacher, Streicher, Urrutia, Jarno,
  P\'{e}contal-Rousset, Bacon, \&
  B\"{o}hm}]{Weilbacher:DesignAndCapabilitiesOfTheMuseData:2012}
Weilbacher, P.~M., Streicher, O., Urrutia, T., {et~al.} 2012, Design and
  capabilities of the MUSE data reduction software and pipeline, Vol. 8451

\bibitem[{Williams {et~al.}(1996)Williams, Blacker, Dickinson, Dixon, Ferguson,
  Fruchter, Giavalisco, Gilliland, Heyer, Katsanis, Levay, Lucas, McElroy,
  Petro, Postman, Adorf, \& Hook}]{Williams:TheAstronomicalJournal:1996}
Williams, R.~E., Blacker, B., Dickinson, M., {et~al.} 1996, The Astronomical
  Journal, 112, 1335

\bibitem[{Wisotzki {et~al.}(2016)Wisotzki, Bacon, Blaizot, Brinchmann, Herenz,
  Schaye, Bouch\'{e}, Cantalupo, Contini, Carollo, Caruana, Courbot, Emsellem,
  Kamann, Kerutt, Leclercq, Lilly, Patr\'{i}cio, Sandin, Steinmetz, Straka,
  Urrutia, Verhamme, Weilbacher, \&
  Wendt}]{Wisotzki:AstronomyAndAstrophysics:2016}
Wisotzki, L., Bacon, R., Blaizot, J., {et~al.} 2016, Astronomy and
  Astrophysics, 587

\bibitem[{Zhan(2006)}]{Zhan:JournalOfCosmologyAndAstroParticlePhysics:2006}
Zhan, H. 2006, Journal of Cosmology and Astro-Particle Physics, 8

\end{thebibliography}

\bibliographystyle{aa}

\appendix

\section{Adding Spitzer IRAC and ground-based photometry}
\label{sec:adding-ground-based}

The paper focuses on photometry from R15, but this is limited to HST
imaging and while up to 11 bands are available, the reddest filter is
F160W and there are no intermediate width filters. This can be
contrasted with the study by S14 which has
photometry in a total of 44 bands, including the $K_s$-band and
Spitzer IRAC photometry in 3.6, 4.5, 5.8, and 8 $\mu$m. We have not
used this catalogue as our master catalogue for three main reasons: 1)
it does not go as deep as R15, thus many more galaxies with a MUSE
redshift have no corresponding photometry in the catalogue; 2) the
source extraction and redshift determinations were done on the R15
catalogue (see Paper I and II for details); and 3) the S14 catalogue
does not include the UV photometry from
\citet{Teplitz:TheAstronomicalJournal:2013}. 

\begin{figure}
  \centering
  \includegraphics[width=84mm]{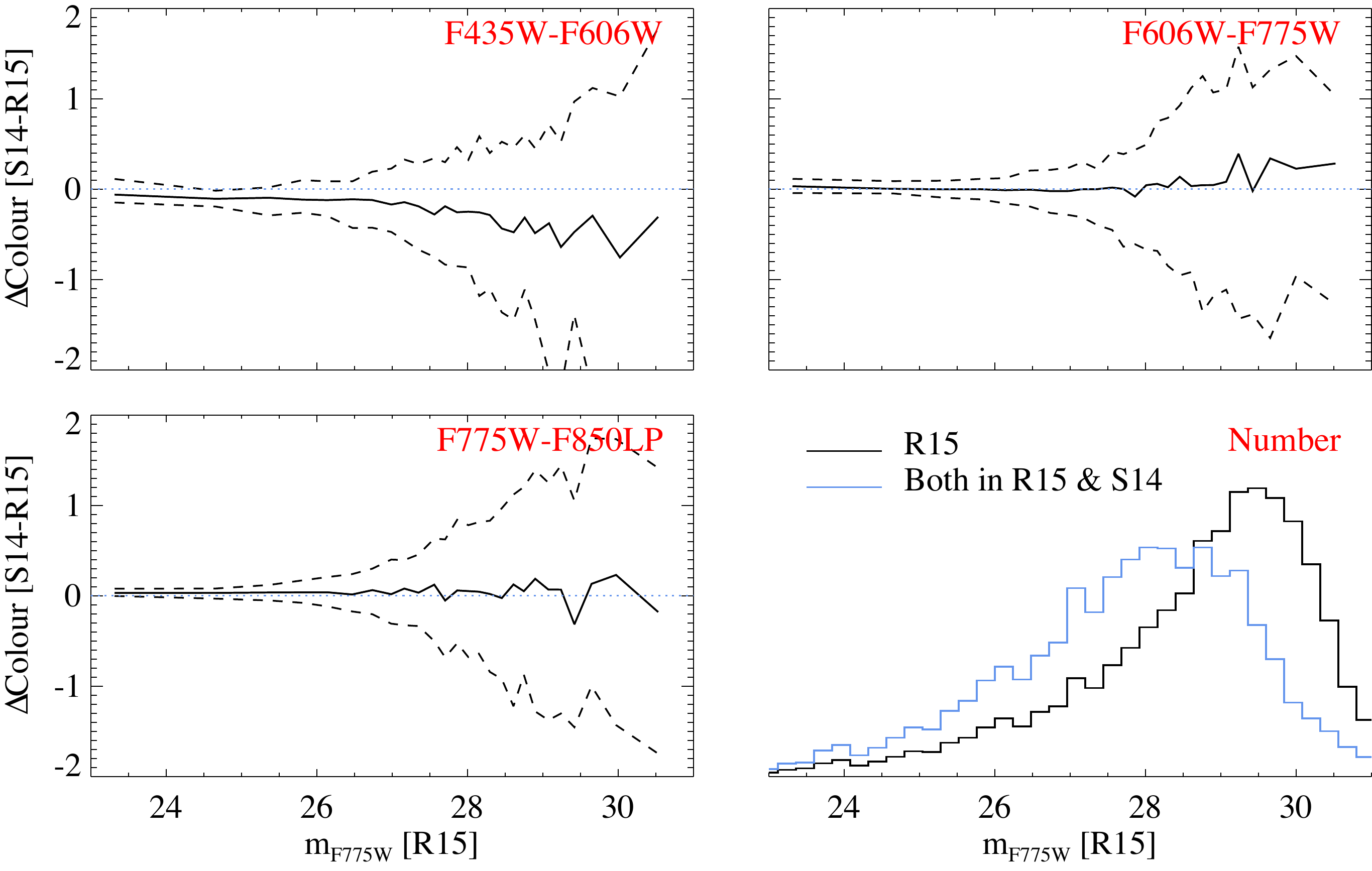}
  \caption{Top left: The difference in $\mhst{F435W}-\mhst{F606W}$
    colour between S14 and R15 as a function of the F775W magnitude in
    R15. The solid line shows the median in bins containing 151
    objects each, with the dashed lines enclosing
    68\% of all objects.The horizontal dotted line shows the zero level
    to help comparisons. Top right: the same, but for the
    $\mhst{F606W}-\mhst{F775W}$. Bottom left: the same for the
    $\mhst{F775W}-\mhst{F850lp}$ colour. Bottom right: in black the number of
    objects in the R15 catalogue over the MUSE \mosaic\ field as a
    function of \mhst{F775W}, while the blue histogram shows the
    number of objects that are in the matched catalogue of R15 and
    S14. The histograms are both normalised to unit area to ease
    comparison. }
  \label{fig:s14_vs_r15_color_comp}
\end{figure}

However the S14 catalogue offers the possibility to test how the
photo-z estimates are influenced by the photometric catalogue used. To
do this, we have matched the photometric catalogues from
S14\footnote{Downloaded from
  \url{http://3dhst.research.yale.edu/Data.php}} to the master
catalogue used here, requiring that the catalogue positions differ by
less than 0\farcs 2. This results in 3,956 matches for which 3,899
have five or more colours, so we can calculate reliable photometric
redshifts. For these 3,899 we have MUSE redshifts with confidence
$\ge 2$ (1) for 905 (1071). It is important to keep in mind that due
to difference in image segmentation, two matched detections might
differ rather significantly although for most objects we expect good
correspondence. This is borne out in a direct comparison of photometry
with a slight difference in F435W photometry with S14 being slightly
brighter than R15. This is shown indirectly in
Figure~\ref{fig:s14_vs_r15_color_comp} which compares the
$\mhst{F435W}-\mhst{F606W}$, $\mhst{F606W}-\mhst{F775W}$, and
$\mhst{F775W}-\mhst{F850lp}$ colours calculated using the S14
photometry to that calculated using the R15 photometry (the
conclusions are unchanged when using flux ratios rather than
magnitudes). The solid line shows the median trend and it can be seen
that as a function of the \mhst{F775W} magnitude from R15, the colours
are broadly in good agreement with execption of the aforementioned
offset in F435W which could be caused by the iterative adjustment of
photometry done by S14. The scatter is substantial at faint magnitudes
but consistent with that expected from the photometric
uncertainties. The final panel in the figure shows the number of
objects in the R15 catalogue within the MUSE \mosaic\ field of the
view (black) as a function of F775W magnitude. This can be contrasted
to the joint catalogue of S14 and R15 (blue) which one can see is
relatively deficient in the faintest objects since both histograms are
normalised to unit area.

\begin{figure}
  \centering
  \includegraphics[width=84mm]{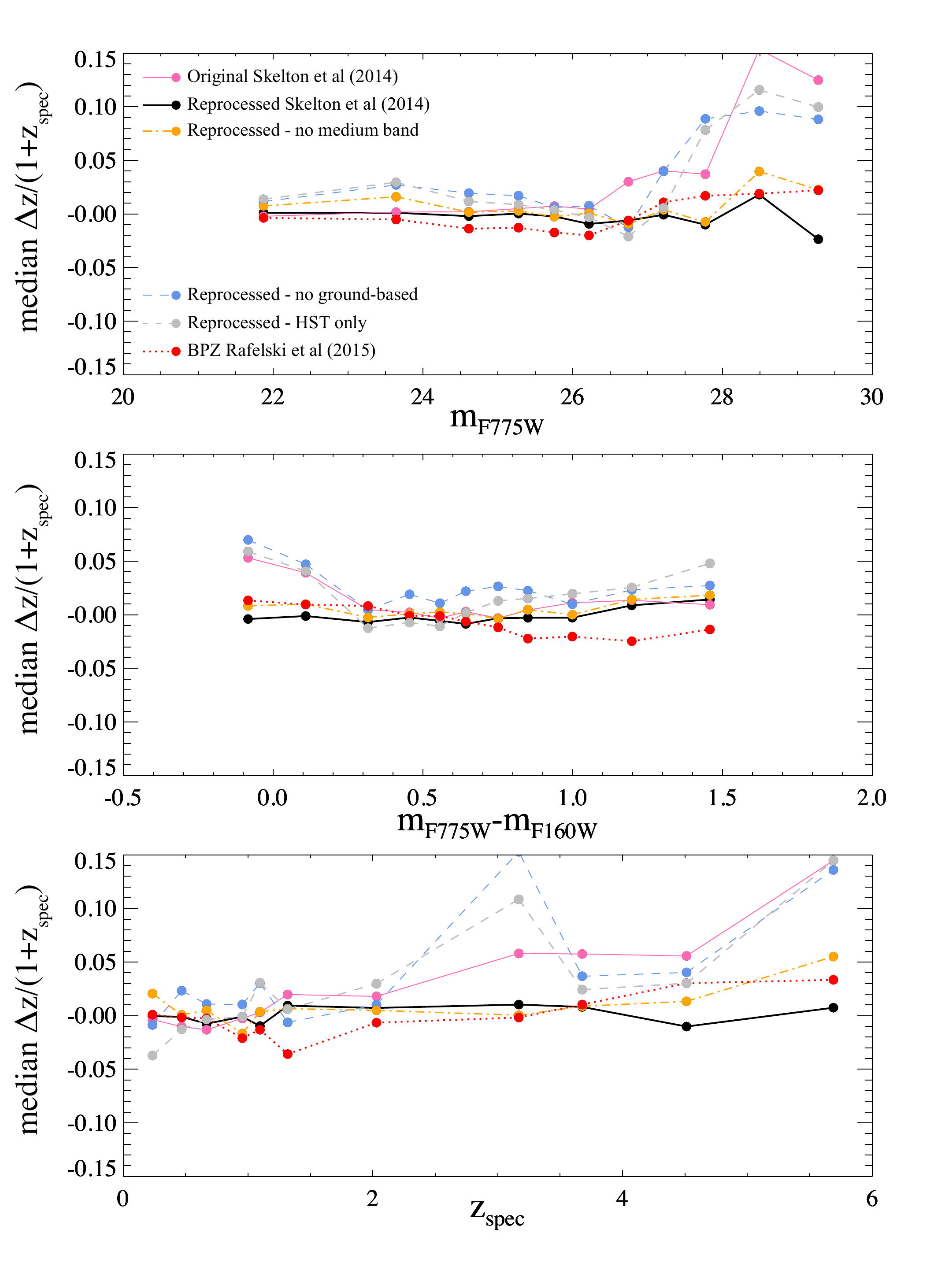}
  \caption{Bias, defined as the median of \dzn\ as a function of
    \mhst{F775W} magnitude (top), $\mhst{F775W}-\mhst{F160W}$ colour
    (middle), and \zMUSE\ (bottom) from fitting EAZY to the S14
    photometry. The pink thin, solid line shows the results using the
    photo-zs from
    \citet{Skelton:TheAstrophysicalJournalSupplementSeries:2014},
    while the thicker black line shows the results after using the
    optimised EAZY settings described in the text. The dash-dotted
    orange line shows the results when excluding the medium-band
    filters, the dashed blue line the result when excluding all
    ground-based photometry, and then short-dashed grey line shows the
    result when only HST photometry is included in the fit. The
    results from the BPZ code run on the R15 photometry is included
    for reference as the dashed red line (BEAGLE, and EAZY run on R15
    photomety gives similar results).}
  \label{fig:sk14_bias}
\end{figure}

\begin{figure}
  \centering
  \includegraphics[width=84mm]{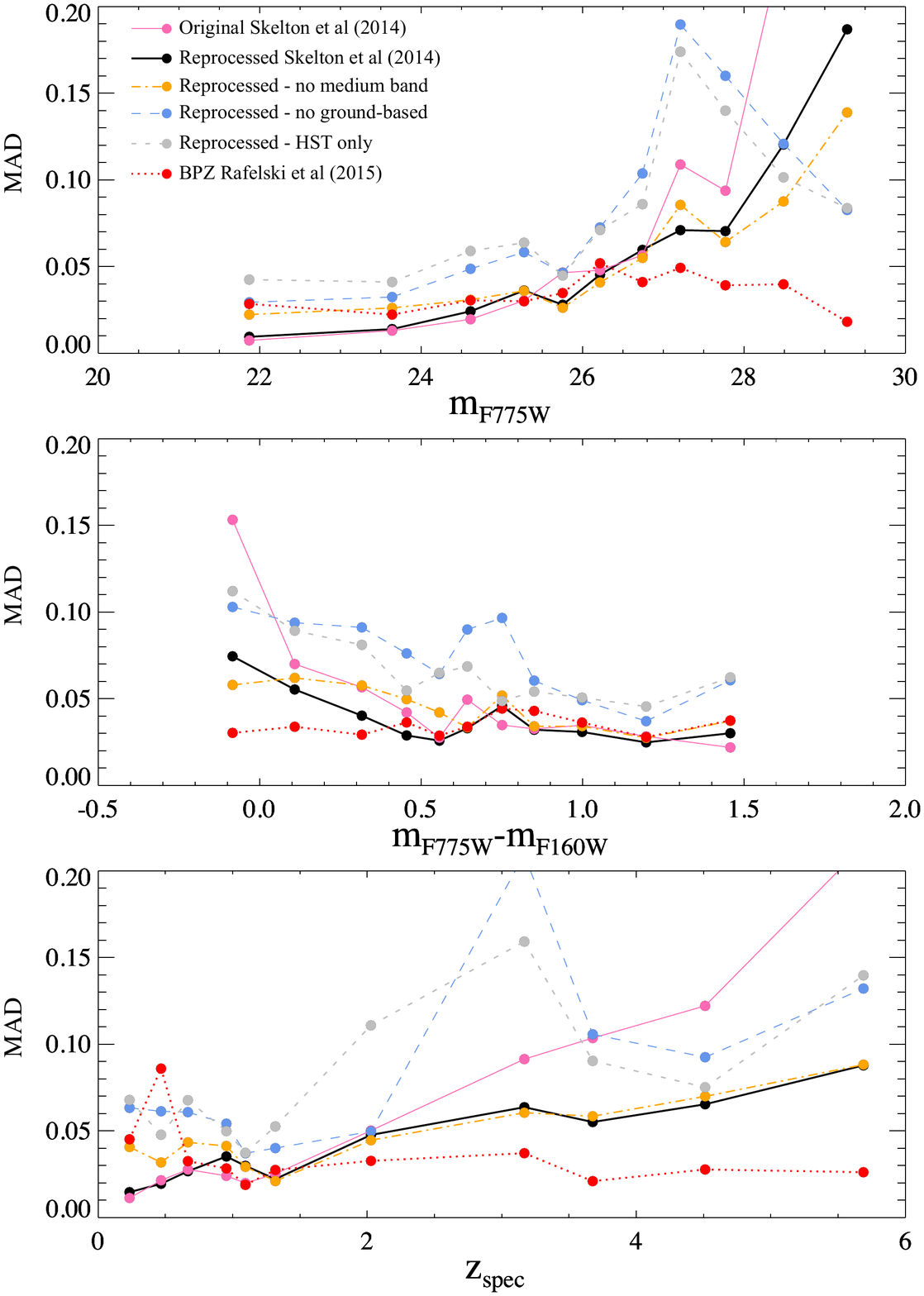}
  \caption{Scatter, defined as $\mathrm{MAD}\left(\dzn\right)$, for
  the S14 photometric catalogue. The symbols are the same as in
  Figure~\ref{fig:sk14_bias}.}
  \label{fig:sk14_scatter}
\end{figure}

\begin{figure}
  \centering
  \includegraphics[width=84mm]{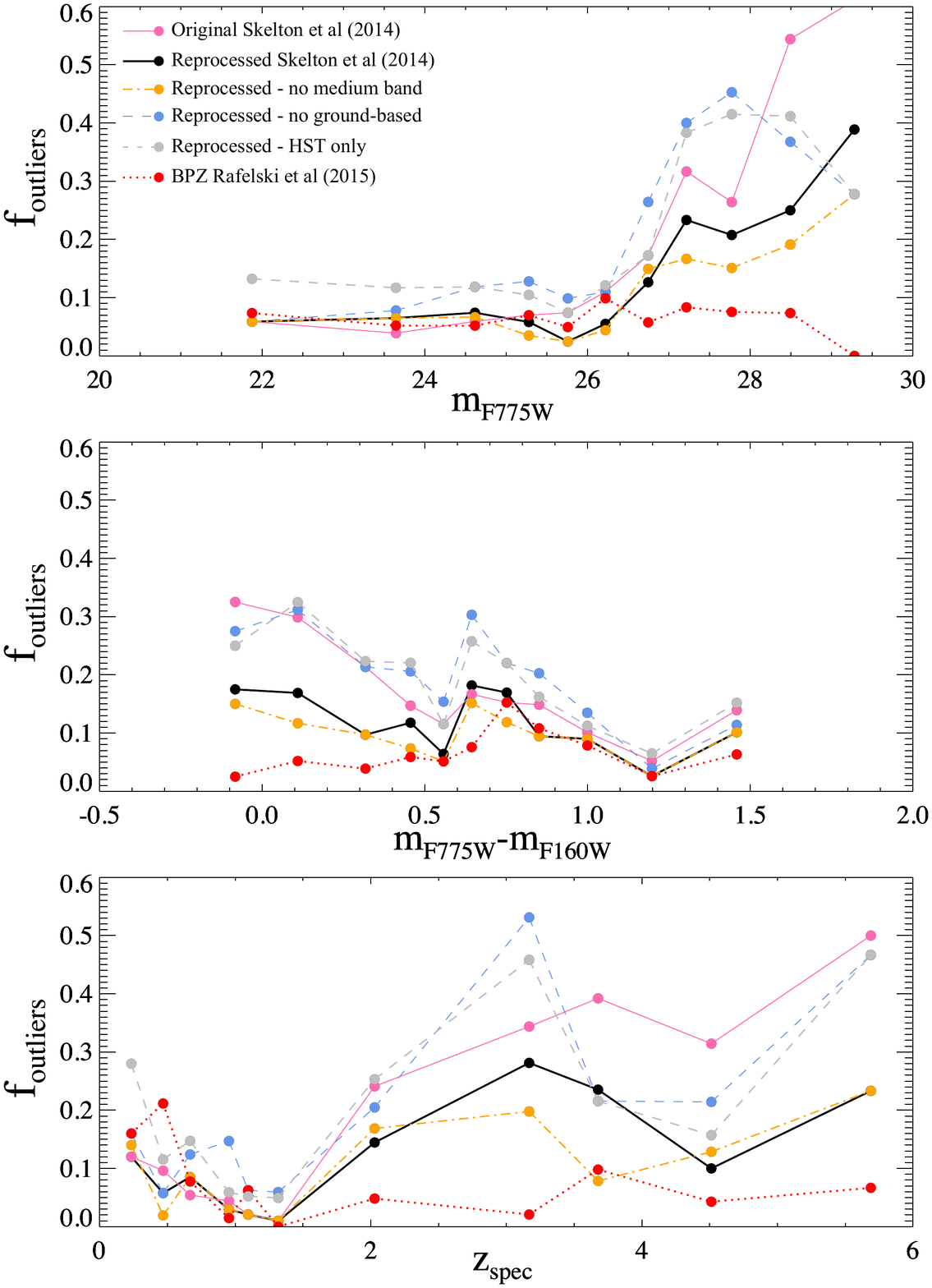}
  \caption{Fraction of outliers, defined as
    $\left|\dzn\right| > 0.15$, for the S14 photometric catalogue. The
    symbols are the same as in Figure~\ref{fig:sk14_bias}.}
  \label{fig:sk14_outliers}
\end{figure}

We run EAZY using the same settings that we found gave the best
results in section~\ref{sec:reduc-bias-phot} but we will also show the
results from S14 below. We will refer to these re-runs of EAZY as the
reprocessed S14 photo-zs. Our reference run uses all 44 filters and
is plotted as a thicker black line in
Figures~\ref{fig:sk14_bias}--~\ref{fig:sk14_outliers}. To test the
sensitivity of the predictions to the photometry available, we also
run EAZY excluding the 14 medium band filters (dash-dotted orange line
in the figures), and excluding all ground-based photometry (dashed
blue line). To first order, removing filters can be viewed as removing
information, one would therefore expect that the runs with a larger
number of filters will result in photo-zs that better match the
spectroscopic redshifts.

This basic assumption is borne out in Figure~\ref{fig:sk14_bias},
where we show the median $\dzn$ as a function of \mhst{F775W}
magnitude (top), $\mhst{F775W}-\mhst{F160W}$ colour (middle), and
\zMUSE\ (bottom). 

The first thing to note is that while they perform well, the original
S14 photo-zs (thin pink line) are considerably worse than the BPZ
(dotted red line) and the reprocessed photo-zs. This demonstrates the
same result that we found in section~\ref{sec:reduc-bias-phot} and
underlines the importance of the choice of template set. We also see
that the reprocessed photo-zs that use all photometry perform best in
terms of median bias and shows no magnitude and very weak redshift
trends. It is also clear that removing the medium band photometry
reduces performance, and removing all ground-based photometry makes
the photo-zs considerably worse.

Finally, it is also noteworthy that the BPZ photo-zs using the RAF
catalogue which only use HST photometry, perform much better than the
corresponding reprocessed photo-zs that only uses space-based
(HST+Spitzer) photometry. This might at first glance be somewhat
surprising, but there are two main reasons why this might be so:
firstly, as mentioned above, the R15 catalogue uses HST UV photometry
of the UDF that were not used by S14. This is likely to improve
photometry redshift performance, particularly at lower
redshift. Secondly, the assignent of MUSE redshift to an HST object
was based on the RAF catalogue and their segmentation map, but we have
not done a similar scrutiny of the S14 catalogue so some of the S14
photometric objects may refer to an object that has a different
\zMUSE, or which includes several distinct objects at different
redshift in the RAF catalogue.

The same broad trends can also be seen in
Figure~\ref{fig:sk14_scatter}, which shows the scatter around the
median (calculated as $\mathrm{MAD}\left[\dzn\right]$). The most
notable point here is that the scatter rises sharply towards the
faintest magnitudes for all, except the R15 based BPZ photo-zs. We
should note that just as for R15, we have used all S14 photometry
blindly despite S14 suggesting this is not a good idea, but the
scatter at the bright end matches very well the $\mathrm{MAD}=0.010$
found for this field in
\citet{Bezanson:TheAstrophysicalJournal:2016}. Thus the increased
scatter at fainter magnitudes might simply be a reflection of the
challenge of providing reliable IRAC and ground-based photometric
measurements for these sources. This is also indicated by the fact
that the reprocessed photo-zs that omit all ground-based photometry
perform as well or better than the run with all 44 bands at the very
faintest magnitudes ($\mhst{F775W}>28$). To show this explicitly, we
have run EAZY on the catalogue with only HST photometry, that is
excluding IRAC photometry as well. The results of this is shown as the
short-dashed grey line in
Figures~\ref{fig:sk14_bias}--~\ref{fig:sk14_outliers}, which should be
compared to the dashed blue line. The top panel of
Figure~\ref{fig:sk14_scatter}, for instance, shows that the scatter is
generally lower when IRAC fluxes as included until
$\mhst{F775W}\approx 26$, where the two curves cross over. The bias is
less affected and the inclusion of IRAC fluxes improves the bias
slightly down to $\mhst{F775W}\approx 28$. At brighter magnitudes the
removal of IRAC data worsens the performance, in good agreement with
the findings of \citet{Bezanson:TheAstrophysicalJournal:2016}. These
findings are not particularly surprising: when IRAC photometry is of
good quality it will help constrain the stellar populations of a
galaxy, particularly at high redshift, thus adding IRAC fluxes should
improve photo-z estimates as long as the quality of the photometry is
sufficient. For faint galaxies it is much more challenging to extract
reliable IRAC fluxes and hence including them in the photo-z fit might
lead to incorrect solutions.

Finally, Figure~\ref{fig:sk14_outliers} shows the fraction of outliers
for the same photo-z runs as in the previous two figures. The outlier
fraction has here been defined as galaxies having
$\left|\dzn\right|>0.15$ and we see a strong increase in the outlier
fraction towards fainter magnitudes for the S14 original and
reprocessed photo-zs, in good quantitative agreement with
\citet{Bezanson:TheAstrophysicalJournal:2016}'s study using grism
redshifts. This appears to be a reflection of the increased scatter at
fainter magnitudes seen in Figure~\ref{fig:sk14_outliers}.

In summary then, we can see that adding more bands can lead to a
smaller bias, but the increased scatter and related high outlier
fraction at faint magnitudes means that we prefer the R15 catalogue
for the present study as this performs better with 11 bands than the
S14 catalogue with 44 at faint magnitudes. At bright magnitudes we are
limited by the number of objects with \zMUSE, but data are consistent
with the statement that the increased number of filters in the S14
catalogue leads to better performing photo-zs than the R15
catalogue. This would also agree well with the findings of 
\citet{Bezanson:TheAstrophysicalJournal:2016} for the same field.

\section{The significant (``catastrophic'') outliers}
\label{sec:serial-outliers}

Finally, it of interest to comment on the galaxies for which the
photo-zs are in strong disagreement with the spectroscopic
redshift. Here we will define two different classes of these
outliers. We will refer to the galaxies for which
$\left|\dzn\right| > 0.5$ as catastrophic absolute outliers and those
galaxies for which $\left|\dzn\right| > 5\sigma_{\mathrm{MAD}}(x)$,
where $\sigma_{\mathrm{MAD}}(x)$ is the local standard deviation at
auxiliary quantity $x$, as catastrophic relative outliers.

%
%
\begin{table*}
  \centering
  \caption{Fraction of significant outliers for the three
    different photo-z codes}
\begin{tabular}{rrrrrrrrr}
\multicolumn{9}{c}{$\left|\dzn\right| > 0.5$ absolute outliers} \\ \hline
  & \multicolumn{2}{c}{EAZY} &  \multicolumn{2}{c}{BPZ} &  \multicolumn{2}{c}{BEAGLE} &  \multicolumn{2}{c}{All} \\ 
       &  N  & \%  &  N  &  \%  &  N & \% & N & \% \\ 
$\mhst{F775W}<30$   &   47 &   4.34 &   17 &   1.57 &   20 &   1.85 &    6 &   0.55 \\
$\mhst{F775W}<27$   &   18 &   2.87 &    7 &   1.11 &    8 &   1.27 &    2 &   0.32 \\  
  \hline\hline

\multicolumn{9}{c}{$5\sigma$ relative outliers \rule[-0.5ex]{0pt}{2.5ex}} \\ \hline
  & \multicolumn{2}{c}{EAZY} &  \multicolumn{2}{c}{BPZ} &  \multicolumn{2}{c}{BEAGLE} &  \multicolumn{2}{c}{All} \\ 
       &  N  & \%  &  N  &  \%  &  N & \% & N & \% \\
vs \mhst{F775W}  &   78 &   7.20 &   56 &   5.17 &   60 &   5.54 &   27 &   2.49 \\ 
vs \zMUSE  &    75 &   6.92 &   61 &   5.63 &   56 &   5.17 &   24 &   2.21 \\ 
vs $\mhst{F775W}-\mhst{F160W}$  &   68 &   6.64 &   55 &   5.37 &   50 &   4.88 &   21 &   2.05 \\ 
\end{tabular}
  \label{tab:significant_outliers}
\end{table*}

The set of relative outliers does depend on the quantity which you
calculate the local scatter against, $x$ in the defintion above. In
practice, however, this is not a strong effect,

In Table~\ref{tab:significant_outliers} we provide a summary of the
two types of catastropic outliers. For the absolute outliers, we took
all galaxies brighter than $\mhst{F775W}=30$ and 27 respectively,
whereas for relative outliers we used all galaxies to
$\mhst{F775W}=30$ and we have excluded the 11 galaxies that have no
BEAGLE photometric redshift.  It is clear from the table that BEAGLE
and BPZ perform best with a catastrophic outlier fraction of $<1.3$\%
for galaxies brighter than $\mhst{F775W}=27$ although we caution that
low-number statistics start to play a significant role here. The last
two columns show the number and fraction of outliers where all
codes report photo-zs strongly in disagreement with the spectroscopic
redshift and it is clear that this is lower by a factor of $\sim 4$
than the absolute outlier fraction for each code. The implication of
this is that combining different estimators for identification of
catastrophic outliers ought to be a good way to identify these.

The catastrophic relative outliers present much the same
picture. Again BEAGLE and BPZ perform clearly better than BPZ and
clearly better than EAZY, and again we see a clear potential for a
gain from combining methods.

\begin{figure*}
  \centering
  \includegraphics[height=0.90\textheight]{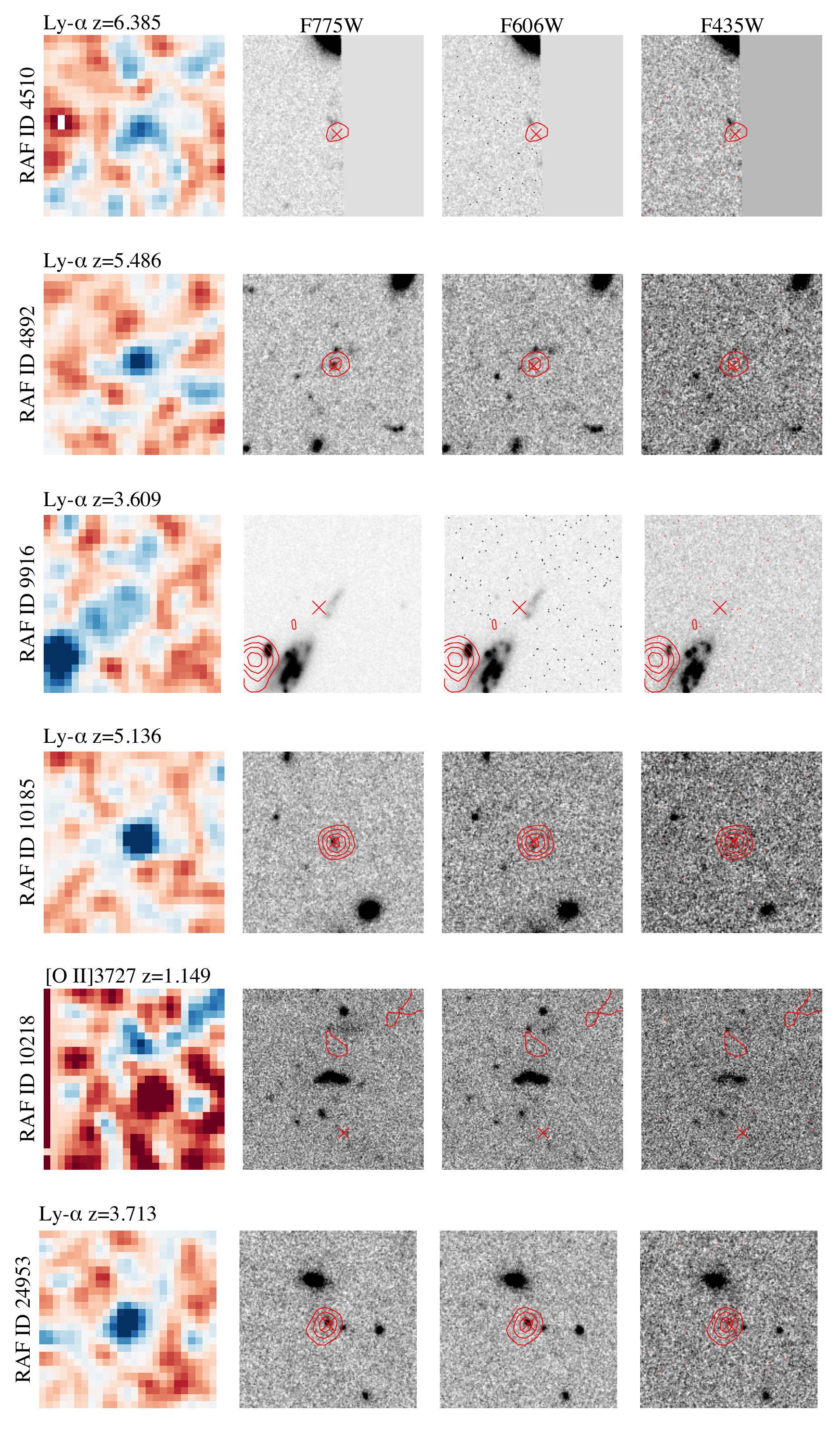}
  \caption{The six objects that are catastrophic absolute and
    relative outliers. The left column shows a smoothed narrow-band
    image over \lya, where the colour coding shows a formal S/N per
    pixel and goes from $-2$ to $3$. The subsequent images shows the
    F775W, F606W, and F435W images with the contours from the \lya\
    image overlaid, with the lowest contour at a formal S/N of 2 per
    pixel. The red cross marks the light-weighted centroid of the
    narrow-band image within the 1D spectrum extraction mask.}
  \label{fig:outliers}
\end{figure*}

Finally, let us briefly discuss the nature of the outliers. For
concreteness, we will focus on the most problematic objects, namely
those that are always both catastrophic relative and absolute
outliers, regardless of code. There are a total of 6 of these objects
and we are interested in exploring whether the failure of all codes to
fit these are due to problems with the data, or to intrisically add
properties of the sources. The narrow-band images over the strongest
line and the HST images are shown in Figure~\ref{fig:outliers}.

\textbf{RAF 4510, MUSE ID 7213}. This is a $\mhst{F775W}=27.9$
object with a clear narrow-band image over \lya\ and
$\zMUSE=6.385$. The photometric redshift estimates from BPZ
($\zBPZ=1.28$), EAZY ($\zEAZY=1.28$), and BEAGLE ($\zBEAGLE=1.36$) are
all in close agreement but far away from the spectroscopic
redshift. The \lya\ source is however not overlapping with the
photometric object, and the only photometric source (RAF 4510)
close to the \lya\ peak is visible also in \mhst{F435W}, ruling out a
$z>6$. Thus in this case, the most reasonable explanation is that the
photometric object is not associated to the fairly clear \lya\ line.

\textbf{RAF 4892, MUSE ID 7285}. This object has $\mhst{F775W}=29.0$
and $\zMUSE=5.496$. This is in strong disagreement with the
photometric redshifts which are $\zBPZ=0.76$, $\zEAZY=0.86$, and
$\zBEAGLE=0.99$. In this case the photometric object co-incident with
the \lya\ narrow-band image is clearly seen in the F775W image but is
not convincingly seen in the F606W image, although it has a catalogue
flux of $\mhst{F606W}=30.31$, and it is entirely absent in F453W. The
source is in a crowded region and the F606W photometry might be is
incorrect, leading to the mismatched photometric redshift.

\textbf{RAF 9916, MUSE ID 1504}. This $\mhst{F775W}=26.24$ object
has $\zMUSE=3.609$ and $\zBPZ=0.25$, $\zEAZY=0.41$, and
$\zBEAGLE=0.46$. The most prominent feature in the narrow-band image
over \lya\ is the strong emitter with MUSE ID 6878 (RAF 7843) in
the lower left. There is a clear line also over RAF 9916 and it is
aligned with the object seen in the HST image. The bright source in
the image is MUSE ID 6877 (RAF 9958) at $\zMUSE=0.734$ In this case
it is entirely possible that the \lya\ seen over RAF 9916 belongs
to an extended emission region around MUSE ID 6678 and is unrelated to
RAF 9916. 

\textbf{RAF 10185, MUSE ID 399}. This object has
$\mhst{F775W}=28.89$ and the \lya\ narrow-band image is well aligned
with the HST source. It has $\zMUSE=5.136$ and $\zBPZ=0.61$,
$\zEAZY=0.56$, and $\zBEAGLE=0.59$. The line is strong with a
pronounced asymmetric shape and seems highly unlikely to be \oii{3727}
(which also would be in disagreement with the photo-zs). There is
residual light in the F606W image which in the rest-frame of the
source would have a central wavelength of 960\AA. The emission seen in
F606W, assuming the spectroscopic redshift is correct, would seem to
require a very transparent IGM for this source and this might be the
cause of the photometric redshift discrepancy.

\textbf{RAF 10218, MUSE ID 83}. This object has
$\mhst{F775W}=26.09$ and has been given a redshift of $\zMUSE=1.149$ in the 
\udft\ spectrum. The photometric redshifts all cluster near $z=3$, with 
$\zBPZ=2.86$, $\zEAZY=3.02$, and $\zBEAGLE=2.87$. The narrow-band
image over the putative \oii{3727} line is not convincing and upon
inspection of the spectrum, it seems to have  \lya\ absorption just at
the blue edge of the spectrum and a tentative detection of
Si\,\textsc{ii} $\lambda 1526$ and C\,\textsc{iv} $\lambda
1548,1550$ which would give $z=2.937$ in much better agreement with
the photo-zs. Thus in this case it is arguably the spectroscopic
redshift that is incorrect. 

\textbf{RAF 24953, MUSE ID 6702}. This object has $\zMUSE=3.713$
and  $\zBPZ=0.12$, $\zEAZY=0.10$, and $\zBEAGLE=0.86$. It is fairly
faint at $\mhst{F775W}=28.86$ and has a slightly asymmetric emission
line that has been identified as \lya. Identifying this line as
\oiii{5007} would bring the spectroscopic redshift in reasonable
agreement with BPZ and EAZY but there is no sign of \oiii{4959} nor
any other lines and they should be easily observed given the strength
of the one line seen, this seems highly unlikely. Thus the
spectroscopic redshift is almost certainly correct and it is unclear
what has caused this strong discrepancy.

\end{document}